 \documentclass[prb,twocolumn,showpacs,preprintnumbers]{revtex4}

\usepackage{graphicx,bm}
\usepackage{amsmath,amssymb}

\usepackage[dvips]{color}

\begin{document}
\title{Kondo effects in a triangular triple quantum dot: \\ 
NRG study in the whole region of the electron filling
}

\author{Takahide Numata$^1$, Yunori Nisikawa$^1$, Akira Oguri$^1$, 
and Alex C. Hewson$^2$}

\address{$^{1}$ Graduate School of Science, Osaka City University, 
Osaka 558-8585, Japan}

\address{$^{2}$ Department of Mathematics, Imperial College, 
 180 Queen's Gate, London SW7 2BZ, UK}


\date{\today}

\begin{abstract}
We study the low-energy properties of 
a triangular triple quantum dot connected to two non-interacting leads 
in a wide parameter range, using the numerical renormalization group (NRG). 
Various kinds of Kondo effects take place in this system 
depending on the  electron filling $N_\mathrm{tot}$, 
or the level position $\epsilon_d$ of the triple dot. 
The SU(4) Kondo behavior is seen 
 in the half-filled case $N_\mathrm{tot} \simeq 3.0$ 
 at the dip of the series conductance, 
and it causes a charge redistribution between 
the even and odd orbitals in the triangle. 
We show generally that the quasi-particle 
excitations from a local Fermi-liquid ground state 
acquire a channel symmetry at zero points 
of the two-terminal conductance, in the case 
the system has time-reversal and inversion symmetries. 
It causes the SU(4) behavior at low energies,
while the orbital degeneracy in the triangle 
determines the high-energy behavior.
At four-electron filling  $N_\mathrm{tot} \simeq 4.0$, 
a local $S=1$ moment emerges at high temperatures 
due to a Nagaoka ferromagnetic mechanism. 
It is fully screened by the electrons from the two conducting channels  
via a two-stage Kondo effect, which is caused by 
a difference in the charge distribution in the even and odd orbitals.

\end{abstract}


\maketitle

\section{Introduction}

The Kondo effect in quantum dots 
is an active field of current research.
\cite{GlatzmanRaikh,NgLee,Goldharber,Cronenwett} 
It has been studied in various situations, 
such as Aharanov-Bohm (AB) rings,\cite{Akera,Izumida1,Hofstetter,Kobayashi} 
tunneling junctions coupled to superconducting hosts, 
\cite{GlazmanMatveev,ClerkAmbegaokar,RozhkovArovas,Schoenenberger1,Johannes,Karrasch,Yoichi} 
and a number of multi-dot and multi-orbital systems.
\cite{Izumida2,Izumida3,ao99,aoQuasi}
Due to the variety of combinations of quantum dots that can be fabricated, 
and their high tunability, 
 it is possible to study novel 
interplays between the strong electron correlation and quantum-mechanical 
interference effects. These were once studied as different topics
in the field of the condensed matter physics.

The triangular triple quantum dot (TTQD) is an interesting system,
 which can demonstrate various types of the Kondo effects 
depending on an external field, 
the electron filling $N_\mathrm{tot}$, and so on.\cite{TKA1,ONTN} 
For instance, a closed path along the triangle under a magnetic field can work 
as an AB interferometer.\cite{TKA1} 
Also in the case four electrons occupying the cluster of the triple dot,
$N_\mathrm{tot} = 4.0$,
the closed loop  induces a local $S=1$ moment  
due to a Nagaoka ferromagnetic mechanism.\cite{ONTN}
Furthermore, the orbital degrees of freedom cause 
a sharp conductance dip appearing 
at half-filling $N_\mathrm{tot}=3.0$,\cite{ONTN}
which links to the SU(4) Kondo effects.
These features distinguish the TTQD from a linear chain 
of three quantum dots,\cite{OH,ONH,NO,TKA2,Zitko,Zitko_Bonca} 
and from the other three-level 
systems.\cite{Eto,Sakano,LeoFabrizio,Kita,Hecht_3,Paaske}
Experimentally, triple quantum-dot systems have been realized 
in AlGaAs/GaAs heterostructure,\cite{Vidan-Stopa,Canada,Amaha2} 
self-assembled InAs,\cite{Amaha1} and  
a single wall carbon nanotube.\cite{Lindelof_3dot} 
The triangular trimmer of Cr atoms placed upon an Au surface 
is also a related system.\cite{Jamneala,Affleck3a,Zarand}

The TTQD has a big parameter space to be explored, 
and the number of the conducting channels which are coupled to 
the triangle also gives interesting variations in 
low-energy properties.\cite{Numata,Zitko2,Logan3,Ulloa}
In Refs.\ \onlinecite{Zitko2} and \onlinecite{Logan3} 
the TTQD's connected, respectively, 
to two leads\cite{Zitko2} and one lead\cite{Logan3} 
were studied.  
The main interest in these two recent works was 
the behavior at half-filling $N_\mathrm{tot} \simeq 3.0$, 
but the effects of the gate voltage $\epsilon_d$ which 
varies the electron filling in the TTQD were not examined.  
The other recent report Ref.\ \onlinecite{Ulloa} addresses 
the effects of $\epsilon_d$ in the cases of the two and three leads, 
but the parameter region examined was restricted in 
a small interaction $U$ and a large dot-lead coupling $\Gamma$. 
In the previous works 
we studied the TTQD connected to two leads,\cite{ONTN,Numata}  
and showed that the various types of the Kondo behavior clearly 
take place at different values of the gate voltage. 
The calculations were carried out, however, 
for a typical but just one parameter set 
which  describes a large $U$ and small $\Gamma$ situation.

The purpose of the present work is to study 
the feature of the Kondo behavior in the TTQD connected 
to two leads in a wide parameter range. 
To this end, we calculate the conductance and scattering phase 
shifts at zero temperature, 
making use of the numerical renormalization group (NRG).
Furthermore, we calculate also the TTQD contributions to 
the entropy and spin susceptibility, 
and from their temperature dependence 
we deduce the Kondo energy scale. 
Our results reveal the precise features of the Kondo behavior 
in a wide parameter region of the electron filling $N_\mathrm{tot}$, 
interaction $U$, and the hybridization energy scale $\Gamma$. 
For instance, the $S=1$ moment is fully screened at low temperatures 
via two separate stages by the conduction electrons 
from the two non-interacting leads,  
which break the $C_{3v}$ symmetry of the triangle. 
The  two-stage screening process reflects 
the charge distribution in the even and odd orbitals which are 
classified according to the parity.
 It is confirmed that the $S=1$ Kondo behavior 
can be seen in a wide parameter region for $N_\mathrm{tot}=4.0$. 
It is robust against the perturbations  
the typical energy of which is less 
than the finite energy separation between the Nagaoka 
and the first excited states for the triangular cluster. 
Specifically, the $S=1$ Kondo behavior 
is not sensitive to the deformations of the TTQD 
caused by the inhomogeneities in the level positions 
and the inter-dot hopping matrix elements.

We show also that the fixed-point Hamiltonian, 
which describes the low-lying quasi-particle excitations 
from a local Fermi-liquid ground state, 
acquires an SU(2) symmetry between the even and odd channels 
at zero points of the two-terminal conductance,
in the case that the system has 
 time-reversal and inversion symmetries.
It means that if 
the phase shifts for the even and odd channels satisfy the condition 
$\delta_\mathrm{e}-\delta_\mathrm{o} = n \pi$ 
for $n = 0,\pm 1, \pm 2, \ldots$,  
then the low-energy behavior 
can be characterized by an SU(4) symmetric Fermi-liquid theory 
with the channel and spin symmetries, 
even if the original Hamiltonian does not 
have the SU(4) symmetry on a global energy scale.  
This happens quite generally for multi-dot and multi-orbital systems. 
In the case of the TTQD connected to two conducting channels, 
there are three zero points at most. 
Particularly, the one at half-filling $N_\mathrm{tot}=3.0$ 
shows a pronounced behavior as a sharp dip 
in the Kondo plateau.\cite{ONTN,Numata}
At this zero point the system has an SU(4) symmetry 
not only at low energies but also at high energies, 
which is caused the orbital degeneracy in the cluster 
of the regular triangular. 
These symmetric properties cause the SU(4) Kondo behavior
\cite{Borda,Logan_2dotA,Mravlje_Ramsak_Rejec,Anders_Logan} 
seen in a wide temperature range.\cite{Numata}
Away from this zero point at the dip 
the NRG results of the thermodynamic quantities 
show that the channel symmetry is broken at low temperatures, 
and we observe the SU(2) Kondo behavior due to the spin. 
We find also that the sharp dip structure of the series conductance 
does not disappear from the Kondo plateau,  
even in the presence of infinitesimal deformations 
which break the $C_{3v}$ symmetry, 
although the position of the dip shifts from 
the middle of the plateau as the deformation increases.

We study also the characteristic temperature $T^*$, 
at which a crossover to a singlet ground state take place, 
for a wide range of the electron filling $N_\mathrm{tot}$. 
The result shows that $T^*$ becomes high 
at the conductance dip for $N_\mathrm{tot} =3.0$ 
due to the SU(4) Kondo behavior.
It is enhanced also at a mixed-valence point, 
where the ground-state energy for $N_\mathrm{tot}=3.0$ 
and that for $N_\mathrm{tot}=4.0$ coincide. 
Furthermore, the screening temperature $T^*$ increases 
as the hybridization energy scale $\Gamma$ increases, 
which would raise the experimental accessibility to 
the low-temperature Fermi-liquid region.

The paper is organized as follows.
In Sec.\ \ref{sec:formulation}, 
we give the outline a description of the model and the
formulation.
In Sec.\ \ref{sec:zero_point_SU4}, 
we describe the low-energy channel symmetry that  
the fixed-point Hamiltonian acquire  
at the zero point of the two-terminal conductance. 
The NRG results for the ground-state properties are
presented in Sec.\ \ref{sec:results_I}. 
The results for  
temperature dependence of the TTQD contributions 
of entropy and spin susceptibility 
are presented in Sec.\ \ref{sec:thermo}. 
The effects of the deformations 
which breaks the triangular symmetry 
are discussed in Sec.\ \ref{sec:results_deformation}.
A summary is given in Sec.\ \ref{sec:summary}.

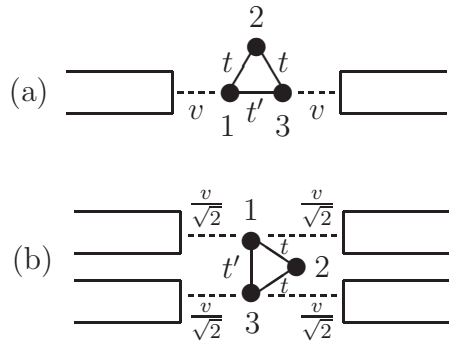
\begin{figure}[t]
\begin{center}
\setlength{\unitlength}{0.7mm}

\begin{minipage}[t]{7cm}


\begin{picture}(110,20)(0,0)
\hspace{-0.8cm}
\thicklines

\put(8.0,7){\makebox(0,0)[bl]{\large (a)}}

\put(19,6){\line(1,0){20}}
\put(19,14){\line(1,0){20}}
\put(39,6){\line(0,1){8}}

\put(71,6){\line(1,0){20}}
\put(71,14){\line(1,0){20}}
\put(71,6){\line(0,1){8}}

\multiput(40.0,10)(2,0){4}{\line(1,0){1}}
\multiput(62.9,10)(2,0){4}{\line(1,0){1}}

\put(55,18.6){\circle*{3.3}} 
\put(50,10){\circle*{3.3}} 
\put(60,10){\circle*{3.3}} 

\put(51,10){\line(1,0){8}}
\put(49.5,10){\line(3,5){6}}
\put(60.5,10){\line(-3,5){6}}

\put(42,5){\makebox(0,0)[bl]{\large $v$}}
\put(65,5){\makebox(0,0)[bl]{\large $v$}}

\put(53,4.5){\makebox(0,0)[bl]{\large $t'$}}
\put(48.5,14){\makebox(0,0)[bl]{\large $t$}}
\put(59,14){\makebox(0,0)[bl]{\large $t$}}

\put(48,1.8){\makebox(0,0)[bl]{\large $1$}}
\put(58.5,1.8){\makebox(0,0)[bl]{\large $3$}}
\put(53.5,22.5){\makebox(0,0)[bl]{\large $2$}}

\end{picture}
\end{minipage}
%
%

\setlength{\unitlength}{0.7mm}
\begin{minipage}[t]{7cm}

\vspace{0.2cm}

\begin{picture}(110,30)(-4,0)
\hspace{-0.8cm}
\thicklines

\put(4.5,13){\makebox(0,0)[bl]{\large (b)}}

\put(16.5,4.5){\line(1,0){20}}
\put(16.5,12.5){\line(1,0){20}}
\put(36.5,4.5){\line(0,1){8}}

\put(16.5,17.5){\line(1,0){20}}
\put(16.5,25.5){\line(1,0){20}}
\put(36.5,17.5){\line(0,1){8}}

\put(67.5,4.5){\line(1,0){20}}
\put(67.5,12.5){\line(1,0){20}}
\put(67.5,4.5){\line(0,1){8}}

\put(67.5,17.5){\line(1,0){20}}
\put(67.5,25.5){\line(1,0){20}}
\put(67.5,17.5){\line(0,1){8}}

\multiput(38.0,21)(2,0){5}{\line(1,0){1}}
\multiput(38.0,9.5)(2,0){5}{\line(1,0){1}}
\multiput(53.2,9.5)(2,0){7}{\line(1,0){1}}
\multiput(53.2,21)(2,0){7}{\line(1,0){1}}

\put(58.6,15){\circle*{3.3}} 
\put(50,10){\circle*{3.3}} 
\put(50,20){\circle*{3.3}} 

\put(50,10){\line(0,1){10}}
\put(51,10){\line(3,2){7}}
\put(51,20){\line(3,-2){7}}

\put(38,22.5){\makebox(0,0)[bl]{\large $\frac{v}{\sqrt{2}}$}}
\put(59,22.5){\makebox(0,0)[bl]{\large $\frac{v}{\sqrt{2}}$}}

\put(38,1){\makebox(0,0)[bl]{\large $\frac{v}{\sqrt{2}}$}}
\put(59,1){\makebox(0,0)[bl]{\large $\frac{v}{\sqrt{2}}$}}

\put(48.5,24){\makebox(0,0)[bl]{\large $1$}}
\put(48.8,2.5){\makebox(0,0)[bl]{\large $3$}}
\put(62,13.0){\makebox(0,0)[bl]{\large $2$}}

\put(44.5,12.8){\makebox(0,0)[bl]{\large $t'$}}
\put(55.5,17.5){\makebox(0,0)[bl]{$t$}}
 \put(55.5,10){\makebox(0,0)[bl]{$t$}}

\end{picture}
\end{minipage}
\end{center}
\caption{Triangular triple quantum dot 
in (a) series and (b) parallel configurations.
The dot labelled with $i=2$ is referred to 
as the {\em apex\/} site, which has no direct 
connection to the leads.}
\label{fig:system}
\end{figure}


\section{Formulation}
\label{sec:formulation}

In this section we describe the model 
and the formulation which we have employed in the present work. 
We also describe some particular characteristics 
of the triangular triple dot in the two solvable limits, 
$\Gamma=0$ and $U=0$, in order to see 
the essential ingredients that cause 
the various Kondo effects taking place in this system.

\subsection{Model}
\label{subsec:model}

We start with a three-site Hubbard model on a triangle 
that is connected to two non-interacting leads 
on the left ($L$) and right ($R$), 
at the sites labelled by $i=1$ and $i=N_D$ ($\equiv 3$), 
respectively, as illustrated in Fig.\ \ref{fig:system} (a).
The Hamiltonian is given by 
\begin{align}
&\mathcal{H} =  
\mathcal{H}_\mathrm{dot}^0 +
\mathcal{H}_\mathrm{dot}^U 
+  \mathcal{H}_\mathrm{mix}  +  
\mathcal{H}_\mathrm{lead}\;, 
\label{eq:H}
\\
& 
\mathcal{H}_\mathrm{dot}^0  = \,    
 - \sum_{<ij>}^{N_D}\sum_{\sigma} t_{ij} 
 \left(
 d^{\dagger}_{i\sigma}d^{\phantom{\dagger}}_{j\sigma}  
+ d^{\dagger}_{j\sigma}d^{\phantom{\dagger}}_{i\sigma}  
\right) 
 \nonumber
 \\
& \qquad \quad \ 
+ 
\sum_{i=1}^{N_D}\sum_{\sigma}  
\epsilon_{d,i} \,
 d^{\dagger}_{i\sigma}d^{\phantom{\dagger}}_{i\sigma} , 
\label{eq:Hdot^0}
\\
& 
\mathcal{H}_\mathrm{dot}^U  =     
  U\sum_{i=1}^{N_D} 
n_{d,i\uparrow}\,n_{d,i\downarrow} , 
\qquad \ \ 
n_{d,i\sigma} \equiv d^{\dagger}_{i\sigma}d^{}_{i\sigma},
\label{eq:HC^U}\\
&
\mathcal{H}_\mathrm{mix} =   
 \sum_{\sigma}   
 \left(  \,
v_L^{}\,   
d^{\dagger}_{1,\sigma}
 C^{\phantom{\dagger}}_{L \sigma}
   +   
v_R^{}\,  
d^{\dagger}_{N_D, \sigma} C^{\phantom{\dagger}}_{R\sigma}
+\, \mathrm{H.c.}
\, \right)   ,
 \label{eq:Hmix}
\\
& \mathcal{H}_\mathrm{lead} =  
\sum_{\nu=L,R} 
 \sum_{k\sigma} 
  \epsilon_{k}^{\phantom{0}}\,
         c^{\dagger}_{k \nu \sigma} 
         c^{\phantom{\dagger}}_{k \nu \sigma}
\,.
\label{eq:H_lead}
\end{align}
Here,   
$d^{\dagger}_{i\sigma}$ creates an electron with spin $\sigma$ 
at the $i$-th site in the dot, 
 $\epsilon_{d,i}$ the onsite energy in the dot, 
and $U$ the Coulomb interaction.
The hopping matrix elements $t_{ij}$ between the dots are chosen 
to be positive ($t_{ij}>0$). 
The conduction electrons near the dots, 
 $C_{\nu \sigma}^{\phantom{\dagger}} 
\equiv \sum_k c_{k \nu \sigma}^{\phantom{\dagger}}/\sqrt{N}$, 
can tunnel into the triangle via 
the hybridization term Eq.~\eqref{eq:Hmix},
which causes the level 
broadening $\Gamma_{\nu} \equiv \pi  \rho \, v^2_{\nu}$, 
with $\rho$ the density of states of the leads.
In the present work we consider 
the equal-coupling case $\Gamma_L^{} = \Gamma_R^{}$ ($\equiv \Gamma$),  
namely $v_L^{} = v_R^{}$ ($\equiv v$), 
and assume that the system has an inversion symmetry 
taking $\epsilon_{d,1}=\epsilon_{d,3}$ ($\equiv \epsilon_d$), 
$t_{12}=t_{23}$ ($\equiv t$) and $t_{13}$ ($\equiv t'$). 
We shall refer to the dot at $i=2$ in Fig.\ \ref{fig:system} 
as the {\em apex} site  
$\epsilon_{d,2} \equiv \epsilon_\mathrm{apex}$,  
and choose the Fermi energy $E_F$ as the origin of the energy $E_F=0$.

\subsection{Cluster of the triangular triple dot}
\label{subsec:isolated_cluster}

In order to see some characteristic features of the system,
we first of all discuss a cluster of a triangular triple dot,
 described by $\mathcal{H}_\mathrm{dot}^0 + \mathcal{H}_\mathrm{dot}^U$.
The eigenstates and energies of this cluster give us a clue to understand  
the high-energy properties of the system. 
To be specific, we consider a regular triangle case, 
choosing  $\epsilon_\mathrm{apex}=\epsilon_d$ and $t'=t$.

In the non-interacting case 
the one-particle states of  $\mathcal{H}_\mathrm{dot}^0$  
can be labelled by a wavenumber $k$ ($=0, \, \pm 2\pi/3$), and 
the eigenvalues are given 
by $E_k^{(1)} =-2t \cos k + \epsilon_d$,    
\begin{align}
E_{k=0}^{(1)}
 = -2t + \epsilon_d\;, \qquad \quad 
E_{k=\pm \frac{2\pi}{3}}^{(1)} = t + \epsilon_d \;. 
\label{eq:tight_bind_triangle}
\end{align}
The excited states, for $t>0$,  
are degenerate  with respect to $k=\pm 2\pi/3$. 
This degeneracy brings interesting varieties to the Kondo effects
depending on the occupation number $N_\mathrm{tot}$, 
or $\epsilon_d$, which can be 
controlled by the gate voltage.
In the case that the number of electrons 
is $N_\mathrm{tot}=3$, 
namely at half-filling,  
the ground state for the triangular cluster  
has the four-fold degeneracy due to the orbital and spin 
degrees of freedoms. This is because the first two electrons 
occupy the one-particle state for $k=0$, 
and then for the third electron 
there are four possible ways 
to occupy one of the orbitals of $k=\pm {2\pi}/{3}$ 
with the spin $\sigma=\uparrow,\downarrow$.
The four-fold degeneracy emerges also 
for a finite Coulomb interaction $\mathcal{H}_\mathrm{dot}^U$, 
and could cause the SU(4) Kondo behavior.

In the case that one additional electron 
is introduced into the triangular cluster,
the ground state of 
$\mathcal{H}_\mathrm{dot}^0+\mathcal{H}_\mathrm{dot}^U$ 
for a four-electron occupation $N_\mathrm{tot}=4$ 
becomes a triplet with the total spin $S=1$ for $U>0$
(see also appendix \ref{sec:nagaoka_eigenstate}).
This is caused by the Nagaoka ferromagnetic mechanism,\cite{Nagaoka}
and the energy separation 
$\Delta E^{(4)} \equiv E_{S=0}^{(4)} - E_{S=1}^{(4)}$ 
between the ground state  
 $E_{S=1}^{(4)}=U-2t+4\epsilon_d$ and 
the lowest singlet excited state $E_{S=0}^{(4)}$  is given by
\begin{align}
\Delta E^{(4)} 
= \,     
\frac{1}{2} \left[\,
U +3 t - \sqrt{ 9 t^2 +U^2 +2Ut} 
\,\right] 
\;.
\label{eq:diff_Nagaoka}
\end{align}
Note that $\Delta E^{(4)}>0$ for $U>0$, and
in the two extreme limits at small and large $U$ it takes the form 
\begin{align}
\Delta E^{(4)}
\simeq& 
\left\{    
\begin{array}{cl} 
 {U}/{3}
\;, & \ \ \  0 \leq U \ll t \\
t
\;,& \ \ \ 0 <t \ll U   
\\
\end{array}
\right.
\;.
\label{eq:diff_Nagaoka_asym}
\end{align}
For weak repulsions, an infinitesimal $U$ lifts 
the degeneracy of the singlet and triplet states 
in the non-interacting ground state.  
In the opposite limit, for large $U$ 
the circular motion along the triangle favors 
the magnetic $S=1$ ground state, 
and the energy separation  
$\Delta E^{(4)}$ is determined by the hopping matrix element $t$. 
If the cluster of the TTQD is connected to the leads, 
the local magnetic moment will be screened by the conduction 
electrons showing the two-stage Kondo behavior 
described in Sec.\ \ref{sec:thermo}.

As the level position $\epsilon_d$ decreases further,  
the occupation number of the electrons increases.
The ground-state energy  $E_{0}^{(5)}$ for 
a five-electron occupation  
and that for six electrons  $E_{0}^{(6)}$ are given by 
\begin{align}
E_{0}^{(5)} \! - E_{S=1}^{(4)} \,= \, 
E_{0}^{(6)} \! - E_{0}^{(5)}   = U+t+\epsilon_d \;.
\label{eq:E5_E6}
\end{align}
Specifically, at the point where the level in the TTQD 
$\epsilon_d$ takes the value of $\epsilon_d= -U-t$, 
the ground-state energies for the three different fillings,
 $N_\mathrm{tot}=4,\,5$ and $6$,  coincide. 
Thus the occupation number jumps from $N_\mathrm{tot}=4$ to $6$,
as $\epsilon_d$ crosses this value.

\begin{figure}[t]
 \leavevmode
\begin{minipage}{1.0 \linewidth}
 \includegraphics[width=1.0\linewidth]{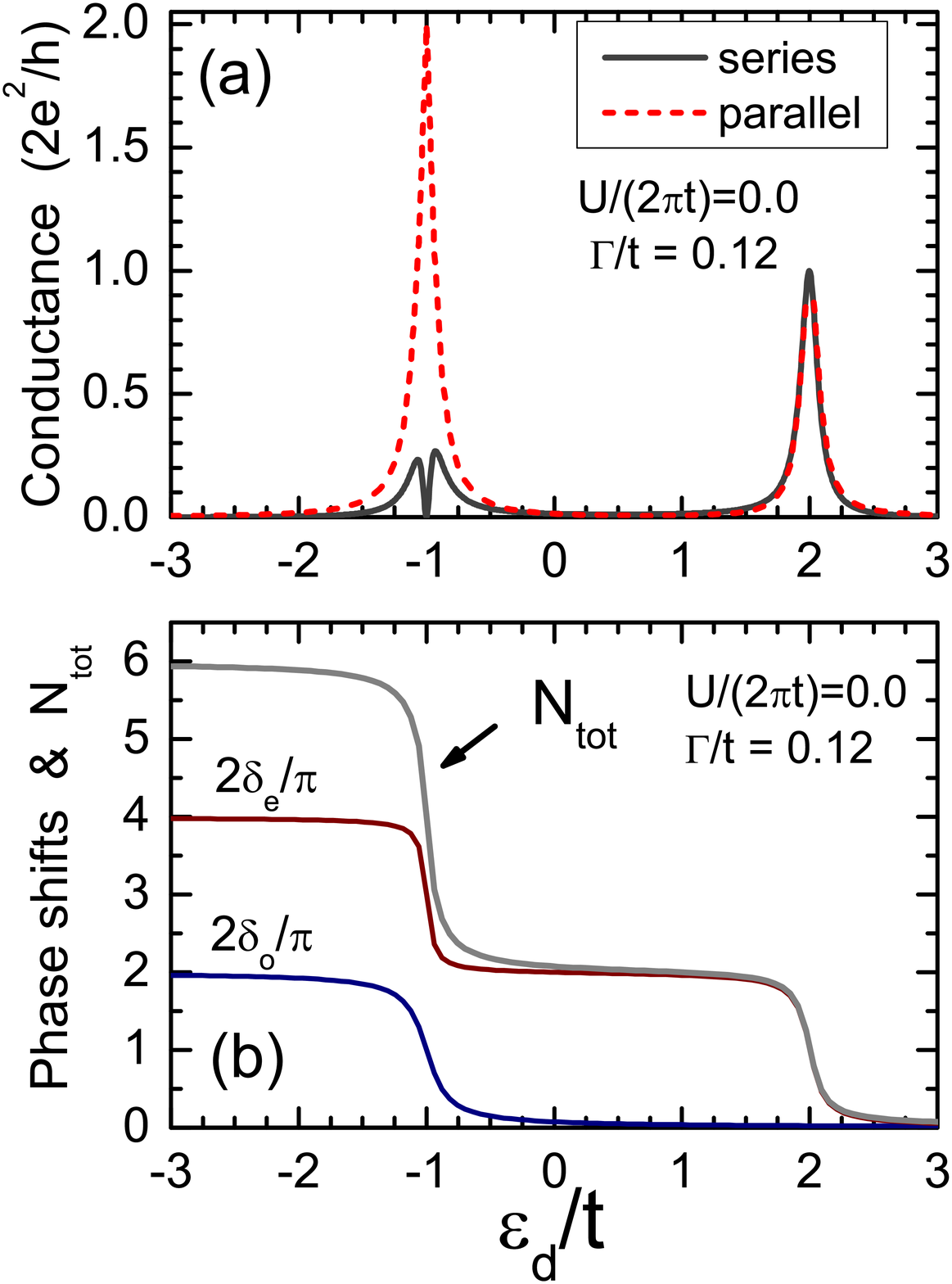}
\end{minipage}
 \caption{(Color online)
Conductances, phase shifts, and the filling $N_\mathrm{tot}$ 
of the triple dot in the non-interacting case $U=0.0$  
are plotted vs $\epsilon_d/t$ 
for a regular triangle $t' = t$,  $\epsilon_\mathrm{apex} = \epsilon_d$, and 
$\Gamma/t = 0.12$. 
 }
 \label{fig:cond_u0}
\end{figure}

\subsection{Phase shifts, conductances \& Friedel sum rule}
\label{subsec:cond_phase_shift}

The charge transfer between the dots and leads 
makes the low-energy states of the whole system  
a local Fermi liquid, 
which can be described by  renormalized quasi-particles. 
Particularly, in the case where  the system has an inversion symmetry 
$\Gamma_L^{} = \Gamma_R^{}$, 
the ground-state properties are characterized 
by the two phase shifts, 
$\delta_\mathrm{e}$ and $\delta_\mathrm{o}$, 
for the quasi-particles with 
the {\em even\/} and {\em odd\/} parities. 
At zero temperature, the series conductance $g_\mathrm{s}$ in   
 the two-channel configuration, 
which is shown in Fig.\ \ref{fig:system} (a), 
and the average number of electrons 
 in the triple dots can be expressed,
respectively, in the form of a Landauer formula
and the Friedel sum rule,\cite{Kawabata,ONH} 
\begin{align}
 g_\mathrm{s} \,= & \   \frac{2e^{2}}{h} 
\sin^{2} \left(\delta_\mathrm{e} -\delta_\mathrm{o} \right) \;,
\label{eq:gs}
\\
 N_\mathrm{tot}
\equiv &  \  \sum_{i=1}^{3}\sum_{\sigma} \,
\langle d_{i\sigma}^{\dagger}d_{i\sigma}^{\phantom{\dagger}}\rangle
\, = \, \frac{2}{\pi}\left(\delta_\mathrm{e}+\delta_\mathrm{o}\right) \:.
\label{eq:N_tot}
\end{align}
Furthermore, the parallel conductance $g_\mathrm{p}$ for the current flowing 
along the horizontal direction in the four-terminal geometry, 
 which is shown in Fig.\ \ref{fig:system} (b),   
can also be obtained from these two phase shifts
\begin{align}
 &
 g_\mathrm{p} \,= \,\frac{2e^{2}}{h} \,
  \left(\sin^{2}\delta_\mathrm{e}  + 
  \sin^{2}\delta_\mathrm{o}\right)\;.
       \label{eq:gp}
 \end{align}

The phase shifts can be expressed 
in terms of the renormalized matrix 
element $\widetilde{t}_{ij}$,\cite{ONH,OH,NO} 
(see also appendix~\ref{sec:app_green})
\begin{align}
 -\widetilde{t}_{ij} 
 \,=& \ -t_{ij} + \epsilon_{d,i}\, \delta_{ij} 
 + \mathrm{Re}\, \Sigma_{ij}^+(0) \;, 
\end{align}
where $\Sigma_{ij}^+(0)$ is the self-energy due 
to the interaction $\mathcal{H}_\mathrm{dot}^U$.
Specifically in the inversion symmetric case, 
the renormalized 
matrix element $\widetilde{t}_{ij}$ for  
$i,j$ ($=1,2,3$) inside the TTQD takes the form 
\begin{align}
 \{-\widetilde{t}_{ij} \} = 
\left[
\begin{matrix}
 \widetilde{\epsilon}_d & -\widetilde{t} & -\widetilde{t}' \cr
 -\widetilde{t} & \widetilde{\epsilon}_\mathrm{apex} & -\widetilde{t}  \cr
 -\widetilde{t}' & -\widetilde{t} &  \widetilde{\epsilon}_d\cr
\end{matrix}
\right] \;,
\end{align}
and the two phase shifts are given, respectively, by 
\begin{align}
\!
\cot \delta_\mathrm{e}  
= \frac{\widetilde{\epsilon}_d \, 
-\widetilde{t}' -2\widetilde{t}^2/\widetilde{\epsilon}_\mathrm{apex}}{\Gamma }
, \quad   
\cot \delta_\mathrm{o} = \frac{\widetilde{\epsilon}_d+\widetilde{t}'}{\Gamma}.
\label{eq:phase_shift_renormalized}
\end{align}

As a simple example, 
the conductances, the phase shifts,   
 and the occupation number $N_\mathrm{tot}$ 
in the non-interacting case  $\Sigma^+_{ij} = 0$ 
are plotted in Fig.\ \ref{fig:cond_u0} as functions of $\epsilon_d$,
for $\epsilon_\mathrm{apex}=\epsilon_d$, 
$t'=t$ and $\Gamma/t = 0.12$. 
We see in the upper panel (a) that the parallel 
conductance  $g_\mathrm{p}$ shows the peaks as the 
 one-particle states defined in Eq.\ \eqref{eq:tight_bind_triangle} 
cross the Fermi level, 
namely at the values of $\epsilon_d$ which satisfy $E^{(1)}_k=0$. 
We can see that the series conductance $g_\mathrm{s}$ 
has a zero point at $\epsilon_d=-t$.
This is due to the destructive interference,
which reflects the value of the phase shifts 
$\delta_\mathrm{e}=3\pi/2$ and 
$\delta_\mathrm{o}=\pi/2$ at this point.
The overall $\epsilon_d$ dependence of 
the phase shifts $2\delta_\mathrm{e}/\pi$, $\, 2\delta_\mathrm{o}/\pi$ 
and local charge  $N_\mathrm{tot}$  
 defined in Eq.\ \eqref{eq:N_tot} 
are shown in the lower panel (b). 
The phase shifts show a step of the height $\pi$ 
as one resonance state crosses the Fermi level.
Correspondingly the local charge also  
shows a staircase behavior. Note that 
two electrons with spin up and down enter  
the resonance state simultaneously in the non-interacting case.
Among the three one-particle states $k=0$ and $\pm 2\pi/3$,   
the lowest one with $k=0$ is an even-parity state,
so that  the even phase shift $\delta_\mathrm{e}$ takes 
the  first step at $\epsilon_d \simeq 2t$.
Then at $\epsilon_d \simeq -t$ electrons enter  
the other two one-particle states  
which can be classified into an even and odd states.


\begin{figure}[t]

\rule{0.3\linewidth}{0cm}
\begin{minipage}[t]{0.7\linewidth}
\setlength{\unitlength}{0.8mm}
\rule{2cm}{0cm}
\begin{picture}(110,30)(0,0)
\thicklines


\put(42.5,4.5){\line(1,0){20}}
\put(42.5,12.5){\line(1,0){20}}
\put(42.5,4.5){\line(0,1){8}}

\put(42.5,17.5){\line(1,0){20}}
\put(42.5,25.5){\line(1,0){20}}
\put(42.5,17.5){\line(0,1){8}}

\put(18.0,21){\line(1,0){12}}

\multiput(33.2,9.5)(2,0){5}{\line(1,0){1}}
\multiput(33.2,21)(2,0){5}{\line(1,0){1}}

\put(18,21){\circle*{3.3}} 
\put(30,9){\circle*{3.3}} 
\put(30,21){\circle*{3.3}} 


\put(19,22.5){\makebox(0,0)[bl]{$\sqrt{2}t$}}
\put(37,23){\makebox(0,0)[bl]{$v$}}

\put(37,11){\makebox(0,0)[bl]{$v$}}

 \put(15,16){\makebox(0,0)[bl]{$\epsilon_{a0}$}}
 \put(27,16){\makebox(0,0)[bl]{$\epsilon_{a1}$}}
 \put(27,12){\makebox(0,0)[bl]{$\epsilon_{b1}$}}


\put(10,20){\makebox(0,0)[bl]{$a_0$}}
\put(29,24){\makebox(0,0)[bl]{$a_1$}}
\put(29,2){\makebox(0,0)[bl]{$b_1$}}

\put(44,20){\makebox(0,0)[bl]{\lq\lq a, even"}}
\put(45,7){\makebox(0,0)[bl]{\lq\lq b, odd"}}

\end{picture}
\end{minipage}

\caption{Even and odd orbitals:
the onsite potential of each orbital is given by 
$\,\epsilon_{a0}=\epsilon_\mathrm{apex}$,
$\,\epsilon_{a1}=\epsilon_d -t'$,
and $\,\epsilon_{b1}=\epsilon_d +t'$. 
}
\label{fig:even_odd}
\end{figure}
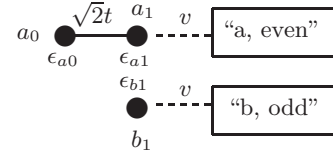

\subsection{Even and odd orbitals}
\label{subsec:even_odd}

\begin{figure}[t]
\begin{minipage}{1.0\linewidth}
 \includegraphics[width=1.0\linewidth]{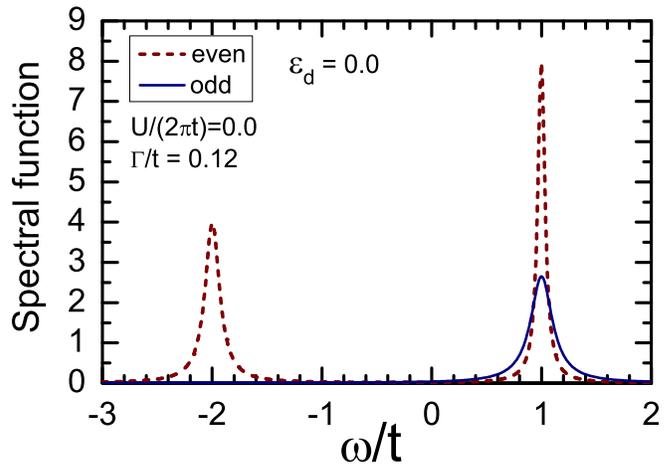}
\end{minipage}
\caption{(Color online)
Spectral functions $A_\mathrm{e}(\omega)$ and 
$A_\mathrm{o}(\omega)$ 
for non-interacting electrons $U=0.0$, 
for $\epsilon_d=0.0$, $t'=t$, $\epsilon_\mathrm{apex}=\epsilon_d$, 
and  $\Gamma/t = 0.12$. 
The even and odd components are 
defined in Eqs.\ \eqref{eq:even_odd_start}--\eqref{eq:A_o}.
 }
\label{fig:spec_u0}

\end{figure}

The non-interacting Hamiltonian $\mathcal{H}_0 \equiv   
\mathcal{H}_\mathrm{dot}^0 
+  \mathcal{H}_\mathrm{mix}  +  
\mathcal{H}_\mathrm{lead}$ can be written in a 
 diagonal form, using the even-odd basis 
\begin{align}
 a^{}_{1 \sigma} \equiv 
\frac{d^{}_{1 \sigma} + \,d^{}_{3 \sigma}}{\sqrt{2}}
\;, \qquad \quad 
 b^{}_{1 \sigma} \equiv 
\frac{d^{}_{1 \sigma} - \,d^{}_{3 \sigma}}{\sqrt{2}}
\;,
\label{eq:even_odd_start}
\end{align}
and $a^{}_{0 \sigma} \equiv d^{}_{2 \sigma}$.
Here,  for the linear combinations 
the labels $0$ and $1$ 
are assigned in the way that 
is shown in Fig.~\ref{fig:even_odd}.
The couplings to the two leads  by $\Gamma$ 
breaks the $C_{3v}$ symmetry of the regular triangle,
and lift the degeneracy of the even and odd states. 
In order to see this, 
we consider the spectral functions defined by
\begin{align}
A_\mathrm{e}(\omega) \,=& \  - \frac{1}{\pi}\,
\mathrm{Im} \left(\,
\langle\!\langle a_{0\sigma}^{\phantom{\dagger}}; 
 a_{0\sigma}^{\dagger} \rangle\!\rangle_{\omega}
 + \langle\!\langle a_{1\sigma}^{\phantom{\dagger}}; 
 a_{1\sigma}^{\dagger} \rangle\!\rangle_{\omega}
\,\right) ,
\label{eq:A_e}
\\
A_\mathrm{o}(\omega) \,=& \  - \frac{1}{\pi}\, 
\mathrm{Im}\  \langle\!\langle b_{1\sigma}^{\phantom{\dagger}}; 
 b_{1\sigma}^{\dagger} \rangle\!\rangle_{\omega} \;,
\label{eq:A_o}
\end{align} 
where $\langle\!\langle \widehat{\mathcal{O}}; 
\widehat{\mathcal{P}} \rangle\!\rangle_{\omega} 
\equiv -i\int_0^{\infty} dt\, e^{i(\omega +i0^+)t}\, 
\langle \{\widehat{\mathcal{O}}(t),
\widehat{\mathcal{P}}\}\rangle$ 
 is the retarded Green's function.
 These spectral functions 
for non-interacting electrons, $U=0$, are plotted 
in Fig.\ \ref{fig:spec_u0} 
as functions of $\omega$ for $\Gamma/t = 0.12$ and $\epsilon_d=0.0$.
The odd component $A_\mathrm{o}(\omega)$ 
has a single peak at $\omega\simeq t$, 
and its width is wider than that of the peak of 
 $A_\mathrm{e}(\omega)$ emerging at the same position.
This is because on the resonance at $\omega \simeq t$ 
the spectral weight of $A_\mathrm{e}(\omega)$ 
is dominated by the component of the apex site $a_0$, 
which is situated away from the leads, as shown in Fig.\ \ref{fig:even_odd}.
In contrast, $A_\mathrm{o}(\omega)$ represents 
the spectral weight of the level $b_1$, 
which is directly connected to one of the leads.
This geometry of the triangle and the two 
leads in the even-odd basis causes 
the difference in the peak width.
If the level position is lowered  
from $\epsilon_d=0.0$ which is the value 
chosen for Fig.\ \ref{fig:even_odd}, 
the peaks move towards the left side in the figure.
As the two resonance states approach the Fermi energy $\omega=0$,
the third electron enters the odd-resonance state first
rather than the even one,
because of the difference in the peak width. 
Then, both the resonance states become singly occupied 
when the two peaks just reach $\omega=0$. 
Correspondingly, the phase 
shifts take the values 
$\delta_\mathrm{e}=3\pi/2$ and $\delta_\mathrm{o}=\pi/2$,
and the series conductance has  
a dip at $\epsilon_d/t=-1.0$ in Fig.\ \ref{fig:cond_u0}.

The interaction Hamiltonian defined in Eq.\ \eqref{eq:HC^U} 
can be rewritten, using the even-odd basis, in the form
\begin{align}
\mathcal{H}_\mathrm{dot}^U  
=& \ U n_{a,0\uparrow}\,n_{a,0\downarrow}
+ \frac{U}{2} \left(
n_{a,1\uparrow}
\,n_{a,1\downarrow}
+n_{b,1\uparrow}\,n_{b,1\downarrow} \right)
\nonumber \\ 
& 
+ \frac{U}{2} \left(
\frac{1}{2} n_{a,1}\,n_{b,1}
-2 \vec{\bm{S}}_{a,1}\cdot\vec{\bm{S}}_{b,1}
\right)
\nonumber \\ 
& 
+ \frac{U}{2} \left(
 a^{\dagger}_{1 \uparrow} a^{\dagger}_{1 \downarrow} 
b^{}_{1 \downarrow} b^{}_{1 \uparrow}
+ 
 b^{\dagger}_{1 \uparrow} b^{\dagger}_{1 \downarrow} 
a^{}_{1 \downarrow} a^{}_{1 \uparrow}
\right).
\label{eq:even_odd}
\end{align}
Here, 
$\vec{\bm{S}}_{a,i}= \sum_{\sigma \sigma'}
a^{\dagger}_{i \sigma} \vec{\bm{\sigma}}_{\sigma\sigma'} 
a^{}_{i \sigma'}/2 $,   $\,\,\vec{\bm{\sigma}}$ the Pauli matrices,
$n_{a,i\sigma}= 
a^{\dagger}_{i \sigma}a^{}_{i \sigma} $, and 
$n_{a,i}= \sum_{\sigma}n_{a,i \sigma} $. 
The operators $n_{b,1\sigma}$ and $\vec{\bm{S}}_{b,1}$ 
for the odd orbital $b_{1\sigma}$ are defined in the same way.
We see in Eq.\ \eqref{eq:even_odd} that the interaction 
contains some long-range components between $a_1$ and $b_1$ orbitals.  
Specifically,  the last two terms in the right hand side 
 describe a charge transfer which breaks 
the charge conservation in each of the even and odd channels.
Thus for $U \neq 0$, each of the phase shifts 
 $\delta_\mathrm{e}$ and $\delta_\mathrm{o}$ 
does not necessarily correspond, respectively, 
to $\langle n_{a,0}+n_{a,1} \rangle \, \pi/2$ and 
$\langle n_{b,1} \rangle\, \pi/2$. 
The Friedel sum rule given in Eq.\ \eqref{eq:N_tot} holds 
in a way such that 
$\, \delta_\mathrm{e}+\delta_\mathrm{o} =
\langle n_{a,0}+n_{a,1} +n_{b,1}\rangle \, \pi/2$.

\section{Low-energy SU(4) symmetry at zero points of conductance}
\label{sec:zero_point_SU4}

In this section we show that 
the low-lying excitations acquire a special symmetry 
between channels at the zero points of the series conductance. 
This is a general property of a Fermi-liquid fixed point  
for a quantum-impurity system connected to two conduction channels. 
To be specific, 
we consider the case  that the system 
has  time-reversal and 
inversion symmetries. 
We use the discretized Hamiltonian $H_N$ 
of NRG\cite{KWW,KWW2} 
(see also appendix \ref{sec:NRG_approach}) 
 in this section 
not for a numerical purpose, 
but for showing explicitly the symmetric properties 
that emerge in a finite-size spectrum.

The two phase shifts $\delta_\mathrm{e}$ and $\delta_\mathrm{o}$ 
for interacting electrons can be deduced from 
the fixed-point eigenvalues 
of the discretized Hamiltonian $H_N$ defined 
in Eq.\ \eqref{eq:H_N}, following the procedure described 
in Ref.\ \onlinecite{ONH}.
Conversely, for some particular values of the phase shifts,
one can deduce that the system captures  
a special symmetry asymptotically at low energies. 
Particularly at the zero points of the series conductance,
the low-lying energy spectrum 
for the even ($s$-wave) and odd ($p$-wave) channels 
become identical near the Fermi energy. 
It means that the quasi-particles of the local Fermi liquid   
acquire the channel degeneracy as well as the spin degeneracy.

In order to see this, we go back briefly to 
a few essential points of the NRG. 
At a Fermi-liquid fixed point the low-lying energy spectrum  
of the many-body Hamiltonian $H_N$ defined in \eqref{eq:H_N} 
can be reproduced by 
a discretized version of free quasi-particle 
Hamiltonian  $H_\mathrm{eff}^{(N)}$,
\begin{align}
H_\mathrm{eff}^{(N)}  
 \, \equiv & \  \Lambda^{(N-1)/2} 
\left( 
\widetilde{\mathcal{H}}_\mathrm{dot}
 +  H_\mathrm{mix}^{\phantom{0}}  +   H_\mathrm{lead}^{(N)}
 \right) 
\nonumber 
\\
 = &
\sum_{\sigma=\uparrow,\downarrow}
\sum_{\mu=\mathrm{e},\mathrm{o}}
\sum_{l}
\varepsilon_{l\mu}^{(N)} 
\, \gamma_{l,\mu\sigma}^{\dagger} \gamma_{l,\mu\sigma}^{\phantom{\dagger}}
\;.
\label{eq:Hqp_even}
\end{align}
Here, the renormalized  TTQD 
part, $\widetilde{\mathcal{H}}_\mathrm{dot}$,  
in the first line is defined in Eq.\  \eqref{eq:H_d_effective}. 
The second line is a diagonal form, 
 for which  
$\mu$ ($=\mathrm{e},\,\mathrm{o}$) is the channel index,
and the label $l$ is a quantum number 
for the eigenstates in each channel.
The ground state for $H_\mathrm{eff}^{(N)}$ can be constructed 
in a such way that all the states 
for $\varepsilon_{l\mu}^{(N)} < 0$  
are filled by the $\gamma_{l,\mu\sigma}^{}$ particles, 
and 
the excitation energy $\varepsilon_{l\mu}^{(N)}$ of 
the quasi-particles near the Fermi level 
can be determined from the fixed-point 
eigenvalues of $H_N$.\cite{KWW,KWW2}

The phase shifts can be expressed as  
$\kappa_\mu = - \cot \delta_\mu$
in terms of the parameter $\kappa_\mu$ 
which determines the value of a Green's function 
at the Fermi energy, as shown in Eq.~\eqref{eq:G_N1_sym}.
The value of this parameter  
$\kappa_\mu$ can be deduced from 
the quasi-particle energy spectrum 
$\varepsilon_{l\mu}^{(N)}$,\cite{ONH} 
\begin{align}
&
\kappa_\mu 
\,= \,
\frac{2 A_{\Lambda}}{\pi}
 \lim_{N\to \infty}
 D \, 
\Lambda^{(N-1)/2}  
 \mbox{\sl g}_N^{\phantom{\dagger}}(\varepsilon_{l\mu}^{(N)}) 
\; .
\label{eq:kappa2}
\end{align}
%
Here, 
 $A_{\Lambda}$ is a coefficient 
which is defined in \eqref{eq:A_lambda},
$D$ is the half width of the conduction band,
and $\mbox{\sl g}_N^{\phantom{\dagger}}(\varepsilon)$ is 
a local Green's function at the interface $n=0$ 
of the discretized  version of the 
non-interacting lead $H_\mathrm{lead}^{(N)}$ 
given in \eqref{eq:H_lead_NRG}.
The equation \eqref{eq:kappa2} represents the 
relation between the phase shift and the 
low-energy quasi-particle spectrum 
 in each channel $\mu$. 
The right hand side of Eq.\  \eqref{eq:kappa2}  
 converges to the same value of $\kappa_\mu$ 
for any $l$, as long as $\varepsilon_{l\mu}^{(N)}$ is 
an excitation energy near the Fermi energy  
of the quasi-particles  $\gamma_{l,\mu\sigma}^{}$.
Specifically, it is not necessary that $\varepsilon_{l\mu}^{(N)}$ 
should be the lowest excitation energy, 
although the lowest one is preferable 
for a numerical purpose to make the corrections due to 
a residual interaction between the quasi-particles 
small.\cite{hewsonEPJ}

 One can deduce conversely the properties of the low-energy spectrum 
from the values 
of $\kappa_\mathrm{e}$ and $\kappa_\mathrm{o}$,
 using Eq.\ \eqref{eq:kappa2}. 
Particularly at zero points of the series conductance,  
the phase shifts satisfy the condition 
 $\delta_\mathrm{e}-\delta_\mathrm{o} = n \pi$ 
for $n = 0,\pm 1, \pm 2, \ldots$. 
This is equivalent to $\kappa_\mathrm{e}=\kappa_\mathrm{o}$,
and thus from Eq.\ \eqref{eq:kappa2} it is deduced 
that the low-lying energy spectrum for 
the even and odd channels become identical for large $N$, 
namely $\varepsilon_{l,\mathrm{e} }^{(N)} 
= \varepsilon_{l,\mathrm{o}}^{(N)}$ for all $l$ near 
the Fermi level of the quasi-particles. 
Thus the quasi-particles acquire  
a rotational symmetry in the channel (flavor) space 
 asymptotically at low energies, 
in addition to the rotation symmetry of the spin.
Therefore, 
 $H_\mathrm{eff}^{(N)}$ has a SU(4) symmetry at the zero 
points of the series conductance.

In the special case   
$\kappa_\mathrm{e}=0$ and  $\kappa_\mathrm{o}=0$,  
which correspond to $\delta_\mathrm{e}=(n+1/2)\pi$ and 
$\delta_\mathrm{o}=(n'+1/2)\pi$  
for integer $n$ and $n'$,  
the low-lying energy states have  
a particle-hole symmetry in addition to the channel symmetry, 
because the lead  Green's function  
in the right hand side of Eq.\ \eqref{eq:kappa2} 
is an odd function of the frequency  
$
\mbox{\sl g}_N^{\phantom{\dagger}}(-\varepsilon) 
=-\mbox{\sl g}_N^{\phantom{\dagger}}(\varepsilon)$, 
due to the particle-hole symmetry of $H_\mathrm{lead}^{(N)}$.

There are some other interesting cases 
in the unitary limit, 
at which the series conductance reaches $g_\mathrm{s} =2e^2/h$. 
If the two phase shifts take the values of  
 $\delta_\mathrm{e} = (n \pm 1/4)\pi$ 
and $\delta_\mathrm{o} = (n' \mp 1/4)\pi$, 
then $\kappa_\mu$ becomes 
$\kappa_\mathrm{e}=\mp 1$ and $\kappa_\mathrm{o}= \pm 1$. 
Therefore in this case the low-energy excitations acquire  
a particle-hole symmetry between a particle (hole) in the even channel 
 and a hole (particle) in the odd channel, 
although there is no channel symmetry in this case.
It can also be deduced from Eq.\ \eqref{eq:kappa2} that  
for the other set of the phase shifts,  
 $\delta_\mathrm{e} = (n+1/2)\pi$ and $\delta_\mathrm{o} = n' \pi$, 
particle-hole symmetry holds 
in each channel at low energies.

\section{Ground-state properties}
\label{sec:results_I}

 We have carried out the NRG calculations in order  
to study the effects of the interaction on
the transport and magnetic properties. 
Specifically  we concentrate 
on the regular triangle case, for which 
$\epsilon_d=\epsilon_\mathrm{apex}$
and $t'=t$ in the following.  
We will examine the effects of the deformations 
of the regular triangle caused by $\epsilon_\mathrm{apex}$ 
and $t'$ in Sec.\ \ref{sec:results_deformation}.

The ratio of the inter-dot hopping matrix element $t$ 
and the half width of the conduction band $D$ are chosen 
typically to be $t/D =0.1$ in most of our calculations. 
The iterative diagonalization has been carried out by using the even-odd basis.
For constructing the Hilbert space at each NRG step, 
instead of adding two orbitals from even and odd orbitals simultaneously, 
we add the one from the even part first and retain 
3600 low-energy states after carrying out 
the diagonalization of the Hamiltonian. 
Then, we add the other orbital from the odd part, 
and again keep the lowest 3600 eigenstates after the diagonalization. 
This truncation procedure preserves the inversion symmetry. 
We have chosen the discretization parameter to be $\Lambda=6.0$.
It has been confirmed that the conductance for noninteracting 
electrons $U=0$ is reproduced sufficiently well 
with this calculation scheme.\cite{OH,ONH,NO}

\begin{figure}[t]
 \leavevmode
\begin{minipage}{1.0\linewidth}
 \includegraphics[width=1.0\linewidth]{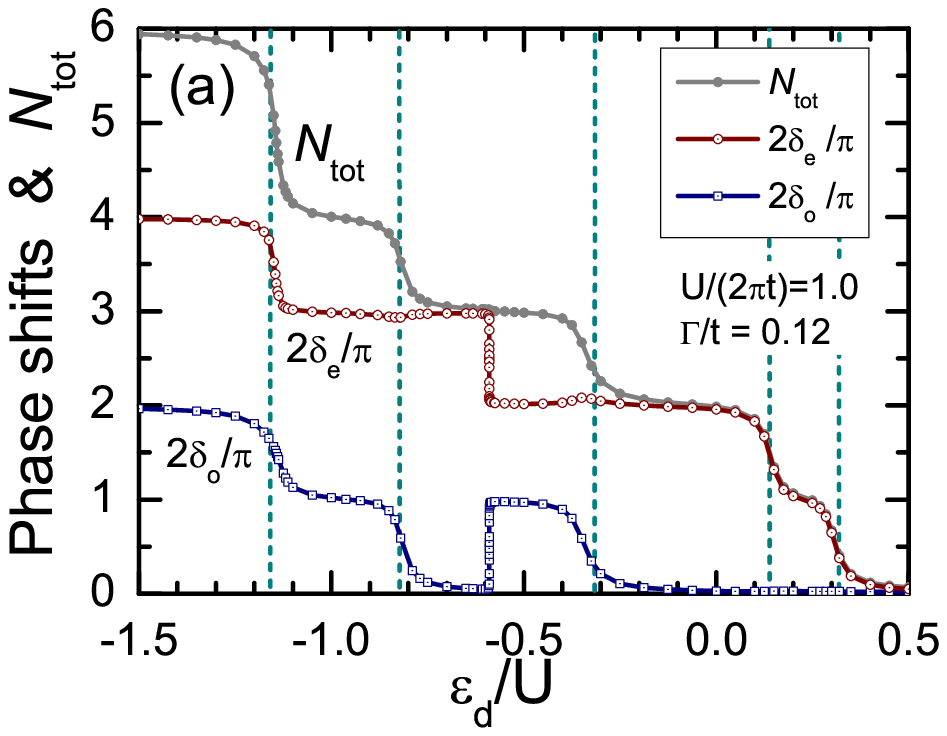}
\end{minipage}
\begin{minipage}{1.0\linewidth}
\includegraphics[width=1.0\linewidth]{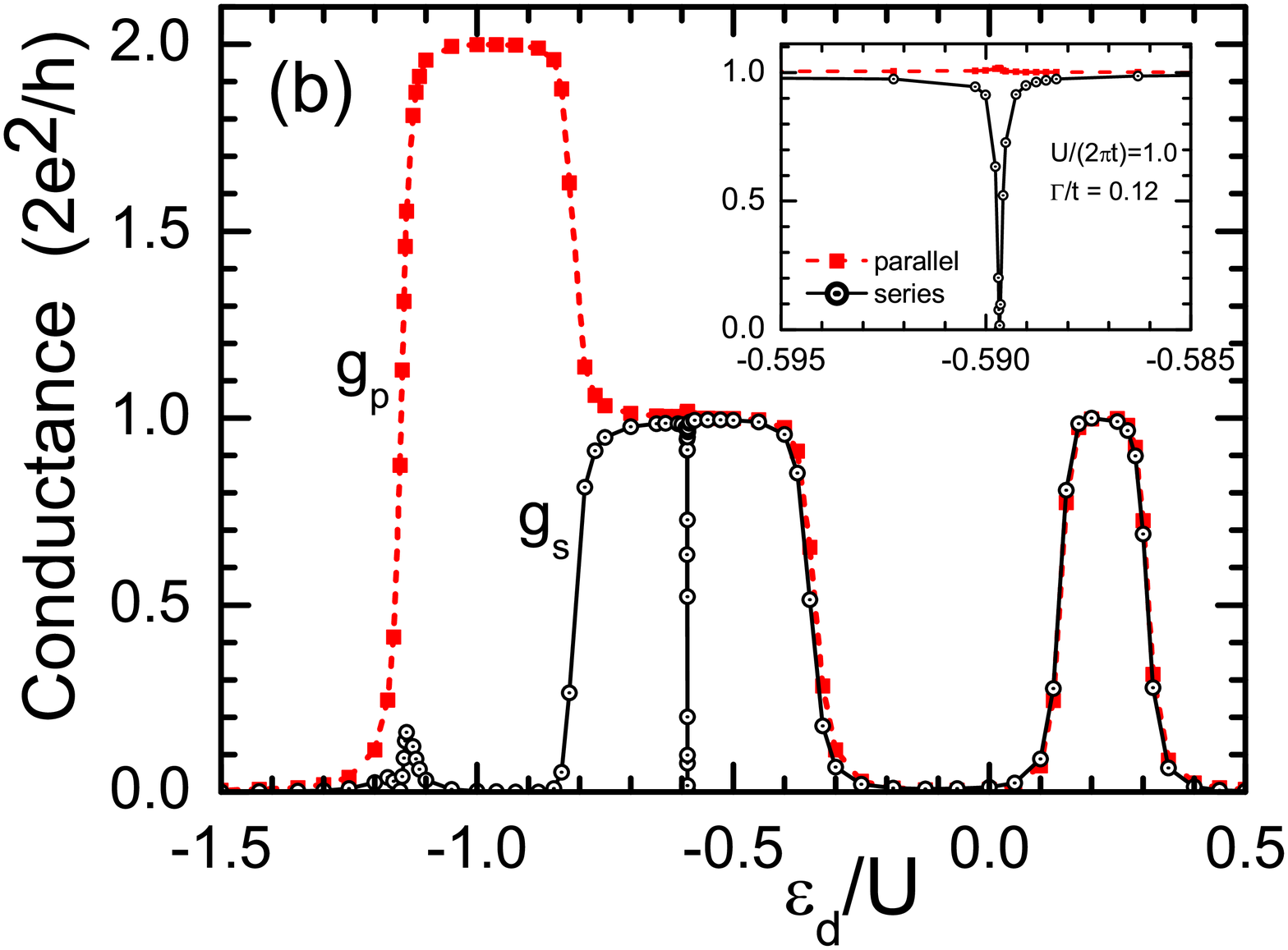}
\end{minipage}
 \caption{(Color online) Plots of NRG 
results as functions of $\epsilon_d/U$ 
for $U/(2\pi t)=1.0$, $\Gamma/t = 0.12$, 
$t'=t$ and $\epsilon_\mathrm{apex}=\epsilon_d$. 
 (a): the local charge $N_\mathrm{tot}$,
the phase shifts  $2\delta_\mathrm{e}/\pi$ and  $2\delta_\mathrm{o}/\pi$. 
 (b): the series ($\odot$) and parallel 
 ({\protect \small {\color{red} $\blacksquare$}}) 
conductances. 
The vertical dashed lines in (a) correspond to $\epsilon_d$, 
at which $N_\mathrm{tot}$ in the limit of $\Gamma = 0$ 
varies discontinuously. At the zero point of the series conductance 
for $\epsilon_d \simeq -0.59U$, the phase shifts take the values 
$\delta_\mathrm{o} \simeq \pi/4$ and $\delta_\mathrm{e} \simeq 5\pi/4$.
Inset in (b) shows the region near the zero point of $g_\mathrm{s}$.} 
 \label{fig:cond_u1}
\end{figure}

\subsection{Results at a small $\Gamma$ and large $U$}
\label{subsec:results_org}

The phase shifts and conductances  
for a relatively weak coupling $\Gamma/t=0.12$ 
and strong interaction $U/(2\pi t)=1.0$ 
are plotted as functions of $\epsilon_d/U$ in Fig.\ \ref{fig:cond_u1}.
The number of electrons $N_\mathrm{tot}$ in the triangle increases 
with decreasing $\epsilon_d$. 
It shows  plateaus for $N_\mathrm{tot} = 1,2,3$ and $4$,
and then jumps to $N_\mathrm{tot}\simeq 6.0$ at 
$\epsilon_d \simeq -1.16 U$ without 
taking a step for $N_\mathrm{tot} \simeq 5.0$. 
The same behavior has already been 
seen in the case of the triangular cluster, 
for which the ground-states energies for the 
three different fillings,  $N_\mathrm{tot} =4,\,5$ and $6$, 
coincide at $\epsilon_d = -t -U$, 
as mentioned in Sec.\ \ref{subsec:isolated_cluster}.
The vertical dashed lines in (a)
correspond to the values of $\epsilon_d$ 
at which the local charge varies discontinuously 
 in the limit of $\Gamma = 0$. 
We can see that the staircase behavior 
of $N_\mathrm{tot}$ in this small $\Gamma$ case 
is essentially determined by 
the one in this isolated limit.

The triangular triple dot has the $S=1$ local moment 
due to the Nagaoka ferromagnetic mechanism 
at the plateau for  $N_\mathrm{tot} \simeq 4.0$,
as mentioned in Sec.\ \ref{subsec:model}. 
This moment is screened at low temperatures 
by the conduction electrons 
from the two leads, and the ground state 
of the whole system becomes a singlet. 
We can see in Fig.\ \ref{fig:cond_u1}(b) that 
the two conductances show a significant 
difference at this filling $N_\mathrm{tot} \simeq 4.0$. 
At $-1.15U\lesssim \epsilon_d \lesssim -0.85U$   
the parallel conductance reaches     
the unitary limit of two conducting channels 
  $g_\mathrm{p} \simeq 4e^2/h$, 
while the series conductance $g_\mathrm{s}$ is suppressed. 
We have seen a similar difference 
between $g_\mathrm{p}$ and $g_\mathrm{s}$ 
for $U=0$ near  
the peaks at  $\epsilon_d=-t$ in Fig.\ \ref{fig:cond_u0} (a). 
We can see in Fig.\ \ref{fig:cond_u1} (a)    
that the two phase shifts are locked at 
the value $\delta_\mathrm{e} \simeq 3\pi/2$ 
and $\delta_\mathrm{o}\simeq \pi/2$ 
in the wide region $-1.15U\lesssim \epsilon_d \lesssim -0.85U$, 
 due to the Coulomb interaction.
Therefore in this region, both the even and odd channels 
give a unitary-limit contribution 
to the four-terminal  conductance $g_\mathrm{p}$, 
while the destructive interference reduces 
the two terminal conductance $g_\mathrm{s}$.
 
\begin{figure}[t]
 \leavevmode
\begin{minipage}[t]{1.0\linewidth}
 \includegraphics[width=1.0\linewidth]{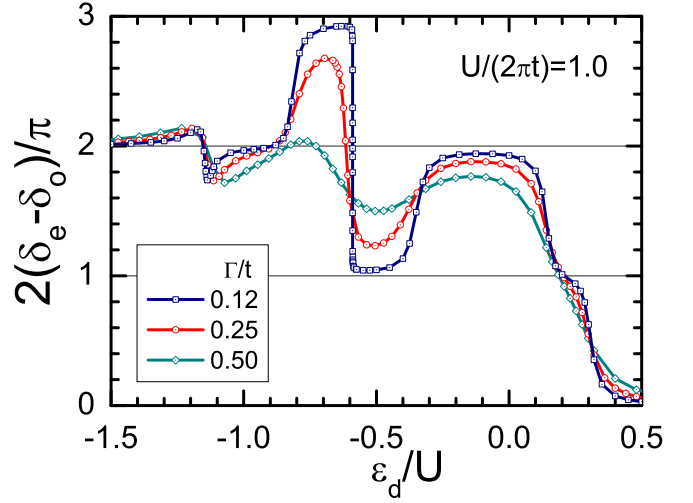}
\end{minipage}
 \caption{(Color online) 
Difference between the two phase shifts 
$2(\delta_\mathrm{e}-\delta_\mathrm{o})/\pi$
is plotted vs $\epsilon_d/U$ for $U/(2\pi t)=1.0$, 
for several values of  $\Gamma/t$  ($\square$) $0.12$
($\odot$) $0.25$, and ({\large $\diamond$}) $0.5$.  
At the points where the lines cross the horizontal line 
of $\delta_\mathrm{e}-\delta_\mathrm{e} =\pi$, 
the series conductance $g_\mathrm{s}$ becomes zero.  
 }
 \label{fig:diff_phase}
\end{figure}

The difference between the two phase shifts  
$\delta_\mathrm{e} -  \delta_\mathrm{o}$  
is shown in Fig.\ \ref{fig:diff_phase}.
It crosses the value of $\pi$ at 
$\epsilon_d/U \simeq -1.15,\, -0.88$ and $-0.59$. 
At these three points the series conductance becomes zero.
The series conductance must have  
a maximum between two adjacent zero points,  
and thus a fine structure of $g_\mathrm{s}$    
seen at  $-1.2U \lesssim \epsilon_d \lesssim -1.1U$ 
can be regarded as one such example. 
It should be noted that in the non-interacting case 
only a single zero point 
appears  as $\epsilon_d$ varies
(see Fig.\ \ref{fig:cond_u0} (a)).

In most of the other regions of the electron filling, 
the series and parallel conductances show 
a similar $\epsilon_d$ dependence.
For instance, 
both the two conductances show a plateau 
with the value $2e^2/h$ 
for $0.15U \lesssim \epsilon_d \lesssim 0.3U$ 
in  Fig.\ \ref{fig:cond_u1}(b). 
This Kondo behavior is caused by 
a $S=1/2$ moment of a single electron which occupies  
the lowest even-parity orbital with the energy $E_{k=0}^{(1)}$.
Then, for $-0.3U \lesssim \epsilon_d \lesssim 0.1U$, two electrons 
fill the lowest orbital, 
and both the series and parallel conductances are suppressed.
As $\epsilon_d$ decreases further,
for $-0.58U \lesssim \epsilon_d \lesssim -0.3U$, 
the third electron enters the odd orbital corresponding 
to the site $b_1$ which is illustrated 
in Fig.\ \ref{fig:even_odd}.   
The moment due to the third electron causes 
the SU(2) Kondo effect in this region.

We can see in Fig.\ \ref{fig:cond_u1}(b) 
that there is a sharp dip in the series conductance 
at $\epsilon_d \simeq -0.59U$ 
in the middle of the plateau for $N_\mathrm{tot} \simeq 3.0$.  
This reflects a sudden change of the two phase shifts,
seen in Fig.\ \ref{fig:cond_u1}(a). 
Near the series conductance dip, the difference between the 
two  phase shifts 
$\delta_\mathrm{e} -\delta_\mathrm{o}$ varies 
from $\pi/2$ to $3\pi/2$ continuously, 
taking the middle value of $\pi$ 
at the zero point of $g_\mathrm{s}$.
In contrast, the sum of the two phase shifts 
does not show a significant change, 
and is almost a constant $\delta_\mathrm{e} 
+\delta_\mathrm{o} \simeq 3\pi/2$ which corresponds  
to the occupation number $N_\mathrm{tot} \simeq 3.0$.
Thus the parallel conductance varies very little 
in this narrow region, 
keeping the value $g_\mathrm{p} \simeq 2e^2/h$.
Furthermore, 
the two phase shifts take the value 
$\delta_\mathrm{o}\simeq \pi/4$ 
and $\delta_\mathrm{e}\simeq 5\pi/4$ 
at the zero point of $g_\mathrm{s}$ in 
the center of the dip.
These values of the phase shifts 
are consistent with the one  reported for 
a capacitively coupled double dot 
with a single electron 
in an SU(4) symmetric limit.\cite{Borda}
As mentioned in Sec.\ \ref{sec:zero_point_SU4}, 
the quasi-particle excitations 
have channel degeneracy as well as the spin degeneracy 
generally at zero points of the series conductance,
and the system has an SU(4) symmetry of 
a Fermi-liquid origin at low energies.  
In the case of the TTQD,      
the system also has an SU(4) symmetry of 
a different origin at high energies, 
which is caused by the four-fold degenerate eigenstates 
of the isolated triangular cluster with three electrons,
 mentioned in Sec.\ \ref{subsec:isolated_cluster}.
We will see in Sec.\ \ref{sec:thermo} from 
the results of the thermodynamic quantities that 
the four-fold degenerate local degrees of freedom 
remain free at high temperatures, 
and are screened at low temperatures 
by the conduction electrons keeping 
the SU(4) symmetric behavior.

\begin{figure}[t]
\begin{center}

\begin{minipage}[t]{\linewidth}
\includegraphics[width=1.0\linewidth]{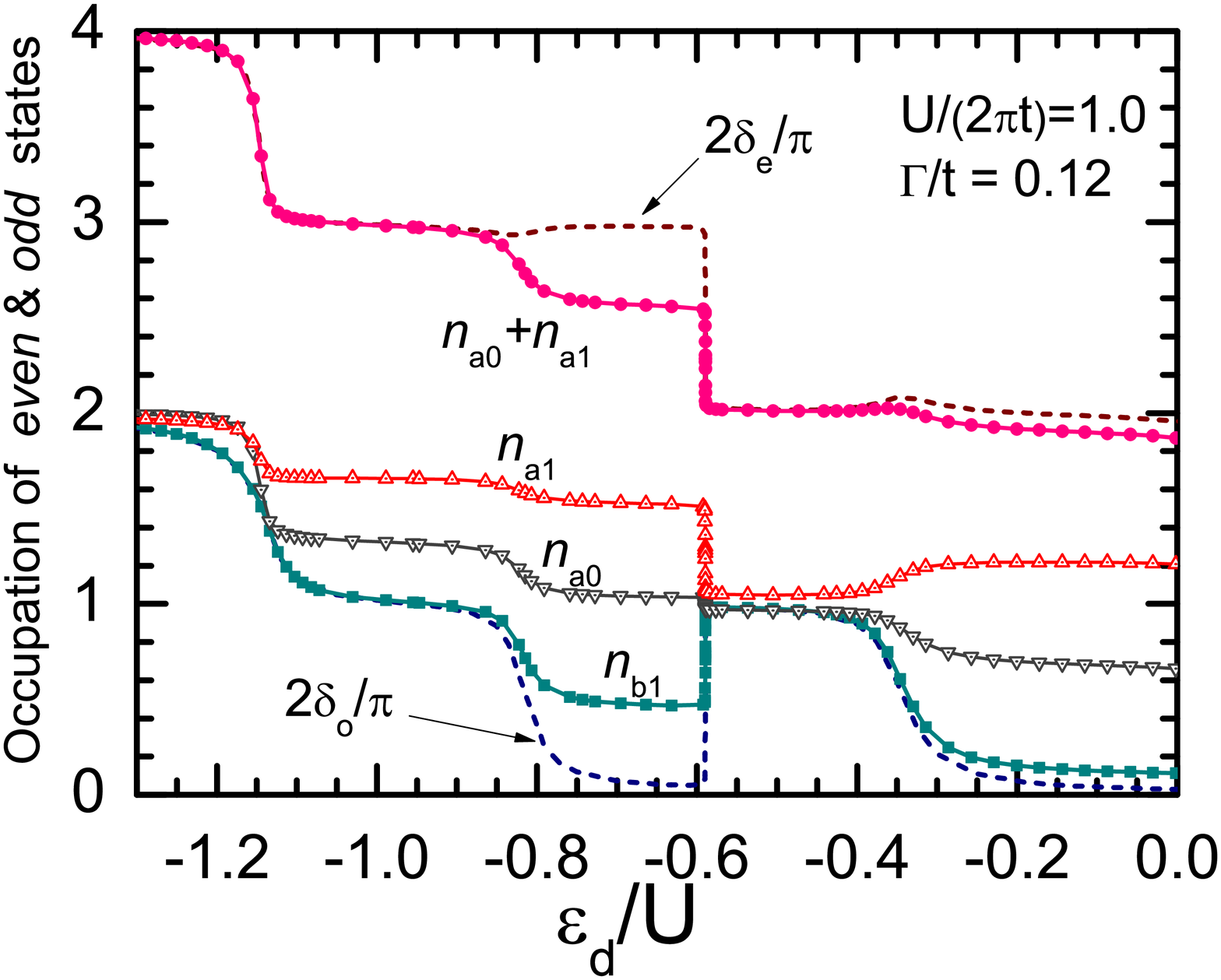}
\end{minipage}

\end{center}
\caption{(Color online) Filling of even and odd orbitals, 
$\langle n_{a,0}\rangle$, $\langle n_{a,1}\rangle$
and $\langle n_{b,1}\rangle$ 
are plotted vs $\epsilon_d/U$ 
for $U/(2\pi t) =1.0$ and $\Gamma/t=0.12$.
The dashed lines correspond to the phase shifts 
$2\delta_\mathrm{e}/\pi$ and $2\delta_\mathrm{o}/\pi$. 
Just at the zero point, $\epsilon_d \simeq -0.59U$, 
of the series conductance in the middle of 
the crossover region 
the phase shifts take the values 
$\delta_\mathrm{o}\simeq \pi/4$, $\delta_\mathrm{e}\simeq 5\pi/4$, 
and three electrons distribute in these orbitals as   
$\langle n_{a,0} \rangle \simeq 1.0$, 
$\langle n_{a,1} \rangle \simeq 1.28$, and 
 $\langle n_{b,1} \rangle \simeq 0.72$.
}
\label{fig:even_odd_occupation}
\end{figure}

The sudden change of the two phase shifts also implies 
that a charge redistribution takes place in the TTQD.  
One might expect from the behavior 
of $\delta_\mathrm{e}$ and $\delta_\mathrm{o}$  that 
the third electron moves away from the odd orbital 
to enter the second even orbital 
for $-0.8U \lesssim \epsilon_d \lesssim -0.6U$. 
Each of the two phase shifts, however, does not necessarily 
correspond to the occupation number of 
the orbitals of each parity, 
$N_\mathrm{even} \equiv \langle n_{a,0} + n_{a,1} \rangle$
 or $N_\mathrm{odd} \equiv \langle n_{b,1} \rangle$,
in the interacting case $U \neq 0$,   
as mentioned in Sec.\ \ref{subsec:even_odd}.
For this reason, 
we have calculated also the expectation value of $\langle n_{a,i} \rangle$ 
and  $\langle n_{b,1} \rangle$ with  the NRG
in order to clarify precisely the charge distribution.  
 The results are presented in Fig.\ \ref{fig:even_odd_occupation}.
We see that an intuitive estimate, 
which assumes $N_\mathrm{even} \approx 2 \delta_\mathrm{e}/\pi$
 and $N_\mathrm{odd} \approx 2 \delta_\mathrm{o}/\pi$, 
approximately works for 
 $\epsilon_d \lesssim -0.85U$ and  $-0.6U\lesssim \epsilon_d $.
For instance, the third electron really enters into 
the odd orbital to take the occupation 
number of the value, $\langle n_{b,1} \rangle \simeq 1.0$,   
for $-0.58U \lesssim \epsilon_d \lesssim -0.4U$  
on the right side of the series conductance dip. 
Such a separation of the Friedel sum rule, however, 
does not make sense in the central region. 
We can see for $-0.75 \lesssim \epsilon_d/U \lesssim -0.6$ 
that the local charges are redistributed between the even and odd orbitals, 
and take the values  
$\langle n_{a,0} \rangle \simeq 1.0$, 
$\langle n_{a,1} \rangle \simeq 1.5$, and 
 $\langle n_{b,1} \rangle \simeq 0.5$. 
Thus  $N_\mathrm{even} \simeq 2.5$ and $N_\mathrm{odd}  \simeq 0.5$,
while the phase shifts take the values  
$2\delta_\mathrm{e}/\pi \simeq 3.0$ 
and $2\delta_\mathrm{o}/\pi \simeq 0.0$. 
Furthermore, at the zero point 
of the series conductance the phase shifts take the values 
$\delta_\mathrm{o}\simeq \pi/4$ and 
$\delta_\mathrm{e}\simeq 5\pi/4$ as mentioned, 
while the occupation number for the even and odd orbitals 
become $\langle n_{a,0} \rangle \simeq 1.0$, 
$\langle n_{a,1} \rangle \simeq 1.28$, and 
 $\langle n_{b,1} \rangle \simeq 0.72$.

The charge distribution in real space can also be deduced,   
using the operator identities 
 $n_{d,2} \equiv n_{a,0}$ and 
 $ n_{d,1} + n_{d,3}=n_{a,1} + n_{b,1}$ 
 which follow from the definition in
\eqref{eq:even_odd_start}, 
\begin{align}
\langle n_{d,1} \rangle = 
\langle n_{d,3} \rangle 
\, = \, \langle n_{a,1} + n_{b,1} \rangle /2
\;.
\end{align}
We have confirmed that 
the numerical value of
$\langle n_{a,0}\rangle$ 
and $\langle n_{a,1} + n_{b,1} \rangle /2$ are 
very close in the whole region 
of $\epsilon_d$ plotted in Fig.\ \ref{fig:even_odd_occupation}.
Thus the occupation number of the apex site, $\langle n_{d,2}\rangle$,  
is not much different from $\langle n_{d,1}\rangle$, 
or $\langle n_{d,3}\rangle$. 
Consequently, for small $\Gamma$,
 the charge distribution is almost homogeneous in  real space.
The charge inhomogeneity in the even-odd basis, however, 
 affects the transport and magnetic properties very much.
This is because the orbitals $a_{1}$ and $b_{1}$ 
are connected directly to the leads as shown in Fig.\ \ref{fig:even_odd},
while the apex site does not have a direct connection 
to the leads. 

We can see also 
at $-1.15U\lesssim \epsilon_d \lesssim -0.83U$ 
in Fig.\ \ref{fig:even_odd_occupation} 
that four electrons in the TTQD 
are distributed in the even and odd orbitals such that   
 $\langle n_{a,0} \rangle \simeq 1.3$, 
$\langle n_{a,1} \rangle \simeq 1.7$, and 
 $\langle n_{b,1} \rangle \simeq 1.0$.
These values are close, respectively, to $4/3$, $5/3$ and $1$, 
which are the occupation numbers in the cluster limit $\Gamma =0$ 
(see appendix \ref{sec:nagaoka_eigenstate}).

\subsection{Large $\,\Gamma$ cases}
 \label{subsec:Gamma_dependence}

\begin{figure}[t]
 \leavevmode
\begin{minipage}[t]{0.49\linewidth}
 \includegraphics[width=1.0\linewidth]{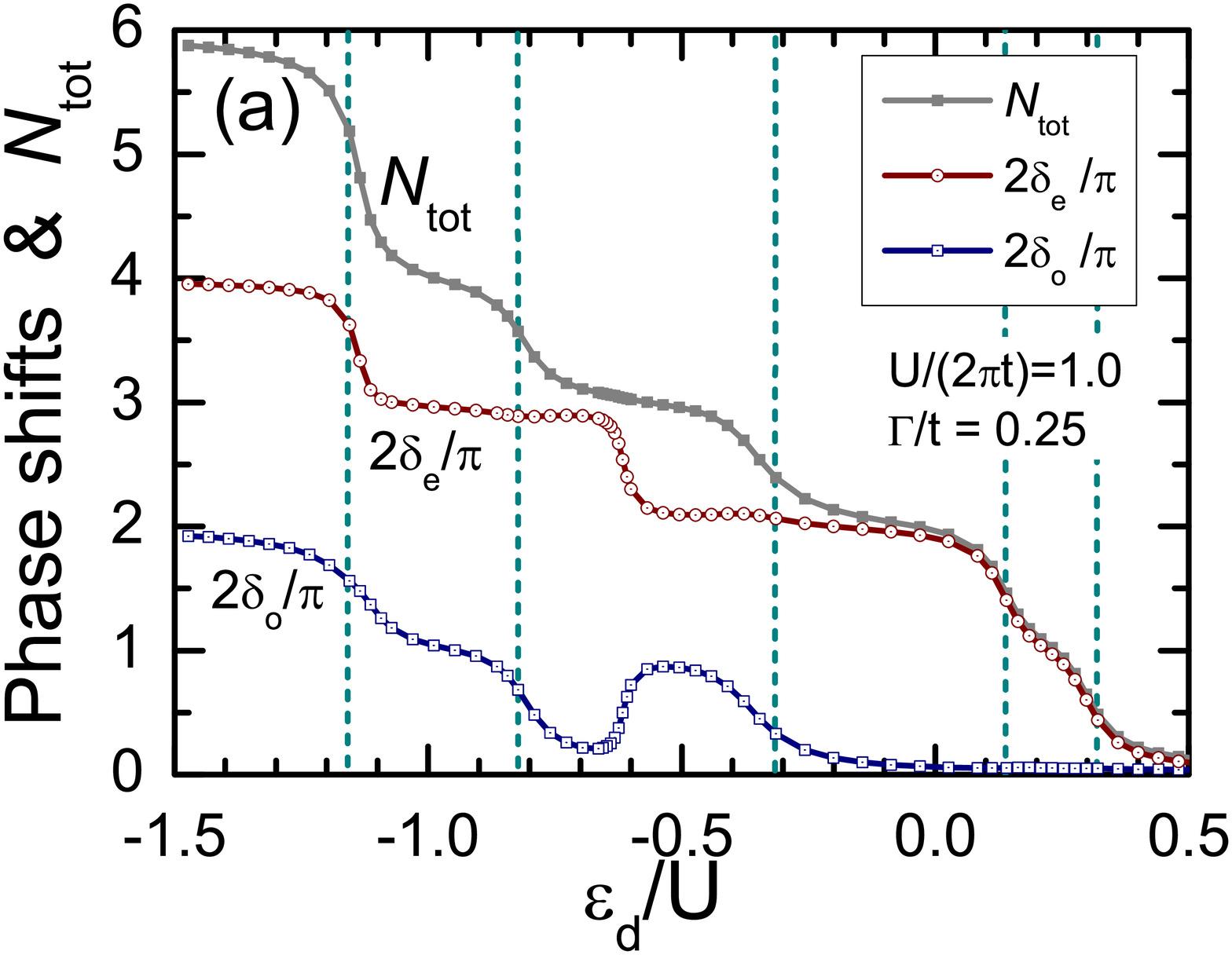}
\end{minipage}
\begin{minipage}[t]{0.49\linewidth}
\includegraphics[width=1.0\linewidth]{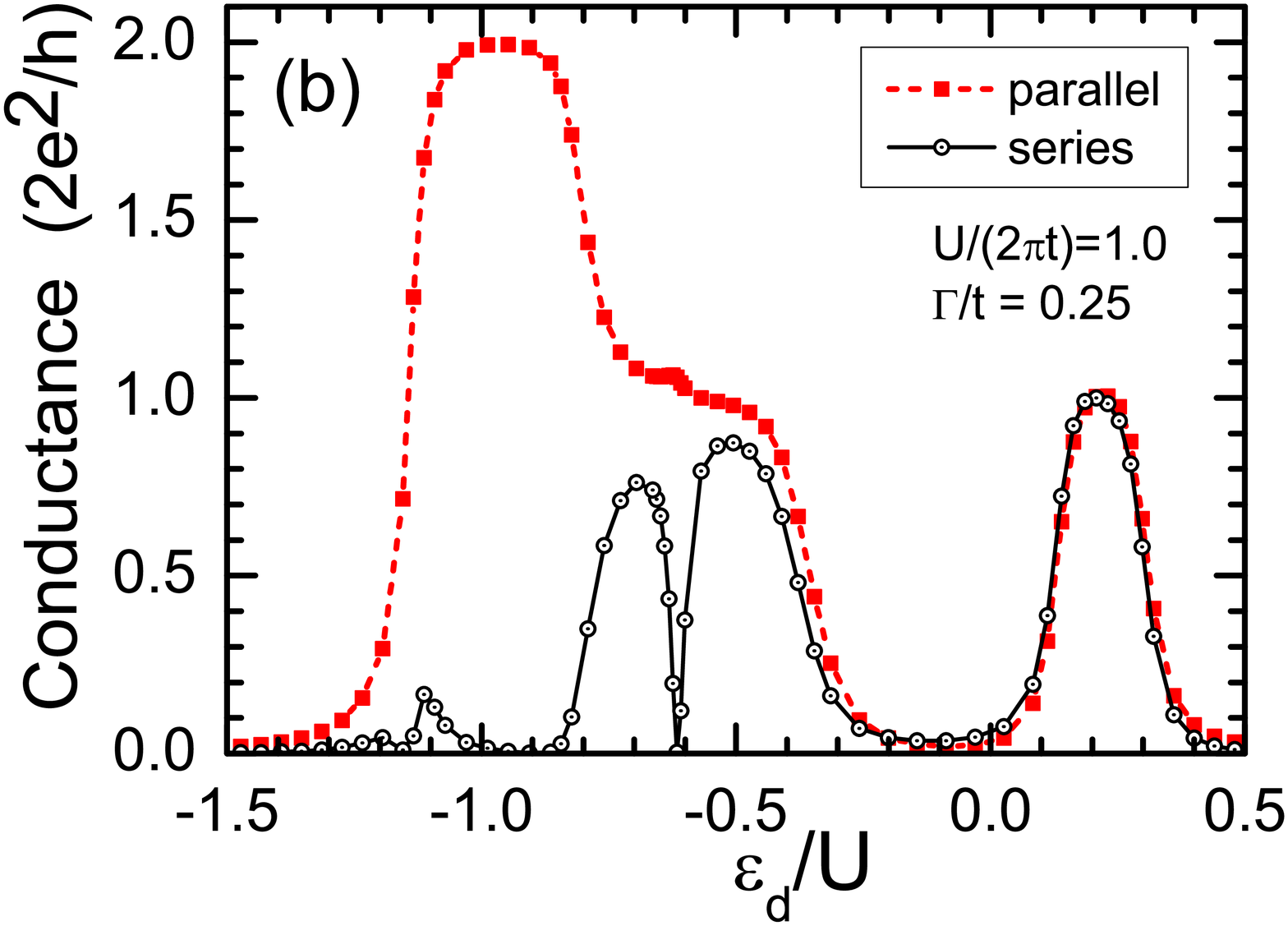}
\end{minipage}

\begin{minipage}[t]{0.49\linewidth}
 \includegraphics[width=1.0\linewidth]{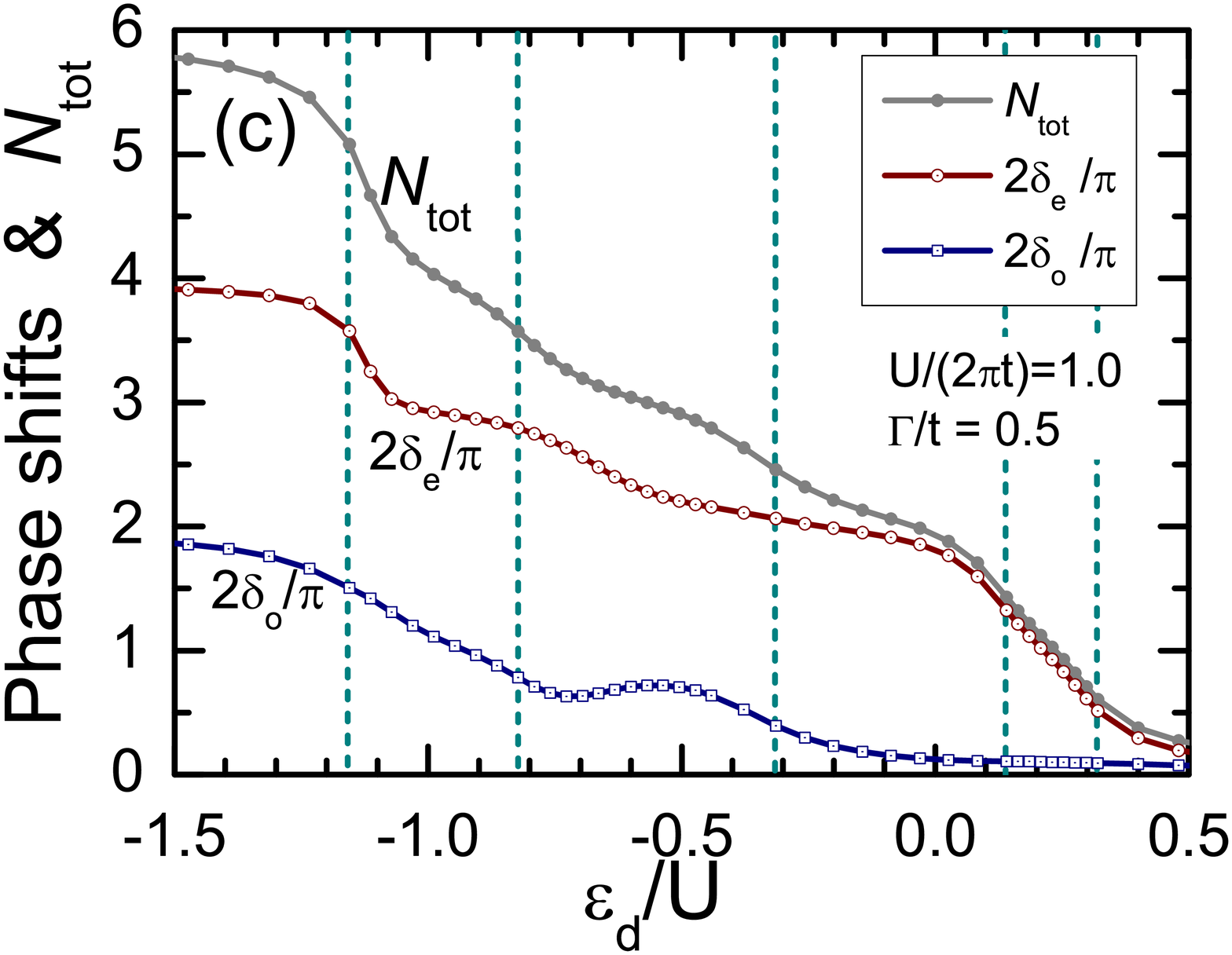}
\end{minipage}
\begin{minipage}[t]{0.49\linewidth}
\includegraphics[width=1.0\linewidth]{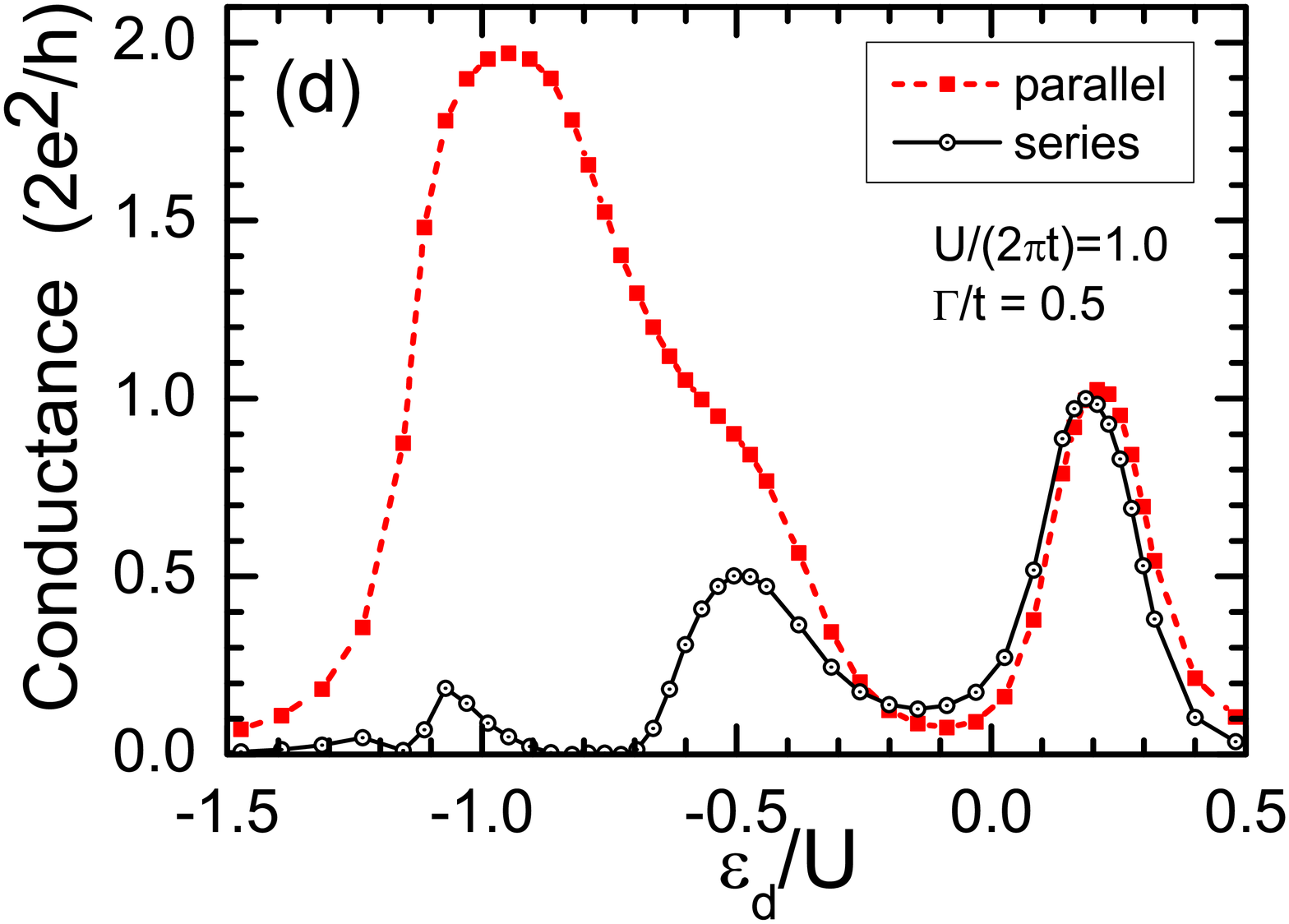}
\end{minipage}

 \caption{(Color online) 
Phase shifts and conductances for large hybridizations  
(upper panels) $\Gamma/t=0.25$ and (lower panels) $\Gamma/t = 0.5$;
$U/(2\pi t)=1.0$ is fixed. 
The vertical dashed lines in (a) and (c) correspond to $\epsilon_d$, 
at which $N_\mathrm{tot}$ in the limit of $\Gamma = 0$ 
varies discontinuously. 
 }
 \label{fig:cond_u1_gam025_05}

\end{figure}

The conductances and phase shifts 
discussed in the above are 
the results obtained for a  
relatively small hybridization $\Gamma/t=0.12$.
For this reason the structures 
of the plateaus and valleys can be seen very clearly. 
We next examine how the strength of the couplings $\Gamma$
between the triangle and two leads affects the ground-state properties.
For comparison with the data shown in Fig.\ \ref{fig:cond_u1},
we have carried out the calculations 
for several larger values of $\Gamma$ 
keeping the interaction $U$ the same value of $U/(2 \pi t) =1.0$. 
The results are shown in Fig.\ \ref{fig:cond_u1_gam025_05}.   
The couplings are chosen such that  $\Gamma/t= 0.25$ 
for (a) and (b),  and  $\Gamma/t= 0.5$ for (c) and (d).


As a general trend,
the structures  seen in the phase shifts and conductances 
are smeared gradually as the couplings get stronger.
We  see in the upper panels (a) and (b) 
of Fig.\ \ref{fig:cond_u1_gam025_05}, however, that 
the structures are still  distinguishable by eye 
in the case of $\Gamma/t=0.25$, 
and the feature of each line can be compared with 
the corresponding line in Fig.\ \ref{fig:cond_u1}. 
Specifically, for $N_\mathrm{tot} \simeq 3.0$  
the broadening makes the dip of the series conductance 
near $\epsilon_d \simeq -0.6 U$ wider. 
This structure reflects the behavior of the phase shifts 
at $-0.8U \lesssim \epsilon_d \lesssim -0.6 U$ 
in Fig.\ \ref{fig:cond_u1_gam025_05} (a). 
In this region the odd phase shift $\delta_\mathrm{o}$ 
still shows a clear concaving valley, 
although it becomes  shallower 
than that in the case of Fig.\ \ref{fig:cond_u1} (a).
This valley is smeared further in the case of $\Gamma/t=0.5$, 
and the fine structure of $g_\mathrm{s}$ 
at $-0.8U \lesssim \epsilon_d \lesssim -0.6 U$ 
is not visible in Fig.\ \ref{fig:cond_u1_gam025_05} (d).
Particularly, the conductance plateau on the left side of the dip 
almost vanishes.
We can see, nevertheless, in Fig.\ \ref{fig:diff_phase} 
that the difference between the two phase shifts  
$\delta_\mathrm{e} -  \delta_\mathrm{o}$  
for $\Gamma/t=0.5$ crosses the value of $\pi$ at 
 $\epsilon_d/U \simeq -0.73$, $-0.82$, and $-1.15$.
It means that there still exist three zero points 
of the series conductance, and 
 $g_\mathrm{s}$ has a weak maximum 
between the two adjacent zero points.

In contrast to the series conductance $g_\mathrm{s}$, 
the parallel conductance  $g_\mathrm{p}$ is less sensitive 
to $\Gamma$.
Particularly, we can see that in  Fig.\ \ref{fig:cond_u1_gam025_05} 
that the broad peaks of $g_\mathrm{p}$  
at $N_\mathrm{tot}\simeq 4.0$ and $N_\mathrm{tot}\simeq 1.0$ 
keep their essential features seen 
in the  small $\Gamma$ case in Fig.\ \ref{fig:cond_u1}. 
This is because the energy gain 
of the Nagaoka state $\Delta E^{(4)}$,
which is defined in Eq.\ \eqref{eq:diff_Nagaoka}, 
is of the order of 
the inter-dot hopping matrix element $t$ for large $U$.
It takes a value of $\Delta E^{(4)}/t = 0.735$ for $U/(2\pi t)=1.0$,
and thus in the present case $\Delta E^{(4)}/t$ is still much 
larger than the energy scale of the hybridization $\Gamma/t=0.5$.   

We can also see another quantitative change  
in Fig.\ \ref{fig:cond_u1_gam025_05} (d) 
that the difference between $g_\mathrm{p}$ and $g_\mathrm{s}$
at $N_\mathrm{tot}\simeq 1.0$  becomes visible 
for large $\Gamma$. A similar tendency was seen also 
for the conductances of a linear triple dot.\cite{ONH}

\begin{figure}[t]
 \leavevmode
\begin{minipage}[t]{0.49\linewidth}
 \includegraphics[width=1.0\linewidth]{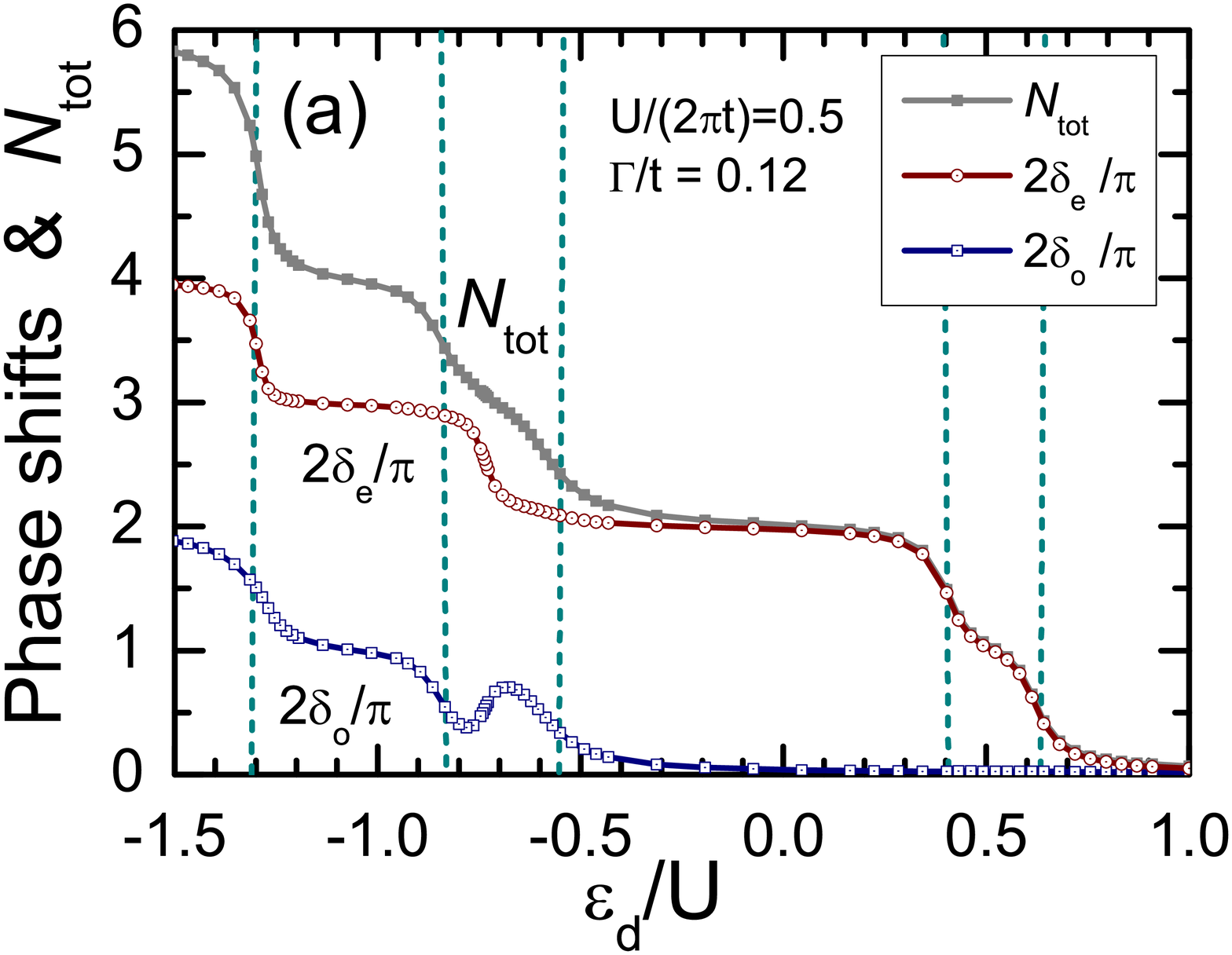}
\end{minipage}
\begin{minipage}[t]{0.49\linewidth}
\includegraphics[width=1.0\linewidth]{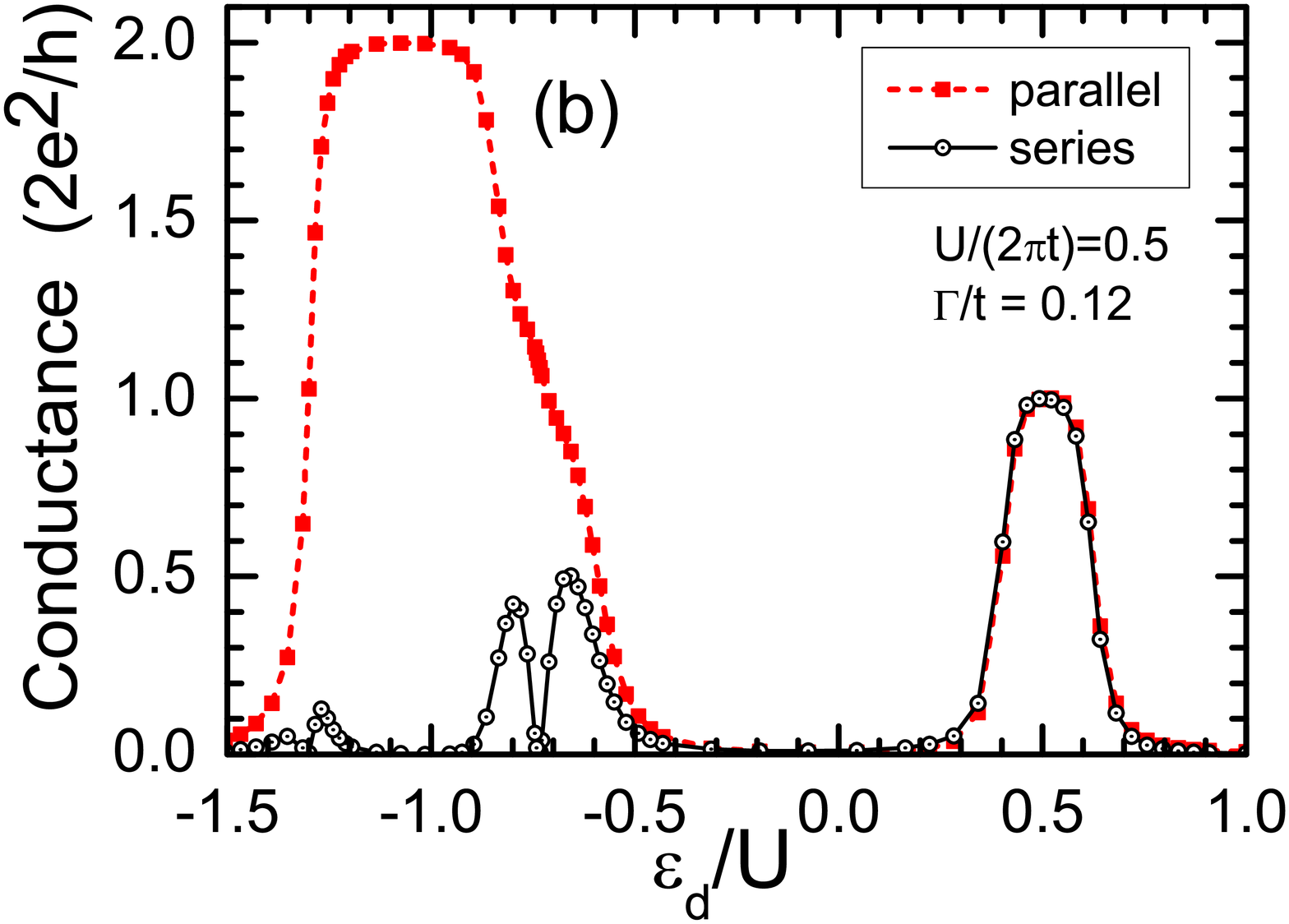}
\end{minipage}

\begin{minipage}[t]{0.48\linewidth}
 \includegraphics[width=1.0\linewidth]{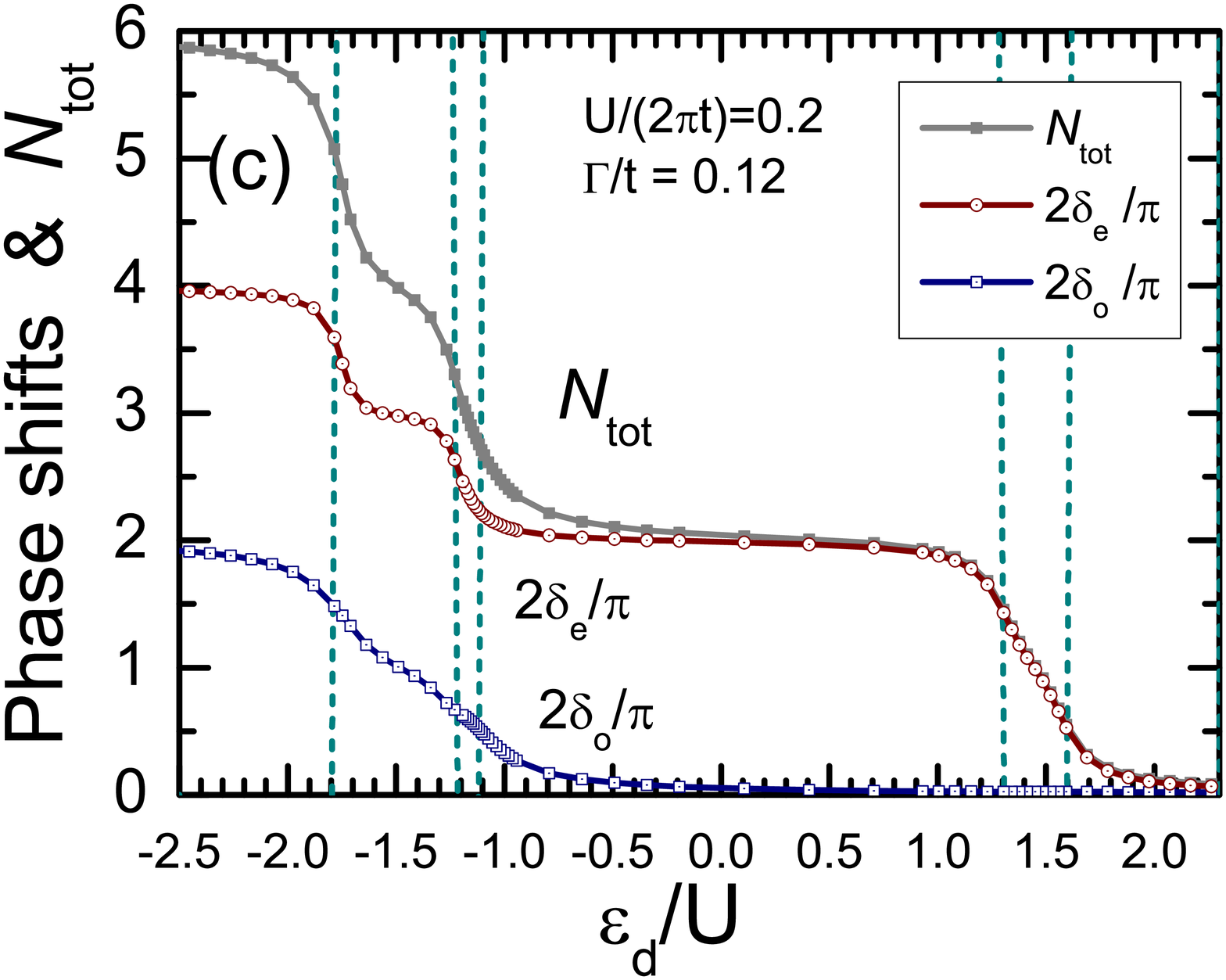}
\end{minipage}
\begin{minipage}[t]{0.49\linewidth}
\includegraphics[width=1.0\linewidth]{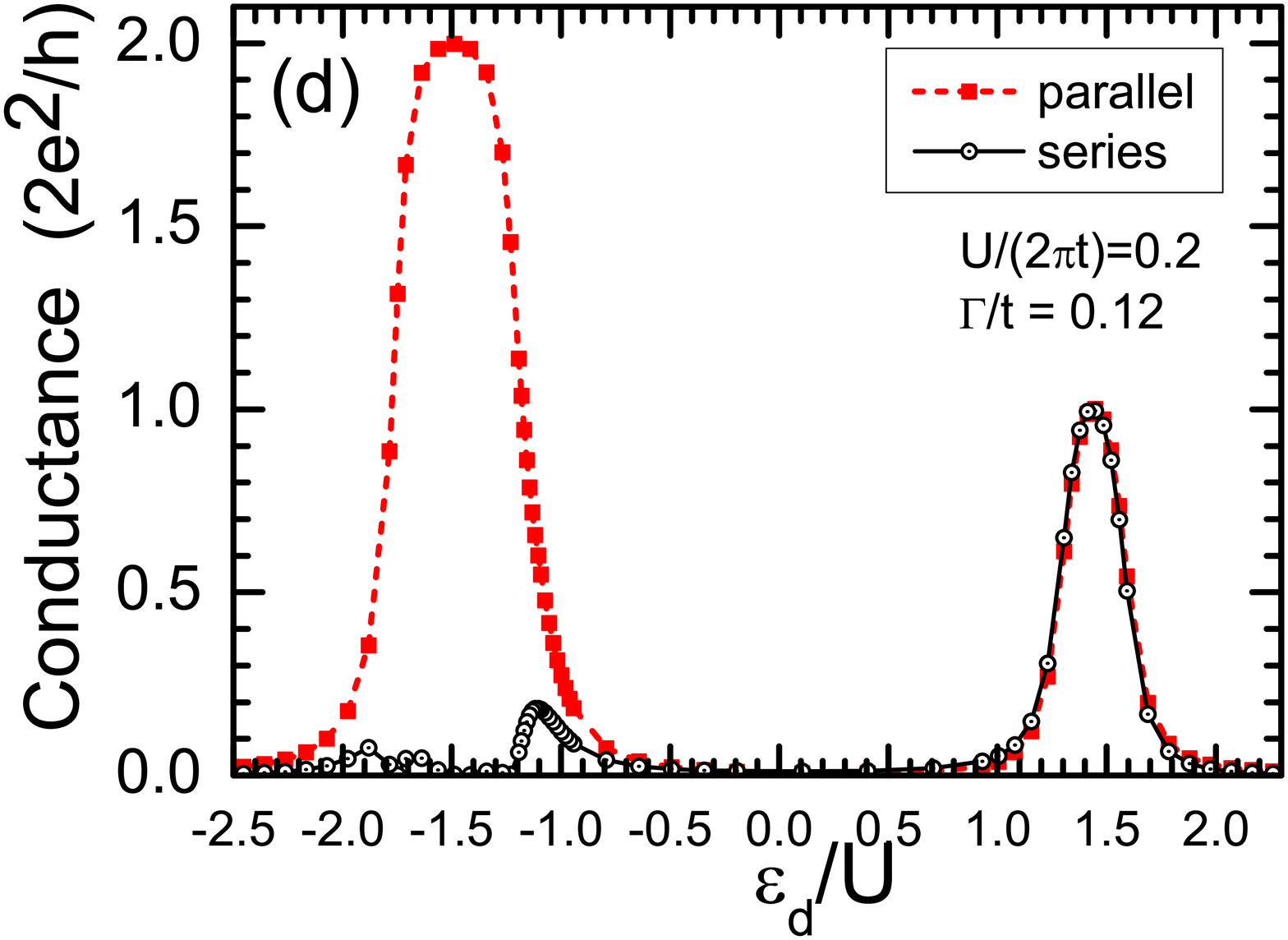}
\end{minipage}

 \caption{(Color online) 
Phase shifts and conductances 
for small values of $U$ 
(upper panels) $U/(2\pi t)=0.5$ and (lower panels) $U/(2\pi t) = 0.2$;
 $\Gamma/t=0.12$ is fixed. 
The vertical dashed lines in (a) and (c) correspond to $\epsilon_d$, 
at which $N_\mathrm{tot}$ for $\Gamma = 0$ varies discontinuously. 
%
 }
 \label{fig:cond_u05_02}
\end{figure}

\subsection{Small $\,U$ cases}
 \label{subsec:U_dependence}

We next consider the ground-state properties 
for weak interactions.
For comparison with the case of Fig.\ \ref{fig:cond_u1},
we have carried out the calculations,   
taking $U$ smaller than  $U/(2\pi t) =1.0$ 
and choosing the same value of $\Gamma/t = 0.12$ for the hybridization.
The results are plotted in Fig.\ \ref{fig:cond_u05_02};   
the upper panels (a)--(b) for $U/(2 \pi t)=0.5$,
and the lower panels (c)--(d) for $U/(2 \pi t)=0.2$. 
The vertical dashed lines in the panels (a) and (c), 
show the values of the $\epsilon_d$  
at which the occupation number $N_\mathrm{tot}$ changes 
in the isolated limit $\Gamma =0$.

We can see that the area for three-electron 
filling  becomes broad as $U$ increases.
Note that in the non-interacting case  
the occupation number $N_\mathrm{tot}$ does not take the steps 
at $N_\mathrm{tot}\simeq 1.0,\,3.0$ and $\,4.0$ 
as shown in Fig.\ \ref{fig:cond_u0}(b). 
Therefore the effects of the Coulomb interaction 
are particularly important for these electron fillings.
The conductance peak at $N_\mathrm{tot}\simeq 1.0$  
becomes wider as $U$ increases, 
and the peak shape deviates from a simple Lorentzian form 
which is seen  in Fig.\ \ref{fig:cond_u0}(b) for $U=0$. 
This is caused by the SU(2) Kondo effects 
due to the moment of a single electron 
occupying the local one-particle state $E_{k=0}^{(1)}$ 
in the triangle, as mentioned.

We see in (a) and (b) of Fig.\ \ref{fig:cond_u05_02}  
that the overall feature of the phase shifts and the conductances 
for $U/(2 \pi t)=0.5$ are similar to the those  
for $U/(2 \pi t)=1.0$ shown in Fig.\ \ref{fig:cond_u1}.
For instance, the dip structure of the series conductance 
is clearly seen for $N_\mathrm{tot}\simeq 3.0$ in (b),
which reflects the behavior of 
the phase shifts at $ -0.8 \lesssim \epsilon_d/U \lesssim -0.6$ 
in Fig.\ \ref{fig:cond_u05_02} (a). 
Particularly, the peak of $g_\mathrm{s}$ on the left side of the dip 
is caused by a concaving behavior of $\delta_\mathrm{o}$. 
The occupation number $N_\mathrm{tot}$, however,  
no longer shows a clear staircase behavior 
at three-electron filling 
for $U/(2 \pi t)=0.5$.
Correspondingly, we see in Fig.\ \ref{fig:cond_u05_02} (b) 
that the parallel conductance  $g_\mathrm{p}$ 
has a shoulder behavior, instead of 
a well-defined plateau,
 at $-0.8\lesssim \epsilon_d/U \lesssim -0.6$.

These plateau and valley structures almost disappear   
for the smaller interaction $U/(2 \pi t)=0.2$, 
as shown in the lower panels (c) 
and (d) of Fig.\ \ref{fig:cond_u05_02}.
Specifically, the region for three-electron filling,
which corresponds to $-1.2 \lesssim \epsilon_d/U \lesssim -1.1$, 
becomes very narrow, as mentioned. 
Nevertheless,  we have confirmed 
that the  difference between the two phase shifts 
$\delta_\mathrm{e} -\delta_\mathrm{o}$  
crosses the value of $\pi$  at 
the points $\epsilon_d/U \simeq -1.2$, $-1.4$ and $-1.7$,
which correspond to the zero points of $g_\mathrm{s}$.
The series conductance has a weak peak  
between the two adjacent zero points, although 
it is not visible on the scale of Fig.\ \ref{fig:cond_u05_02}.

The parallel conductance at 
four-electron filling, which is shown in (b) and (d) 
of Fig.\ \ref{fig:cond_u05_02}, 
keeps the features of the $S=1$ Kondo behavior 
seen for larger $U$ in Fig.\ \ref{fig:cond_u1}.
This is because the Coulomb interaction 
 is still larger 
than the inter-dot hopping matrix element $t$, 
even for $U/(2\pi t) =0.2$. 
Correspondingly, we can see in Fig.\ \ref{fig:cond_u05_02} (c) 
that the even phase shift shows a clear plateau 
behavior for $\delta_\mathrm{e} \simeq 3\pi/2$ 
at $-1.7 \lesssim \epsilon_d/U \lesssim -1.3$,
although the odd phase shift $\delta_\mathrm{o}$ 
 behaves moderately.
Note that the energy gain of the Nagaoka state is given by 
$\Delta E^{(4)} = 0.319\, t$ for $U/(2\pi t)=0.2$, 
and  $\Delta E^{(4)} = 0.563\, t$ for $U/(2\pi t)=0.5$.
Therefore, in both of these two cases 
the energy $\Delta E^{(4)}$ is greater 
than $\Gamma$ ($=0.12t$), and it makes the $S=1$ 
Kondo behavior robust.

\section{Thermodynamic quantities}
\label{sec:thermo}

We present the NRG results for the magnetic susceptibility  
and the entropy of the TTQD in this section.
The Kondo behavior in a wide energy scale can be 
 investigated from the temperature dependence 
of these thermodynamic quantities. 
Particularly, we study how the screening of the local moment 
is completed in the case of the $S=1$ high-spin Nagaoka state 
 at $N_\mathrm{tot} \simeq 4.0$.  
We investigate also the SU(4) Kondo behavior, which takes place  
at the series conductance dip in the middle 
of the plateau at $N_\mathrm{tot} \simeq 3.0$.

 \subsection{Entropy \&  spin susceptibility}

The free energy of the whole system consisting of 
the dots and leads is given by 
 $F= -T\log \left[\mbox{Tr}\, e^{-\mathcal{H}/T}\right]$.
The contribution of the quantum dots on this free energy 
can be extracted, by subtracting the contribution of 
the non-interacting leads, 
$F_\mathrm{lead} = -T 
\log\left[ \mbox{Tr}\, e^{-\mathcal{H}_\mathrm{lead}/T}
\right]$, from the total free energy,\cite{KWW}  
\begin{align}
& \Delta F \, \equiv \, 
F - F_\mathrm{lead} \;.
\label{eq:free_energy}
\end{align}
Then, the entropy $\mathcal{S}$ and the spin susceptibility $\chi$ 
due to the dots can be obtained from    
\begin{align}
\mathcal{S} \equiv - \frac{\partial \Delta F}{\partial T}
\;, \qquad \quad
\chi \equiv -\frac{\partial^2 \Delta F}{\partial H^2}
\;.
\end{align}
Here, the magnetic field $H$ is introduced 
both for the dots and leads, 
adding  
the Zeeman energy $\mathcal{H}_\mathrm{ex}$ to 
the Hamiltonian $\mathcal{H}$ defined in Eq.\ \eqref{eq:H}, 
\begin{align}
\mathcal{H}_\mathrm{ex} 
= & - \sum_{i=1}^{3} \left(
 n_{d,i\uparrow} - n_{d,i\downarrow} \right) H 
\nonumber \\
 & \,  
- \! 
\sum_{\nu=L,R} 
 \sum_{k} \left(
         c^{\dagger}_{k \nu \uparrow} 
         c^{\phantom{\dagger}}_{k \nu \uparrow}
-      c^{\dagger}_{k \nu \downarrow} 
         c^{\phantom{\dagger}}_{k \nu \downarrow}
\right) H . 
\end{align}
These thermodynamic quantities can be 
evaluated using the eigenvalues of 
the discretized Hamiltonian of the NRG.\cite{KWW}

\begin{figure}[t]
 \leavevmode
\begin{minipage}{1\linewidth}
\includegraphics[width=1.0\linewidth]{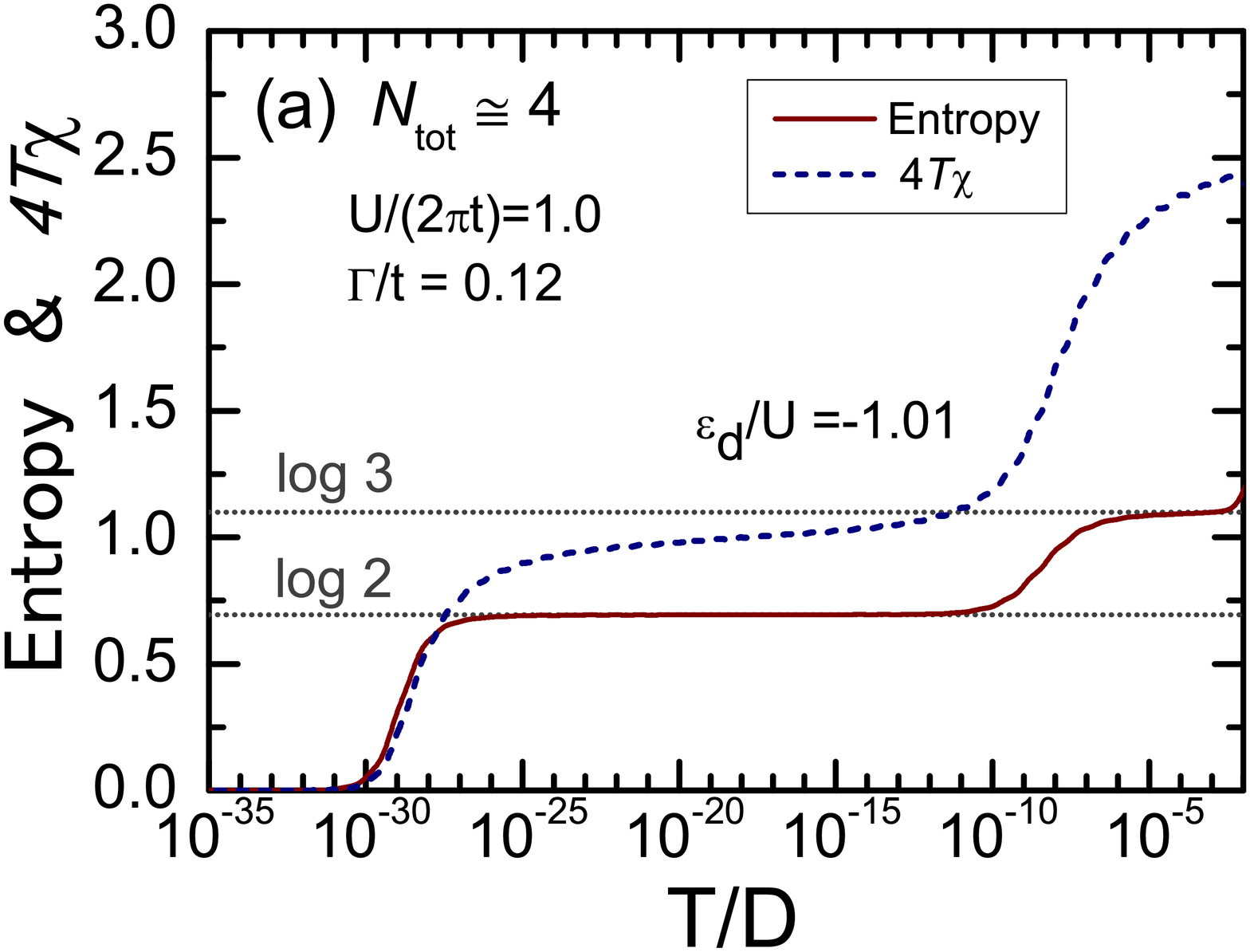}
\end{minipage}
\begin{minipage}[t]{1\linewidth}
 \includegraphics[width=1.0\linewidth]{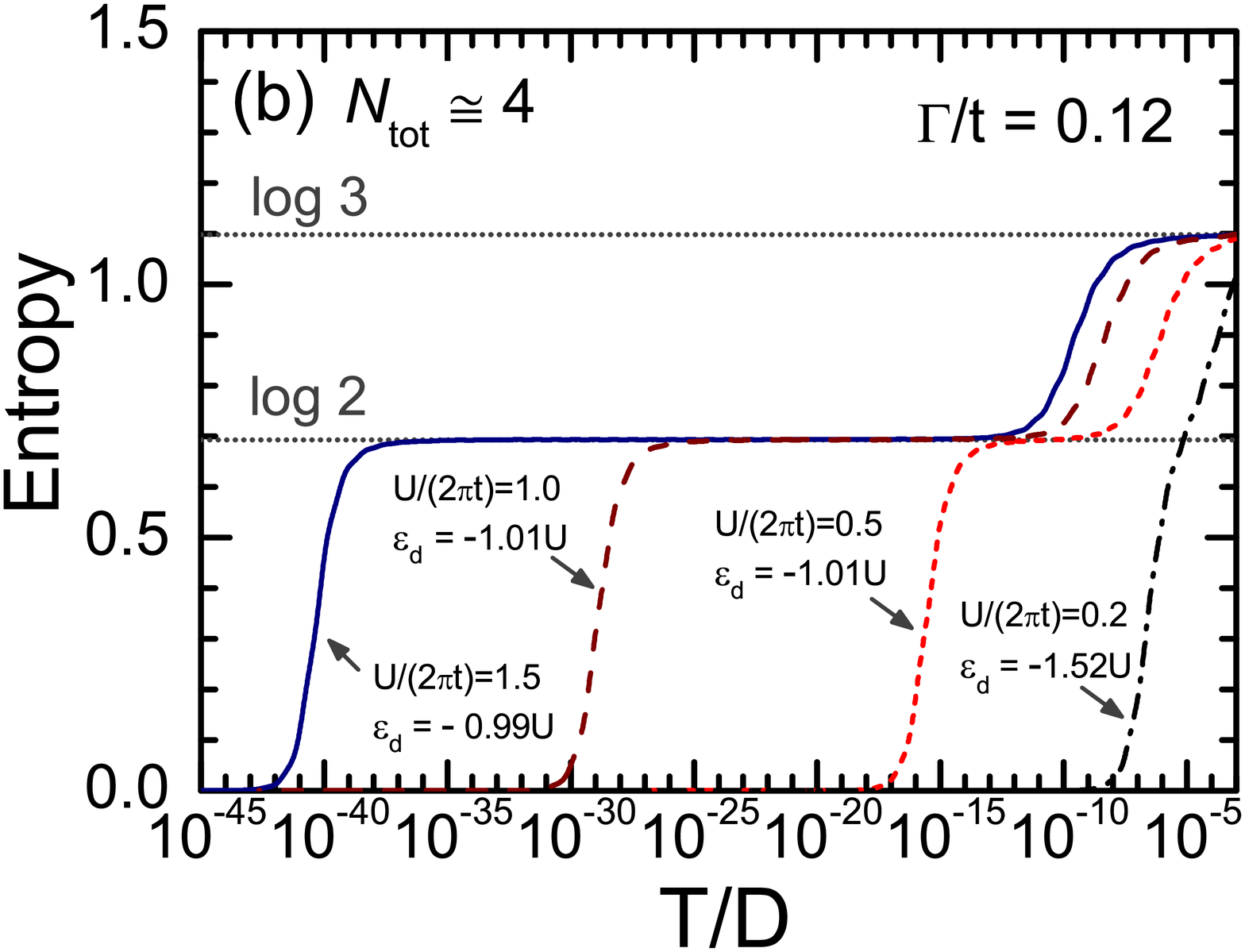}
\end{minipage}
 \caption{(Color online) 
Entropy $\mathcal{S}$ and spin susceptibility $4 T \chi$ 
in the middle of the charge step for $N_\mathrm{tot}\simeq 4.0$ 
are plotted vs $\log (T/D)$ for 
 $t'=t$, $\epsilon_\mathrm{apex} =\epsilon_d$ 
and $\Gamma/t = 0.12$, where
$D$ the half-width of the conduction band. 
In (a) both $\mathcal{S}$ and $4 T \chi$ are plotted  
for $U/(2\pi t)=1.0$ and $\epsilon_d = -1.01U$. 
In (b) entropy is plotted for several values of $U$:  
  (dash-dot line) $U/(2\pi t)=0.2$, $\epsilon_d = -1.52U$,
  (dotted line) $U/(2\pi t)=0.5$, $\epsilon_d = -1.01U$,  
  (dashed line) $U/(2\pi t)=1.0$, $\epsilon_d = -1.01U$, and
  (solid line) $U/(2\pi t)=1.5$, $\epsilon_d = -0.99U$.
%
 }
 \label{fig:S_Tchi4}
\end{figure}

\subsection{Screening of $\,S=1$ Nagaoka high spin 
}
\label{subsec:nagaoka_thermo}

The entropy $\mathcal{S}$ and $4 \chi T$ 
are plotted in Fig.\ \ref{fig:S_Tchi4} (a) as functions of $\log(T/D)$ 
for a parameter set, $U/(2 \pi t) =1.0$ and $\epsilon_d = -1.01U$,  
which corresponds to 
the middle of the charge step for $N_\mathrm{tot} \simeq 4.0$ 
in Fig.\ \ref{fig:cond_u1}.
We can  see that 
the screening of the spin $S=1$ 
local moment is achieved via two separate stages.  
The first and second stages take place  
at $T/D \simeq 10^{-8}$ and  $T/D \simeq10^{-29}$, respectively. 
At high temperatures  $10^{-7} \lesssim  T/D \lesssim 10^{-3}$,   
the entropy shows a plateau of the value of $\mathcal{S}\simeq\log 3$, 
which represents the contribution of a free unscreened $S=1$ moment. 
Then, at intermediate temperatures 
  $10^{-27} \lesssim T/D \lesssim 10^{-10}$,  
the entropy takes the other value $\mathcal{S}\simeq\log 2$ 
corresponding to a spin $1/2$ moment. 
In this region, the spin susceptibility also shows a Curie behavior    
$4 \chi T \simeq 1.0 \Rightarrow 4 s(s+1)/3$ 
with a coefficient $s=1/2$ due to a free spin.
Therefore  in the intermediate temperature region,  
 half of the $S=1$ moment 
is screened by the conduction electrons 
from one of the channel degrees of freedom, 
and the local moment is still in an under-screened 
situation with the remaining free spin of $s=1/2$. 
The full screening is completed 
at low temperature of the order $T/D \simeq 10^{-30}$ 
in the case of Fig.\ \ref{fig:S_Tchi4} (a).
This crossover energy scale to 
the low-energy Fermi-liquid regime 
 can be regarded as 
a Kondo temperature for the $S=1$ local moment 
 in the present case.
We refer to the crossover energy scale to 
the singlet ground state as the screening temperature $T^*$, 
and use this definition 
for a wide range of electron 
filling $\,0 < N_\mathrm{tot} < 6$.

The reason that the two stages are required to  
reach the Kondo singlet state can be understood 
from the distribution of the $S=1$ moment 
in the triangular triple dot.
Specifically in the limit of $\Gamma=0$,
the partial moment $m_{a/b,i}$, 
which constitutes the $S=1$ moment 
as shown in appendix \ref{sec:nagaoka_eigenstate},
distributes in the even-odd orbitals $a_{0}$, $a_{1}$ and $b_{1}$ 
with the weights $1/3$, $1/6$, and $1/2$, respectively.
Among the three orbitals, 
 $a_{1}$ and $b_{1}$ are coupled directly 
to the lead, as illustrated in Fig.\ \ref{fig:even_odd}.
Therefore, the partial moments in these two orbitals 
can be screened directly by the conduction electrons at the first stage. 
In contrast,  the other orbital $a_{0}$ 
has no direct tunneling matrix element to the leads, and thus
 the partial moment in the $a_0$ orbital  
has to be screened indirectly by the conduction electrons 
coming over the $a_{1}$ orbital 
which is already filled up to $\langle n_{a,1} \rangle \simeq 1.7$,
 as mentioned in \ref{subsec:results_org}.
This amount of the occupation number is caused mainly 
by the difference between the onsite potential 
$\epsilon_{a0} = \epsilon_d$ and $\epsilon_{a1} = \epsilon_d -t$ 
in the even-odd basis, namely $\epsilon_{a1}$ is deeper 
than $\epsilon_{a0}$   
(see also appendix \ref{sec:nagaoka_eigenstate}). 
Therefore, in this situation 
the charge and spin fluctuations must be  
nearly frozen in the $a_1$ orbital, 
and the screening temperature 
for the partial moment in the apex site $a_{0}$ becomes low.
These features of the screening at the second stage are 
analogous to those of a super-exchange mechanism.

We see in Fig.\ \ref{fig:S_Tchi4} (a) that 
the essential feature of the screening can be 
observed either in $\mathcal{S}$ or $T \chi$.
At the transient regions, however, 
$\mathcal{S}$ varies more clearly than $T \chi$ does.
Therefore, in order to see how the Coulomb interaction $U$ 
affects the two-stage screening  of the $S=1$ 
moment at $N_\mathrm{tot} \simeq 4.0$,   
the entropy $\mathcal{S}$ is plotted in Fig.\ \ref{fig:S_Tchi4} (b) 
for several values of $U$, choosing 
the parameter set ($U/(2\pi t)$, $\epsilon_d/U$) 
in a way such that it corresponds to   
the middle of the plateau of the parallel conductance,  
 solid line: ($1.5$, $-0.99$), 
 dashed line: ($1.0$, $-1.01$), 
dotted line: ($0.5$, $-1.01$), and 
dash-dot line: ($0.2$, $-1.52$). 
The hybridization is chosen to be $\Gamma/t =0.12$.
Note that the ground-state properties for 
these two small $U$ cases  
were presented in Fig.\ \ref{fig:cond_u05_02}.
We see in Fig.\ \ref{fig:S_Tchi4} (b) 
that the crossover temperature 
decreases as $U$ increases.
Particularly, the second screening stage is more 
sensitive to $U$ than 
the first stage which occurs at higher temperature. 
We can also see that the crossover temperature 
of the first stage and that of the second one 
 become close for small $U$.
For instance, 
the two stages take place almost successively for $U/(2 \pi t) =0.2$. 
Note that the value of the coupling between the TTQD and leads 
chosen for  Fig.\ \ref{fig:S_Tchi4} is relatively small $\Gamma/t=0.12$, 
so that the screening temperature to reach the singlet state 
becomes very low in the present case, especially for large $U$. 
The screening temperature, however, rises as $\Gamma$ increases.
It would make $T^*$ an accessible value in experiments.
We will examine the $\Gamma$ dependence of $T^*$  
in Sec.\ \ref{subsec:gamma_vs_screening}.

\begin{figure}[t]
 \leavevmode
\begin{minipage}[t]{1.0\linewidth}
\includegraphics[width=0.98\linewidth]{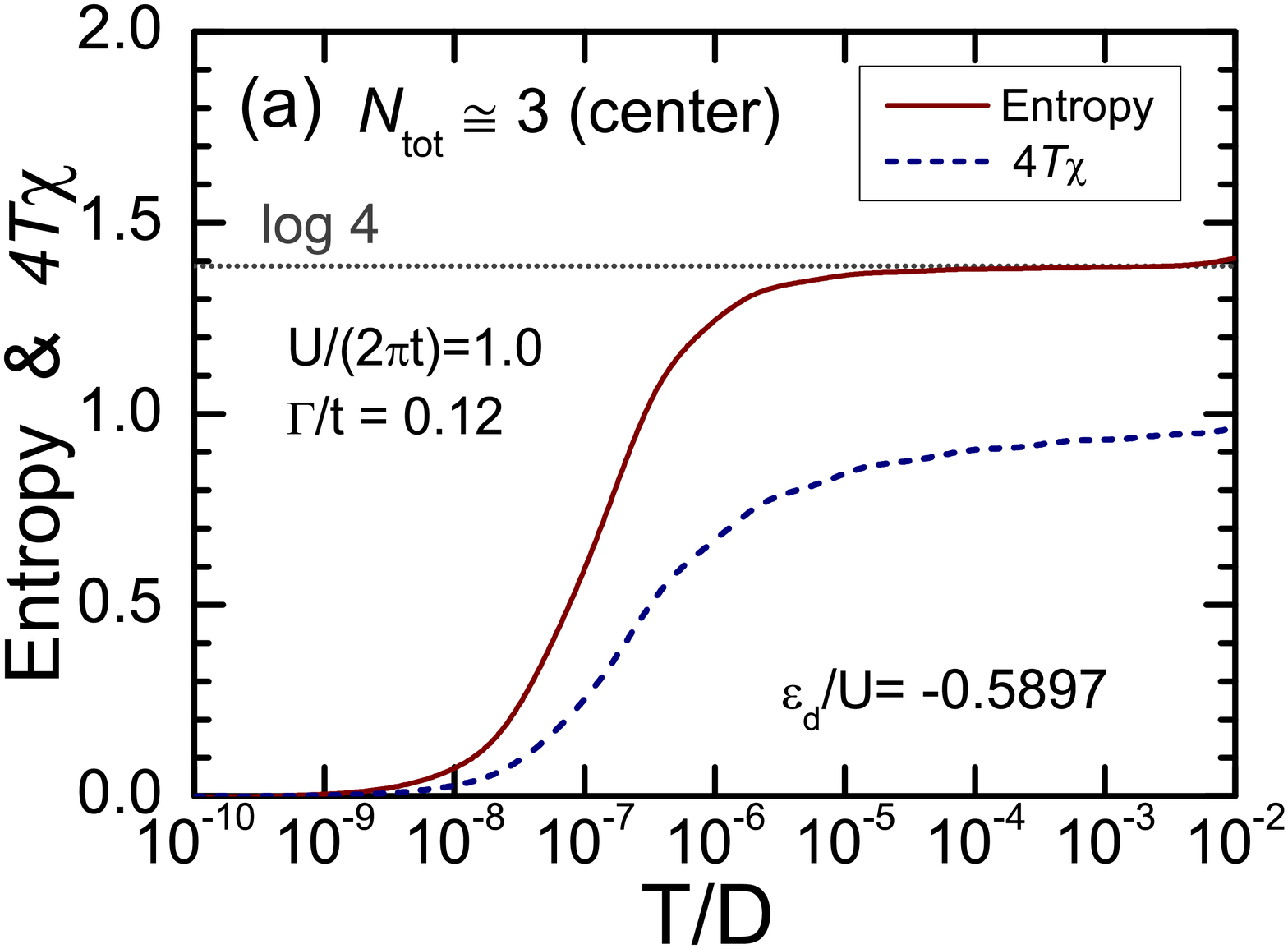}
\end{minipage}
%
\begin{minipage}[t]{1.0\linewidth}
\includegraphics[width=0.96\linewidth]{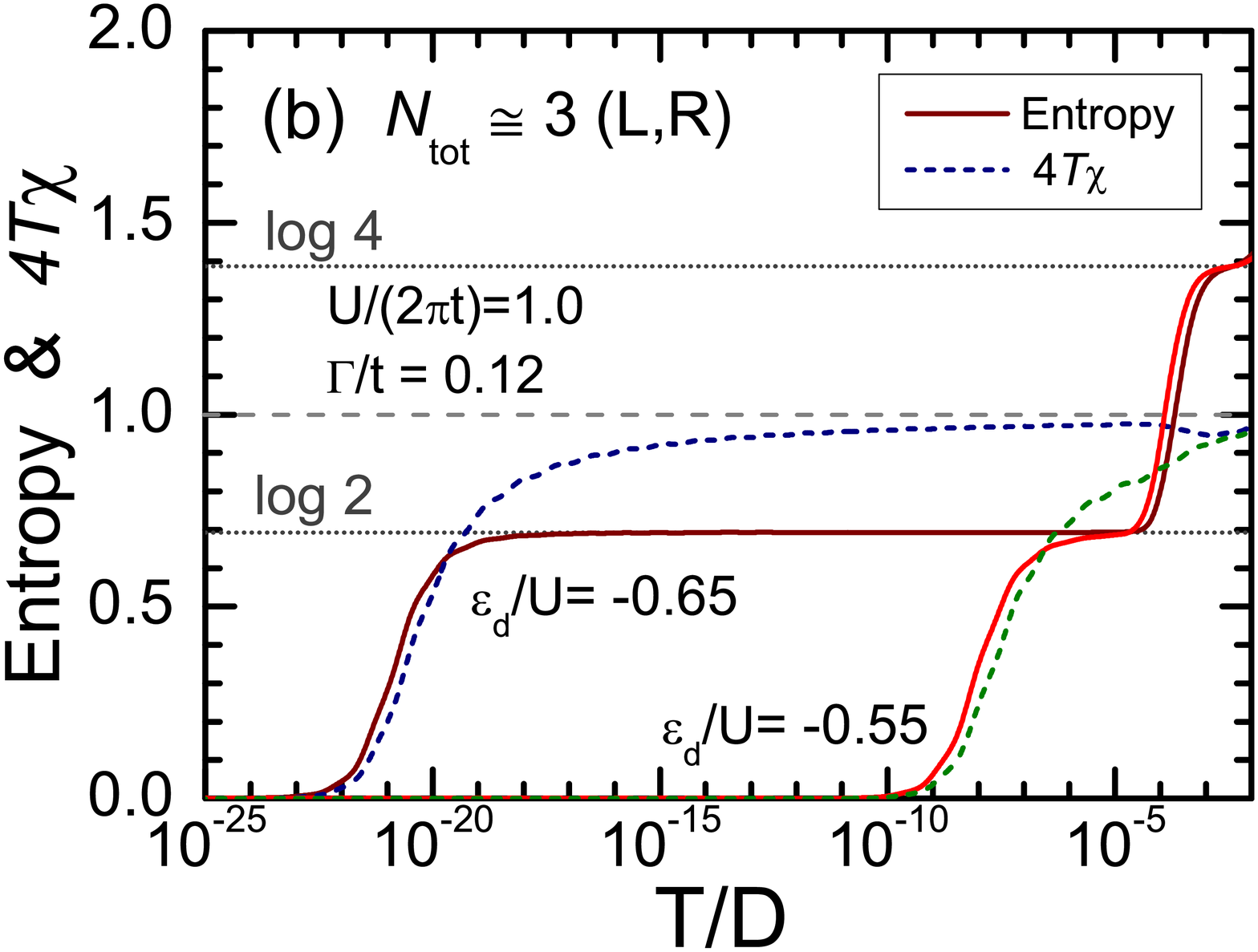}
\end{minipage}
 \caption{(Color online)
Entropy $\mathcal{S}$ and spin susceptibility $4 T \chi$ 
in the step for $N_\mathrm{tot}\simeq 3.0$ are plotted vs $\log (T/D)$  
for  $t'=t$,  $\epsilon_\mathrm{apex} =\epsilon_d$ 
 and $\Gamma/t = 0.12$. 
Impurity level is chosen to be 
 (a) just at the series conductance dip $\epsilon_d = -0.5897U$, 
(b) on the left $\epsilon_d = -0.65U$ and 
the right $\epsilon_d = -0.55U$ sides of the dip.
 }
\label{fig:S_Tchi3}

\end{figure}


\subsection{SU(4) Kondo effect 
at the dip for $N_\mathrm{tot} \simeq 3.0$
}
\label{subsec:su4_thermo}

In this subsection 
we present the results of the thermodynamic quantities 
at three-electron filling $N_\mathrm{tot} \simeq 3.0$.
In Fig.\ \ref{fig:S_Tchi3}, 
 the entropy and  spin susceptibility  
are plotted vs $\log (T/D)$, 
for $U/(2\pi t) = 1.0$, and $\Gamma/t=0.12$.
The impurity level $\epsilon_d$ is chosen 
in a way such that   (a)  $\epsilon_d = -0.5897U$ 
at the zero point of $g_\mathrm{s}$ 
in the middle of the sharp dip seen in Fig.\  \ref{fig:cond_u1}, 
 and (b) $\epsilon_d = -0.65U$ ($-0.55U$) 
at a point on the left (right) of the dip.

We can see in Fig.\ \ref{fig:S_Tchi3} (a) 
that at high temperatures $10^{-5} \lesssim T/D \lesssim 10^{-2}$ 
the entropy is a constant $\mathcal{S} \simeq \log 4$, 
which is caused by the four-fold degeneracy of 
the local states in the isolated TTQD cluster  
 mentioned in Sec.\ \ref{subsec:isolated_cluster}.
Furthermore, the spin susceptibility shows  
the Curie behavior $\chi \simeq s(s+1)/(3T)$ with $s=1/2$ 
at high temperatures.
Then, at low temperatures 
both $\mathcal{S}$ and $\chi T$ decrease with $T$. 
Specifically, the screening of 
 the local four-fold degenerate states is completed 
in a single stage at $T^*/D \simeq 10^{-7}$. 
This behavior is observed just at the zero point,
at which the two phase shifts 
are $\delta_\mathrm{e}\simeq 5\pi/4$ and 
$\delta_\mathrm{o}\simeq \pi/4$. 
At the zero point 
the low-energy excitations from the singlet ground state 
asymptotically have the channel symmetry  
in addition to the global SU(2) symmetry of the spin,
 as shown in Sec.\ \ref{sec:zero_point_SU4}. 
Therefore, the low-temperature behavior of 
these thermodynamic quantities must reflect  
the low-energy SU(4) symmetry of the fixed-point Hamiltonian.
We refer to this zero point of the series conductance 
at $N_\mathrm{tot}\simeq 3.0$ as 
the SU(4) symmetric point in the following.
It should be noted that $T^*$ is enhanced at this point 
due to the channel degeneracy.

For the values of $\epsilon_d$ away from the dip,
the entropy shows the two-stage behavior.
We can see in Fig.\ \ref{fig:S_Tchi3} (b) that 
the entropy still shows the value of  
$\mathcal{S} \simeq \log 4$  
at high temperatures even in the case away from the dip. 
At intermediate temperatures, however, 
it takes another plateau with $\mathcal{S} \simeq \log 2$,
and the spin susceptibility shows 
the Curie behavior with the coefficient of $s=1/2$.
Therefore, the orbital part of the degrees of freedom  
is frozen at intermediate temperatures,
and the spin degrees of freedom still remains free.
 Furthermore, away from dip 
the low-lying energy excited states 
no longer have the channel symmetry.
The full screening of the remaining spin $1/2$ 
 is completed  
at $T^*/D \simeq 10^{-21}$ for $\epsilon_d =-0.65U$, 
and $T^*/D \simeq 10^{-8}$ for $\epsilon_d =-0.55U$. 
These temperatures are lower 
than the screening temperature at the SU(4) symmetric point, 
but still much higher 
than $T^*$  for the $S=1$ Kondo effect  
at $N_\mathrm{tot}=4$, for the same parameter values  
of $U$, $t$, and $\Gamma$.

Furthermore, the screening temperature 
 on the right-hand side of the SU(4) symmetric point 
is higher than that on the left-hand side. 
The difference reflects the charge redistribution, 
taking place at the conductance dip 
as discussed in Sec.\ \ref{subsec:results_org}. 
As we saw in Fig.\ \ref{fig:even_odd_occupation}, 
the distribution is almost homogenous on the right side;
$\langle n_{a,0} \rangle \simeq 
\langle n_{a,1} \rangle \simeq 
\langle n_{b,1} \rangle \simeq 1.0$.
In contrast, it becomes inhomogeneous on the left side; 
$\langle n_{a,0} \rangle \simeq 1.0$, 
$\langle n_{a,1} \rangle \simeq 1.5$ and 
 $\langle n_{b,1} \rangle \simeq 0.5$.
In the second  screening stage
the partial moment in the $a_0$ orbital, 
which is singly occupied on average, 
has to be screened by the electrons coming over the $a_1$ orbital.
Therefore, $T^*$ must be affected by the filling of the $a_1$ orbital. 
Specifically, the charge and spin fluctuations 
at the $a_1$ orbital, which is necessary for the second stage screening, 
tends to be frozen as $\langle n_{a,1} \rangle$ approaches to $2.0$.
The situation is similar to that in the $S=1$ Kondo case, 
mentioned in Sec.\ \ref{subsec:nagaoka_thermo}. 
We will discuss this point further in the next subsection.

The conductances which we discussed 
in Sec.\ \ref{sec:results_I} are 
the results obtained at zero temperature.
If the temperature is raised, 
the series conductance at the SU(4) symmetric point will increase. 
Simultaneously, the series conductance plateaus  
on the left- and right-hand sides  will decrease. 
Then, the dip structure will disappear 
at the temperature of $\,T^*$ on the left side, 
because it is the lowest energy scale near the symmetric point.
Note that the screening temperature $T^*$ itself 
varies depending on the parameters $\Gamma$, $t$ and $U$. 
In order to preserve the typical structure of the dip, however, 
there are upper and lower bounds for $\Gamma/t$ and $U/t$, 
respectively, as we can see 
in Figs.\ \ref{fig:cond_u1_gam025_05} and \ref{fig:cond_u05_02}.

\begin{figure}[t]
 \leavevmode
\begin{minipage}[t]{0.49\linewidth}
\includegraphics[width=0.97\linewidth]{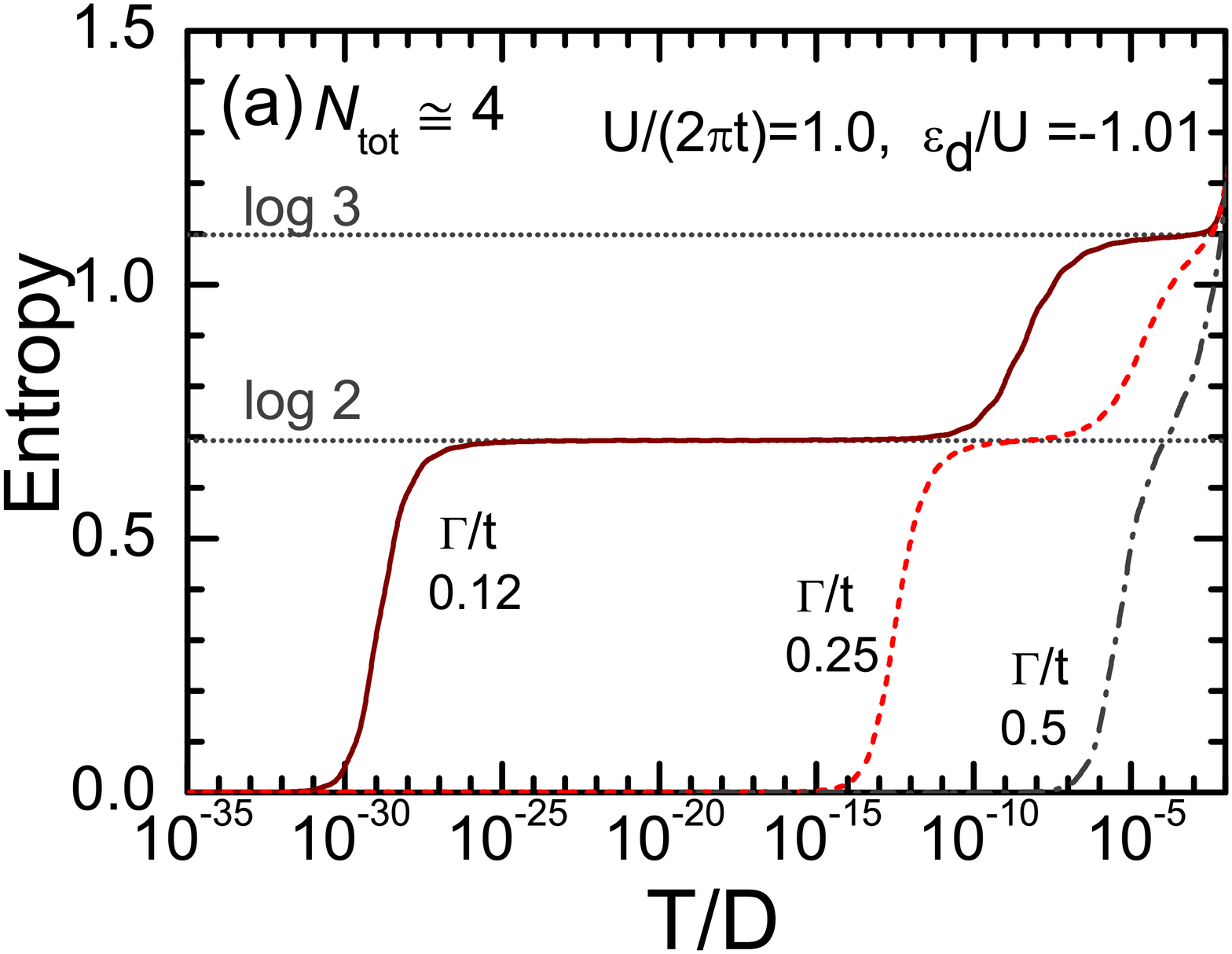}
\end{minipage}
\begin{minipage}[t]{0.49\linewidth}
 \includegraphics[width=1.02\linewidth]{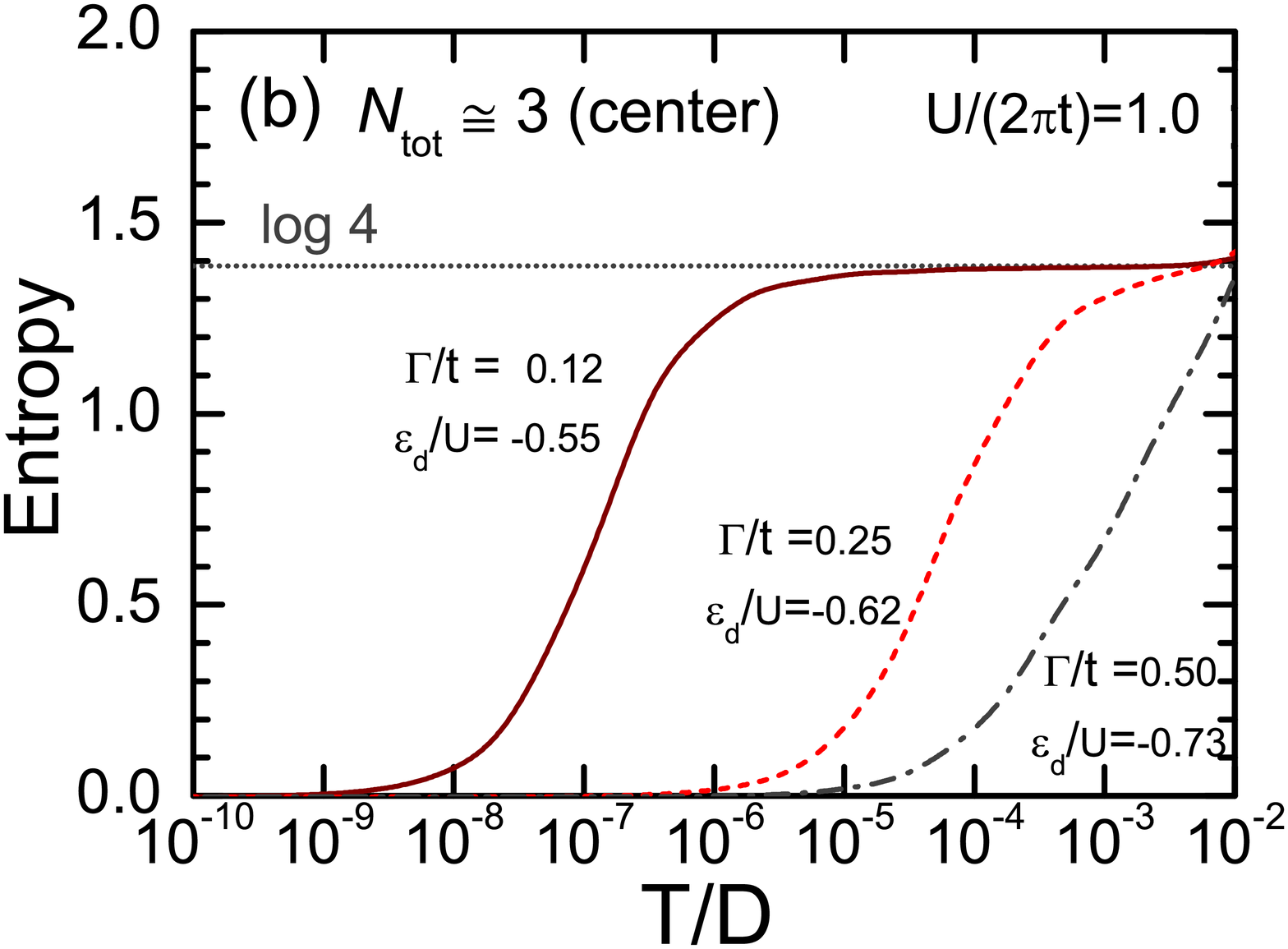}
\end{minipage}
 \caption{(Color online) 
Entropy vs $\log (T/D)$ 
for several values of $\Gamma$;  $U/(2\pi t)=1.0$ is fixed.
Impurity level is chosen to be (a) 
$\epsilon_d /U = -1.01$ which corresponds to $N_\mathrm{tot}\simeq 4.0$, 
and (b) at the dip in the middle of the step 
for $N_\mathrm{tot}\simeq 3.0$ 
($\epsilon_d/U = -0.55,\, -0.62$, and $-0.73$  
for $\Gamma/t = 0.12,\,0.25$, and $0.5$, respectively). 
 }
 \label{fig:S_Tchi_gam}
\end{figure}

\subsection{Dependence of the screening temperature on $\Gamma$ }
\label{subsec:gamma_vs_screening}

We consider in this subsection 
the dependence of the screening temperature $T^*$ 
on the hybridization $\Gamma$ and the level position $\epsilon_d$.
For comparisons, we have calculated the entropy  
for several values of the hybridization 
 $\Gamma/t=0.12$, $0.25$ and $0.5$, 
for a fixed value of the interaction $U/(2 \pi t)=1.0$.
The ground state properties for these parameter sets were 
presented in Figs.\ \ref{fig:cond_u1} 
and \ref{fig:cond_u1_gam025_05}.

Figure \ref{fig:S_Tchi_gam} shows 
the results obtained at 
(a) the parallel conductance plateau 
for $N_\mathrm{tot} \simeq 4.0$ ($\epsilon_d/U =-1.01$),  
and (b) the zero point of the series conductance 
for $N_\mathrm{tot} \simeq 3.0$.
Therefore, we can see in (a) the $\Gamma$ dependence of  
 the two-stage screening of 
the $S=1$ moment of the Nagaoka state, 
and in (b) the single-stage SU(4) Kondo behavior.
In both of these two cases,
the entropy for the intermediate coupling $\Gamma/t = 0.25$ 
retains the structures which we observed 
in the weak coupling case $\Gamma/t=0.12$.
For large hybridization  $\Gamma/t =0.5$, however, 
the charge transfer between the leads and  the TTQD 
brings the system into a mixed-valence regime, 
and the typical structures are smeared.
Furthermore, we can see in (a) and (b) that 
the crossover temperature $T^*$ is 
sensitive to the value of the hybridization, 
and it increases rapidly with the coupling $\Gamma$.


\begin{figure}[t]
 \leavevmode
\begin{minipage}{1.0\linewidth}
\includegraphics[width=1.0\linewidth]{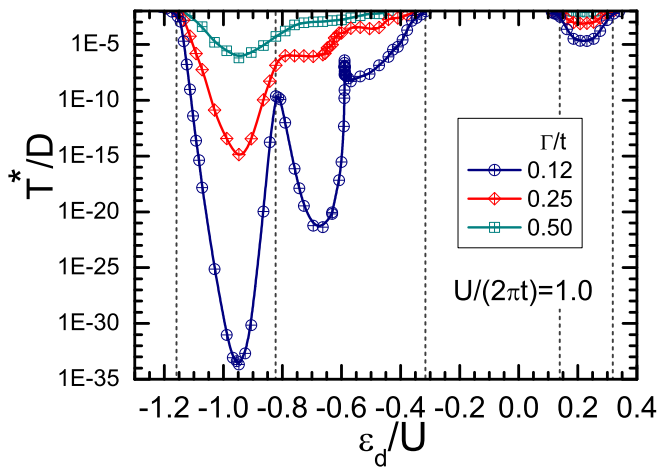}
\end{minipage}
 \caption{(Color online) 
The crossover energy scale $T^*$ as a function of $\epsilon_d$ 
for several values of  $\Gamma/t = 0.12, 0.25$, and $0.5$; 
  $U/(2\pi t)=1.0$ is fixed. 
$T^*$ is defined as the temperature in the middle of 
the low-energy crossover observed in the entropy.
The vertical dashed lines correspond to $\epsilon_d$, 
at which $N_\mathrm{tot}$ in the limit of $\Gamma = 0$ 
varies discontinuously. 
 }
 \label{fig:T_K}

\end{figure}


We have calculated the entropy also for different fillings, 
and  determined the screening temperature $T^*$ from 
the mid-point of a low-energy crossover to a singlet ground state.  
The energy scale $T^*$ defined in this way  
can be regarded as a Kondo temperature 
in the case that the system has a well 
defined local moment at high energies, 
 which is the case of  
$N_\mathrm{tot}\simeq 1.0$, $3.0$, and $4.0$.
The results of $T^*$ are plotted as a function 
of $\epsilon_d$ in Fig.\ \ref{fig:T_K}.
We see that at the single-electron filling 
for $0.1U \lesssim \epsilon_d \lesssim 0.3U$ 
the screening temperature is larger 
than that of the other regions of the filling. 
This is because at this filling the single electron  
enters mainly the $a_1$ orbital which 
is connected directly to one of the leads,
as illustrated in Fig.\ \ref{fig:even_odd}.

In the region of the three-electron filling $N_\mathrm{tot}\simeq 3.0$,
the screening temperature is enhanced 
at the SU(4) symmetric point, $\epsilon_d \simeq -0.59U$,
where the series conductance has a dip. 
We see in Fig.\ \ref{fig:T_K} that 
$T^*$ in the weak hybridization case $\Gamma/t=0.12$ 
has a local minimum at $\epsilon_d \simeq -0.65U$ 
on the left side of the symmetric point,
and $T^*$ becomes very low near the minimum. This is  
because the fraction of the local spin in the $a_0$ orbital 
must be screened indirectly by the conduction electrons 
coming through the $a_1$ orbital which is filled already 
up to $\langle n_{a,1} \rangle \simeq 1.5$. 
As we see in Fig.\ \ref{fig:even_odd_occupation},
there is a significant difference between 
the charge distribution on the left and right sides 
of the symmetric point. 
On the right side, 
at $-0.58\lesssim \epsilon_d/U\lesssim -0.4$,
the local moment is mainly in the odd-parity $b_1$ orbital 
which has a direct tunneling matrix element 
to one of the leads, as discussed in Sec.\ \ref{subsec:results_org}.
For this reason, the screening temperature on the right side of 
the dip becomes larger than that on the left side.

We see also in Fig.\ \ref{fig:T_K} that 
the screening temperature is enhanced significantly 
at $\epsilon_d \simeq -0.82U$ for $\Gamma/t =0.12$.  
This seems to be caused by the charge fluctuations 
between the two different fillings of $N_\mathrm{tot}=3$ 
and $N_\mathrm{tot}=4$.
We note that near this point there is another zero point 
of the series conductance at $\epsilon_d/U \simeq -0.88$, 
as we see in Fig.\ \ref{fig:diff_phase}.
At four-electron filling, $T^*$ becomes small again 
particularly at $\epsilon_d \simeq -0.95U$. 
This is because in the second screening stage of 
the $S=1$ moment the partial moment in the $a_0$ site 
needs to be screened by the electrons from the leads 
which have no direct hopping matrix elements to the apex site,   
as mentioned in Sec.\ \ref{subsec:nagaoka_thermo}.
We see also that the screening temperature  
depends sensitively on $\Gamma$, as well as on $U$.
Therefore, in order to observe the two-stage behavior experimentally, 
$\Gamma$ should be tuned to a large value, 
but still has to be $\Gamma \lesssim \max (U/3,t)$  
in order to retain the typical structures. 
This criterion is determined by the energy scale 
 $\Delta E^{(4)} \simeq \max (U/3,t)$, 
  defined in Eq.\ \eqref{eq:diff_Nagaoka_asym}.

\section{Deformations of triangle}
\label{sec:results_deformation}

So far, we have assumed the full $C_{3v}$ symmetry for the TTQD. 
Although it has already been broken partly by the connection 
to the two leads, the perturbations which  
break the triangular symmetry should be 
examined further, since real TTQD systems must 
be deformed from the regular structure to a certain extent.
We consider in this section how an asymmetry caused by
the inhomogeneity in the level positions   
and the inter-dot hopping matrix elements 
affect the ground-state properties.

In the presence of a deformation 
which is described by $\epsilon_\mathrm{apex}$ and $t'$, 
the one-particle energies of the isolated TTQD 
vary from those for the regular triangle 
given in Eq.\ \eqref{eq:tight_bind_triangle} 
to the following forms
\begin{align}
E_{\mathrm{e},\pm}^{(1)} =& \  
\frac{ \epsilon_\mathrm{apex}+\epsilon_d-t'}{2} 
\pm \sqrt{
\left(\frac{\epsilon_\mathrm{apex}-\epsilon_d+t'}{2}\right)^2 
+2t^2
}
\;,   \\
E_\mathrm{o}^{(1)} =& \ \epsilon_d + t' \;.
\end{align}
Here, 
$E_{\mathrm{e},\pm}^{(1)}$ and 
$E_{\mathrm{o}}^{(1)}$ are 
the eigenvalues for the states with even and odd parities, respectively.  
Note that the system  still has the inversion symmetry in the present case. 
Among these three one-particle states, 
$E_\mathrm{e,-}^{(1)}$ becomes 
the lowest energy for small perturbations.
The other two excited states become degenerate, 
 $E_\mathrm{e,+}^{(1)}=E_\mathrm{o}^{(1)}$,
in the regular triangle case 
with $t'=t$ and $\epsilon_\mathrm{apex} = \epsilon_d$.
The deformations lift this degeneracy, 
and the first order correction with respect to 
$\,\delta \epsilon \equiv \epsilon_\mathrm{apex}-\epsilon_d\,$ 
and $\,\delta t \equiv t'-t\,$ is given by  
\begin{align}
E_{\mathrm{e},+}^{(1)} - E_\mathrm{o}^{(1)} 
\,\simeq \,  \frac{2}{3}\,\delta \epsilon - \frac{4}{3}\, \delta t  
\;.
\label{eq:deformation_1st_order}
\end{align}
Thus, the energy of the even state $E_{\mathrm{e},+}^{(1)}$ 
becomes higher (lower) than that of the odd one $E_\mathrm{o}^{(1)}$ 
for $\delta \epsilon -2 \delta t >0\,$ ($<0$).  
It reflects the fact that the apex site with $\epsilon_\mathrm{apex}$ 
belongs to the even part of the basis. Furthermore, 
$t'$ shifts the onsite potential 
of the $b_1$ orbital upwards, and that of  
the $a_1$ downwards. 
In the following, we examine effects of 
$\,\delta \epsilon\,$ and $\,\delta t\,$ separately, 
in a wide range of the electron filling $N_\mathrm{tot}$.

\begin{figure}[t]
 \leavevmode
\begin{minipage}[t]{0.49\linewidth}
 \includegraphics[width=1.0\linewidth]{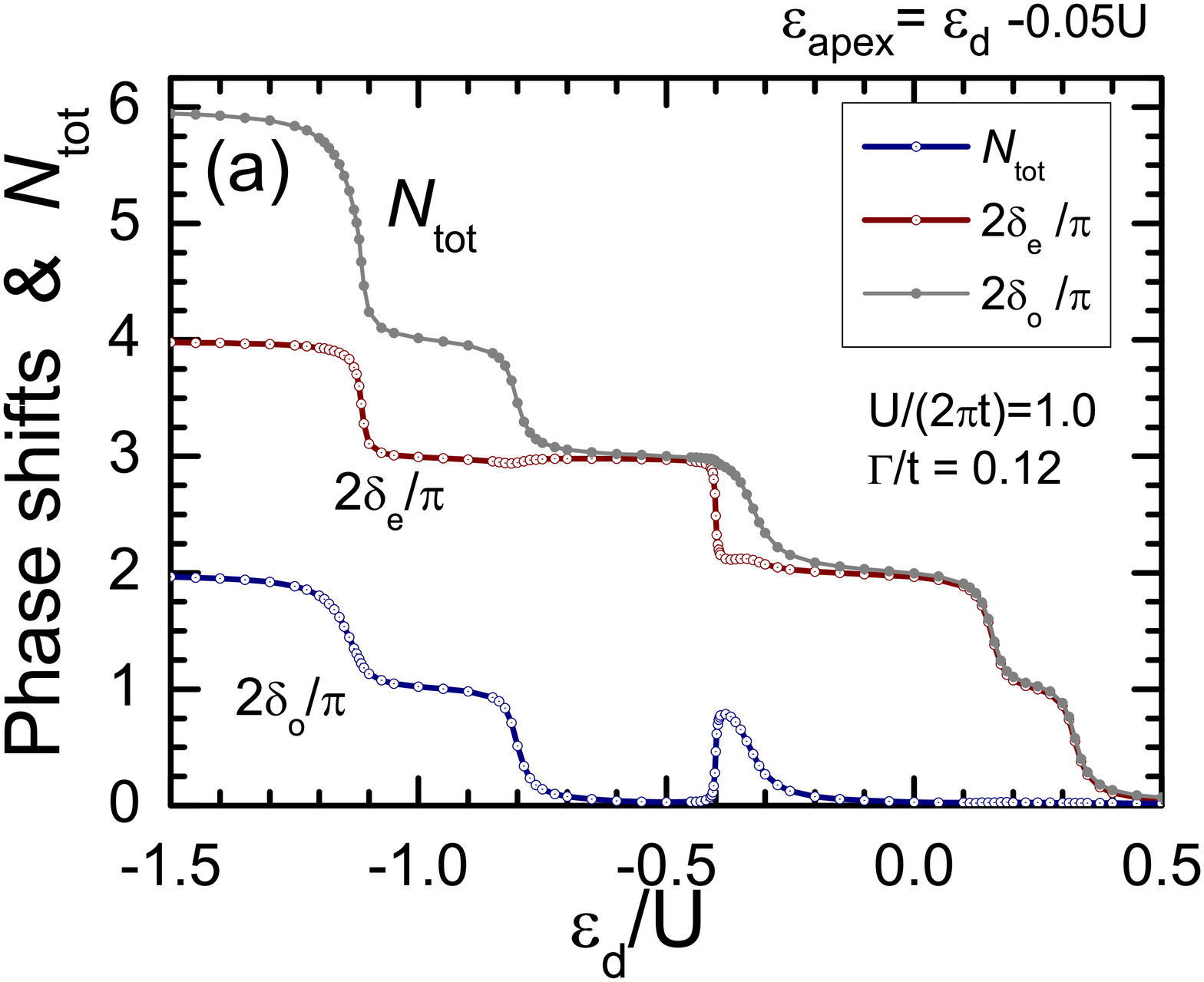}
\end{minipage}
\begin{minipage}[t]{0.49\linewidth}
 \includegraphics[width=1.0\linewidth]{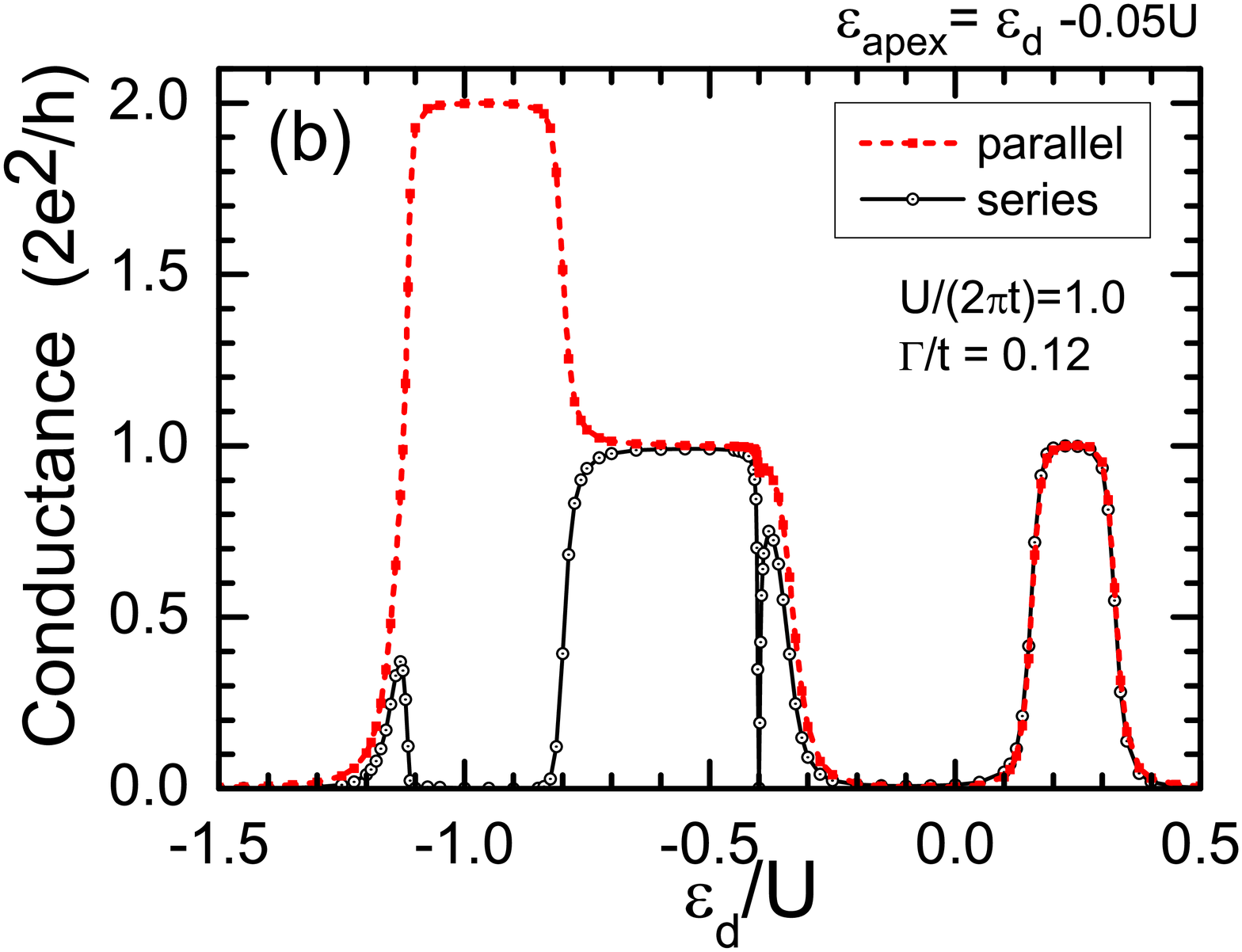}
\end{minipage}

\begin{minipage}[t]{0.49\linewidth}
 \includegraphics[width=1.0\linewidth]{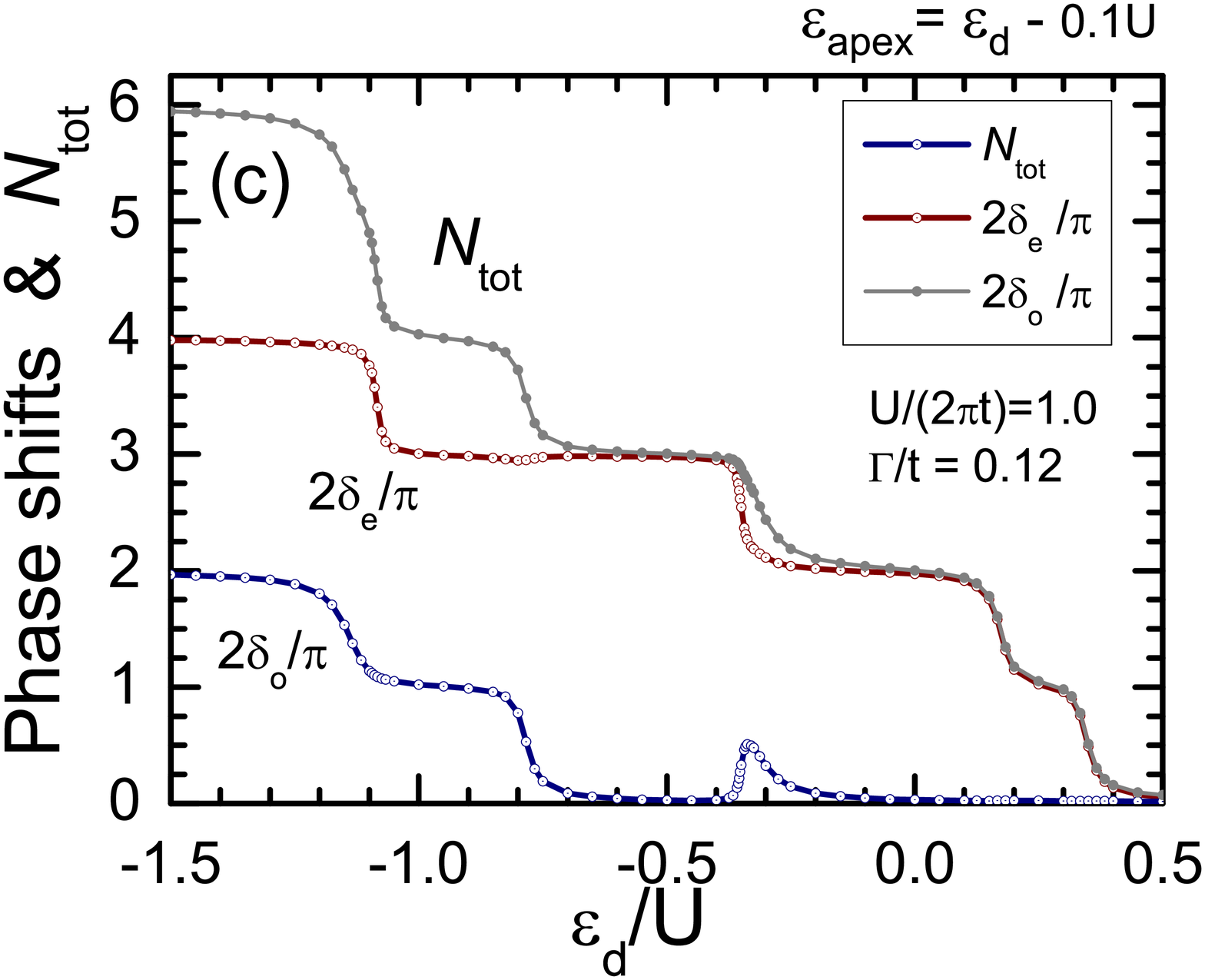}
\end{minipage}
\begin{minipage}[t]{0.49\linewidth}
 \includegraphics[width=1.0\linewidth]{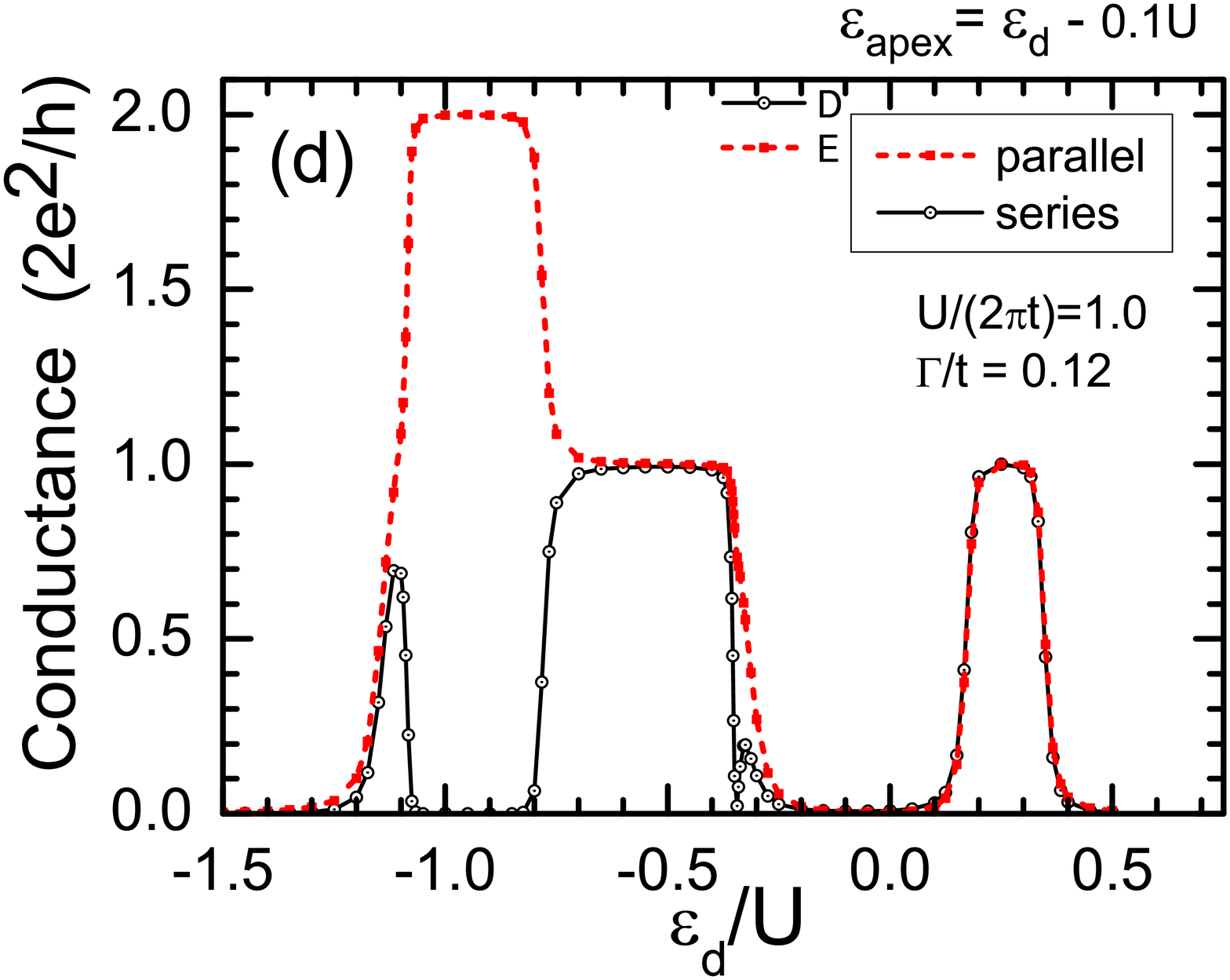}
\end{minipage}

 \caption{(Color online) 
 Phase shifts and conductances are plotted vs $\epsilon_d$, 
keeping the difference between $\epsilon_\mathrm{apex}$ and $\epsilon_d$ 
unchanged; (upper panels) $\epsilon_\mathrm{apex}- \epsilon_d =-0.05U$ and  
  (lower panels) $\epsilon_\mathrm{apex}- \epsilon_d =-0.1U$, 
with  $U/(2\pi t) = 1.0$, $\Gamma/t=0.12$ and $t'=t$.   
 }
 \label{fig:cond_eapex_m005_m01}
\end{figure}

\begin{figure}[t]
 \leavevmode
\begin{minipage}[t]{0.49\linewidth}
 \includegraphics[width=1.0\linewidth]{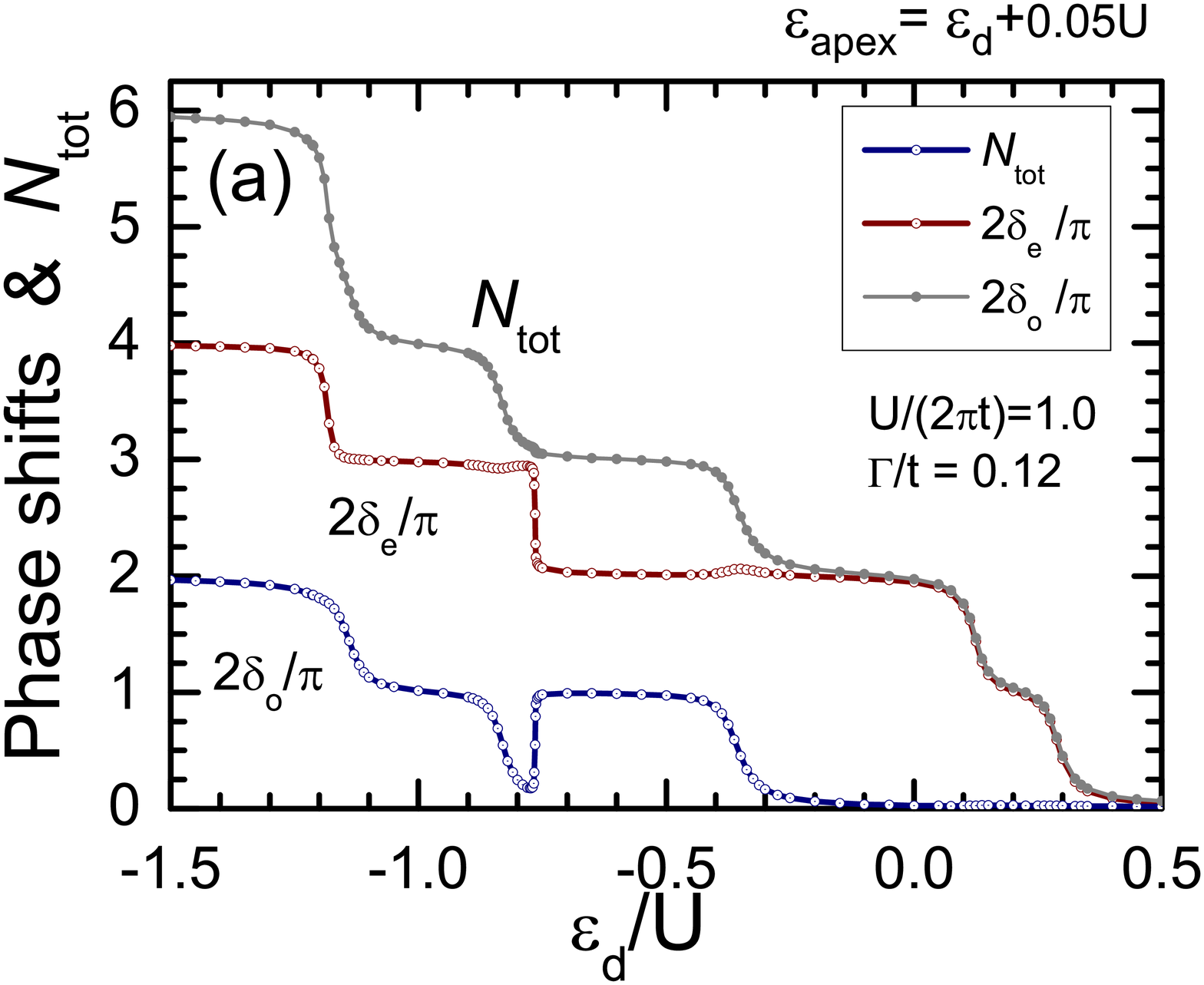}
\end{minipage}
\begin{minipage}[t]{0.49\linewidth}
 \includegraphics[width=1.0\linewidth]{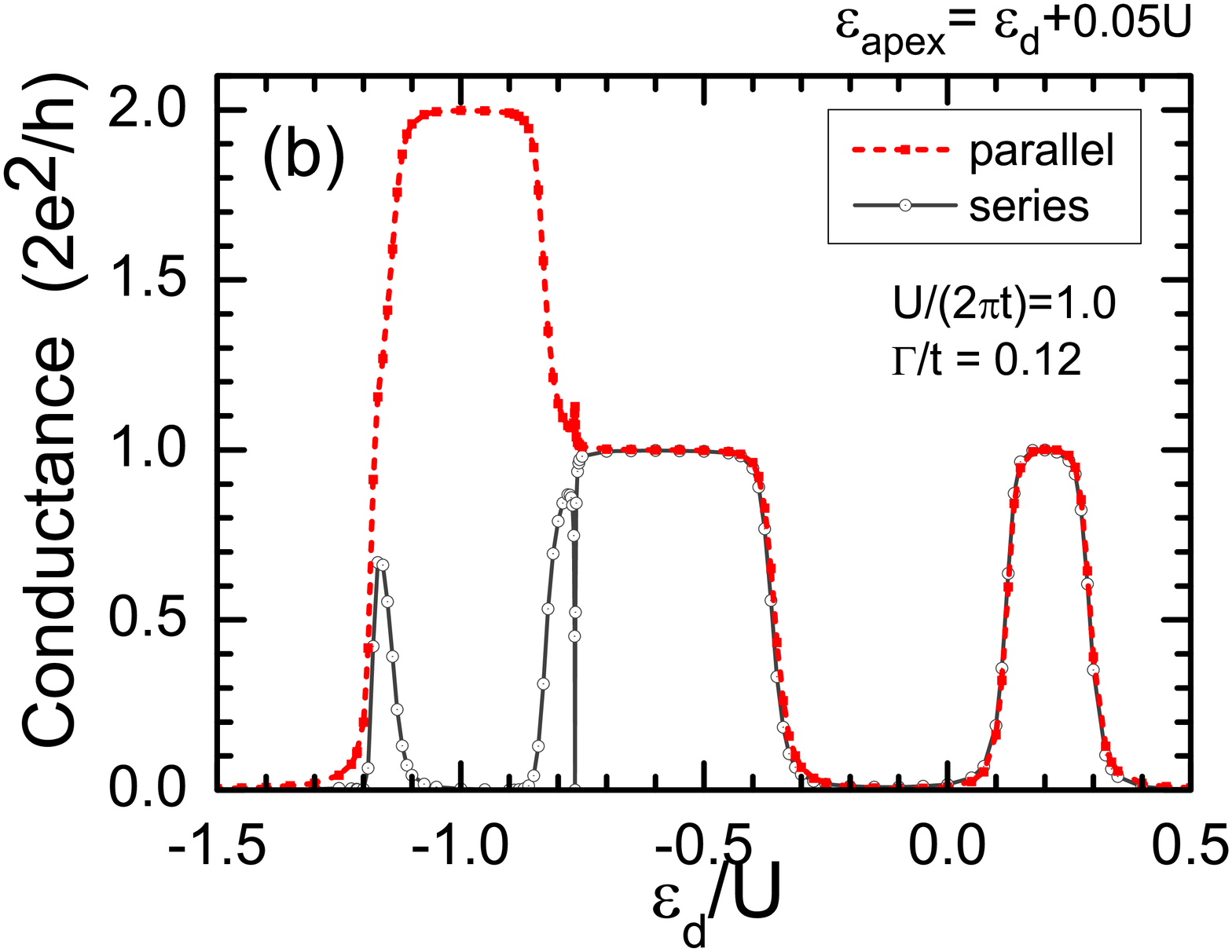}
\end{minipage}

\begin{minipage}[t]{0.49\linewidth}
 \includegraphics[width=1.0\linewidth]{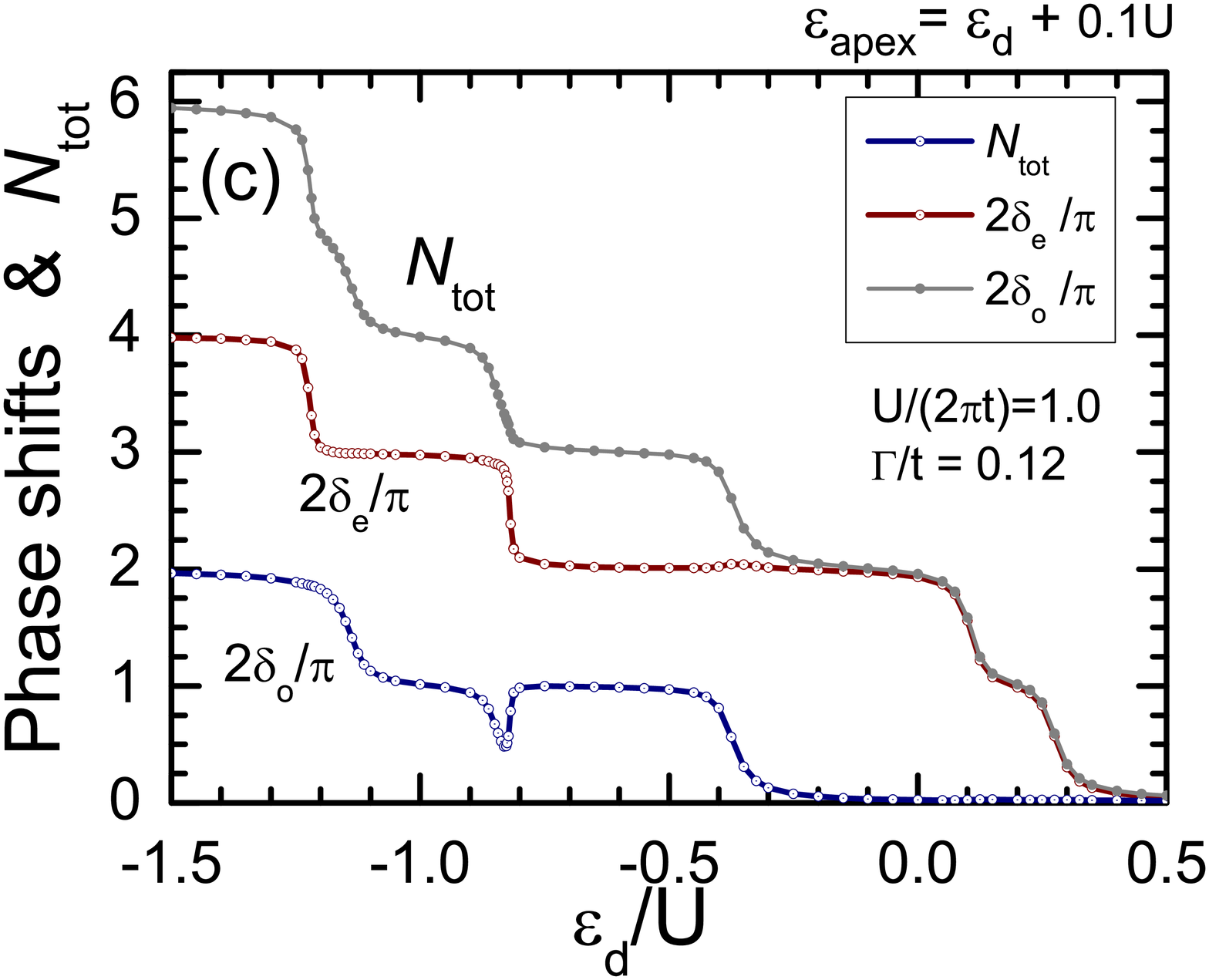}
\end{minipage}
\begin{minipage}[t]{0.49\linewidth}
 \includegraphics[width=1.0\linewidth]{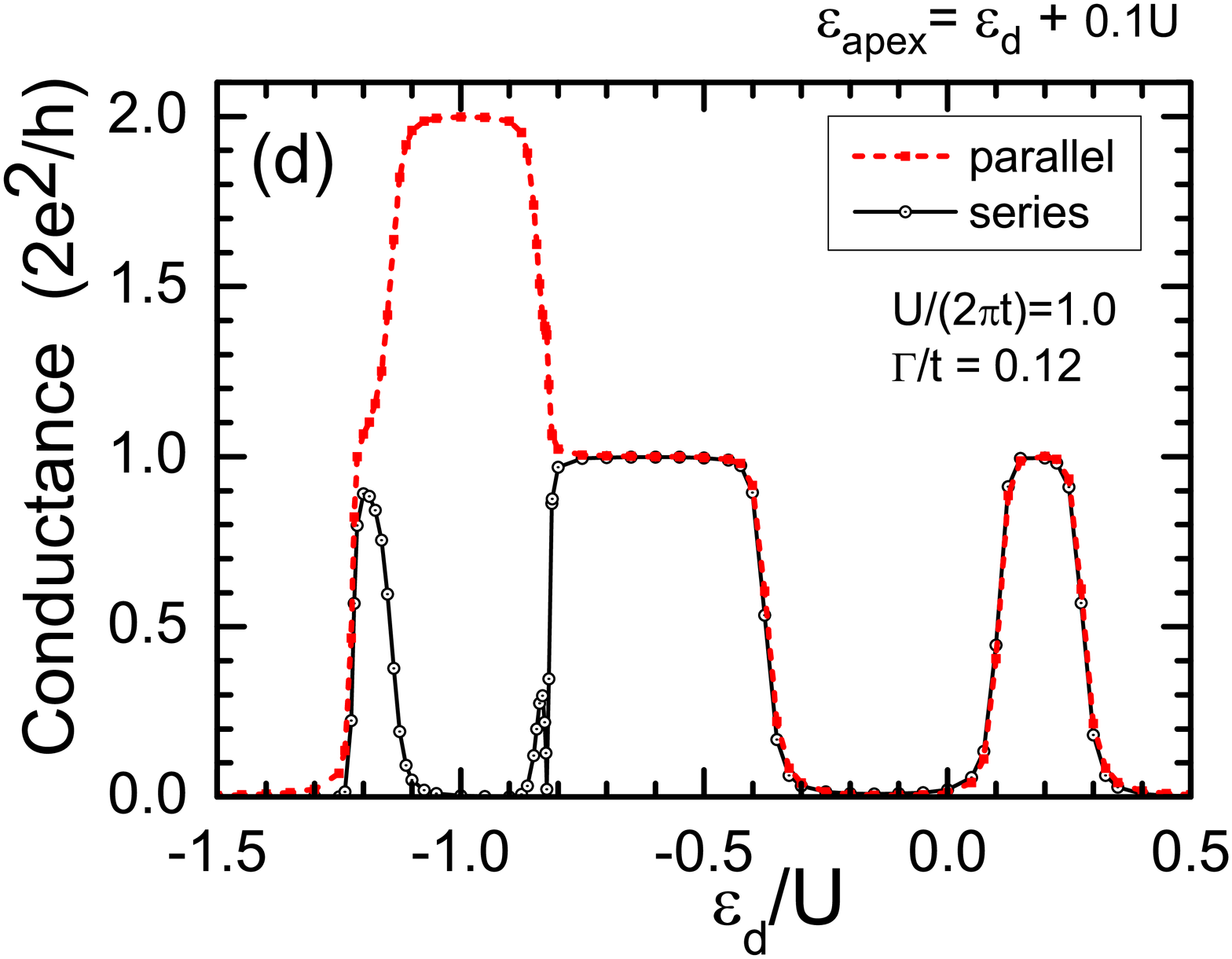}
\end{minipage}
 \caption{(Color online) 
 Phase shifts and conductances are plotted vs $\epsilon_d$, 
keeping the difference between $\epsilon_\mathrm{apex}$ and $\epsilon_d$ 
unchanged; (upper panels) $\epsilon_\mathrm{apex}- \epsilon_d =0.05U$ 
 and (lower panels) $\epsilon_\mathrm{apex}- \epsilon_d =0.1U$, 
with  $U/(2\pi t) = 1.0$, $\Gamma/t=0.12$ and $t'=t$. 
 }
 \label{fig:cond_eapex_p005_p01}
\end{figure}

In Figs.\  \ref{fig:cond_eapex_m005_m01} and \ref{fig:cond_eapex_p005_p01},
the NRG results of the conductances and phase shifts 
are plotted as functions of $\epsilon_d$ 
for several fixed values of $\delta \epsilon$, with $\delta t=0$.
The values of $\epsilon_\mathrm{apex}$ are chosen 
such that $\delta \epsilon/U=-0.05,\, -0.1$ 
for Fig.\  \ref{fig:cond_eapex_m005_m01}, 
and $\delta \epsilon/U = 0.05,\,0.1$ 
for Fig.\ \ref{fig:cond_eapex_p005_p01}. 
The other parameters $U$ and $\Gamma$ are chosen to be 
the same as those for Fig.\ \ref{fig:cond_u1}. 

We see in (b) and (d) of both Figs.\ 
\ref{fig:cond_eapex_m005_m01} and \ref{fig:cond_eapex_p005_p01} 
that the series conductance dip emerges 
 in the Kondo plateau at three-electron filling 
$N_\mathrm{tot} \simeq 3.0$ even in the presence of the small deformations.
Particularly, the typical feature of the sharp dip clearly remains. 
The position of the dip, however, 
shifts sensitively to $\delta \epsilon$,
and it moves towards the right (left) 
for $\delta \epsilon < 0$ ($\delta \epsilon >0$).  
This reflects the behavior of the phase shifts 
at $-0.8U \lesssim \epsilon_d \lesssim -0.3U$ 
in (a) and (c) of these figures. 
Particularly, the odd phase shift $\delta_\mathrm{o}$ 
shows a clear contrast between 
the positive and negative $\delta \epsilon$. 
Correspondingly the kink, which is seen in 
the even phase shift as a step between the values of 
$2\delta_\mathrm{e}/\pi = 2.0$ and $3.0$, 
moves as $\delta \epsilon$ varies.
When the gate voltage $\epsilon_d$ decreases from 
a value at two-electron filling, 
the third electron enters mainly the $b_1$ orbital 
until $\epsilon_d$ reaches the SU(4) symmetric point,  
at which point the charge distribution changes 
significantly as seen in Fig.\ \ref{fig:even_odd_occupation}. 
The redistribution of the charge is protracted 
for $\delta \epsilon > 0$, 
because the odd orbital becomes energetically favorable 
$E_{\mathrm{e},+}^{(1)} > E_\mathrm{o}^{(1)}$.
In the opposite case, for $E_{\mathrm{e},+}^{(1)} < E_\mathrm{o}^{(1)}$, 
the redistribution takes place earlier than the regular triangle case.
As the magnitude $|\delta \epsilon|$ of the deformations  
increases further, 
the conductance dip moves away 
from the Kondo plateau and disappears. 
The range of $|\delta \epsilon|$, in which the dip can be seen, 
will increase with the width of the Kondo plateau.

Another notable change caused by 
the deformations due to $\,\delta \epsilon\,$ 
is the structure at five electron filling $N_\mathrm{tot} \simeq 5.0$.
We see in (b) and (d) of both 
Figs.\  \ref{fig:cond_eapex_m005_m01} and \ref{fig:cond_eapex_p005_p01} 
that a peak of the series conductance 
evolves  at $-1.2U \lesssim \epsilon_d \lesssim -1.1U$, 
as $|\delta \epsilon|$ increases. 
Simultaneously, in this region, 
the parallel conductance has a gentle shoulder. 
Furthermore, we can see also in (c)  of in these figures 
that a weak step for the five-electron occupancy emerges 
for $N_\mathrm{tot}$ 
in the case of $\delta \epsilon = \pm 0.1U$. 

In the other regions of the filling, namely at 
$N_\mathrm{tot} \simeq 4.0$ and $N_\mathrm{tot} \lesssim 2.0$, 
the conductances and phase shifts  are less sensitive 
to the small perturbations due to $\,\delta \epsilon$. 
Specifically, the $S=1$ Kondo behavior is robust,
although the region of $\epsilon_d$ for 
the four-electron filling becomes slightly narrow. 
As mentioned in Sec.\ \ref{subsec:isolated_cluster},
 the Nagaoka state is the ground state of the TTQD for $N_\mathrm{tot}=4$ 
in the limit of $\Gamma = 0$.
We have calculated the cluster eigenstates 
in the presence of the deformations, 
and found that the Nagaoka state remains 
as the ground state near $\epsilon_d/U \simeq -1.0$ 
in a wide range of 
the deformation  $-0.41U < \delta \epsilon  <0.87U$
 for $U/(2\pi t)=1.0$.
This is determined mainly by the finite-size energy separation 
 $\Delta E^{(4)} \equiv E_{S=0}^{(4)}-E_{S=1}^{(4)}$ 
between the Nagaoka and the lowest singlet states, 
mentioned in Sec.\ \ref{subsec:isolated_cluster}.
This energy scale of the TTQD makes the $S=1$ Kondo behavior 
rigid also against the deformations.

\begin{figure}[t]
 \leavevmode

\begin{minipage}[t]{0.49\linewidth}
 \includegraphics[width=1.0\linewidth]{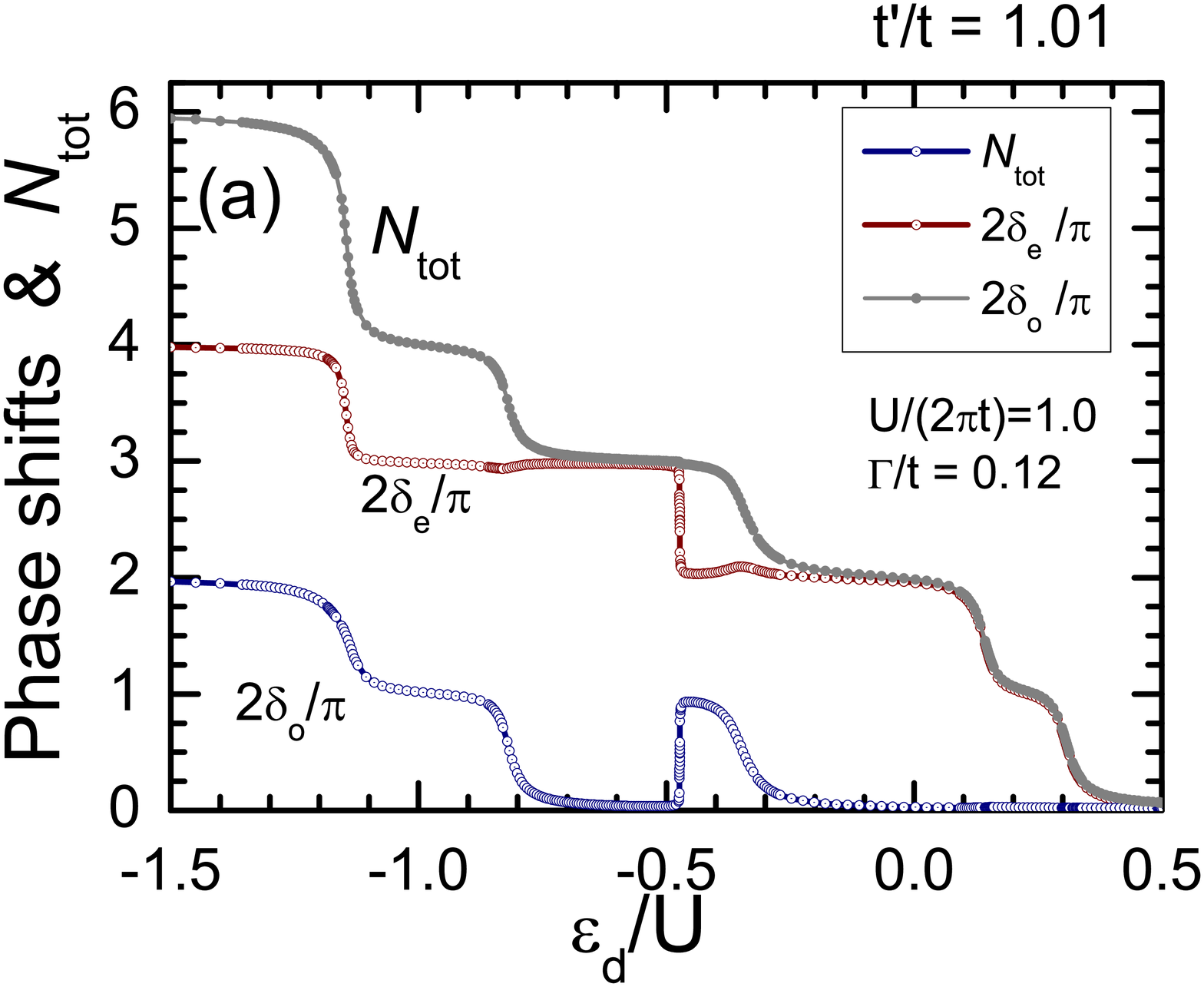}
\end{minipage}
\begin{minipage}[t]{0.49\linewidth}
 \includegraphics[width=1.0\linewidth]{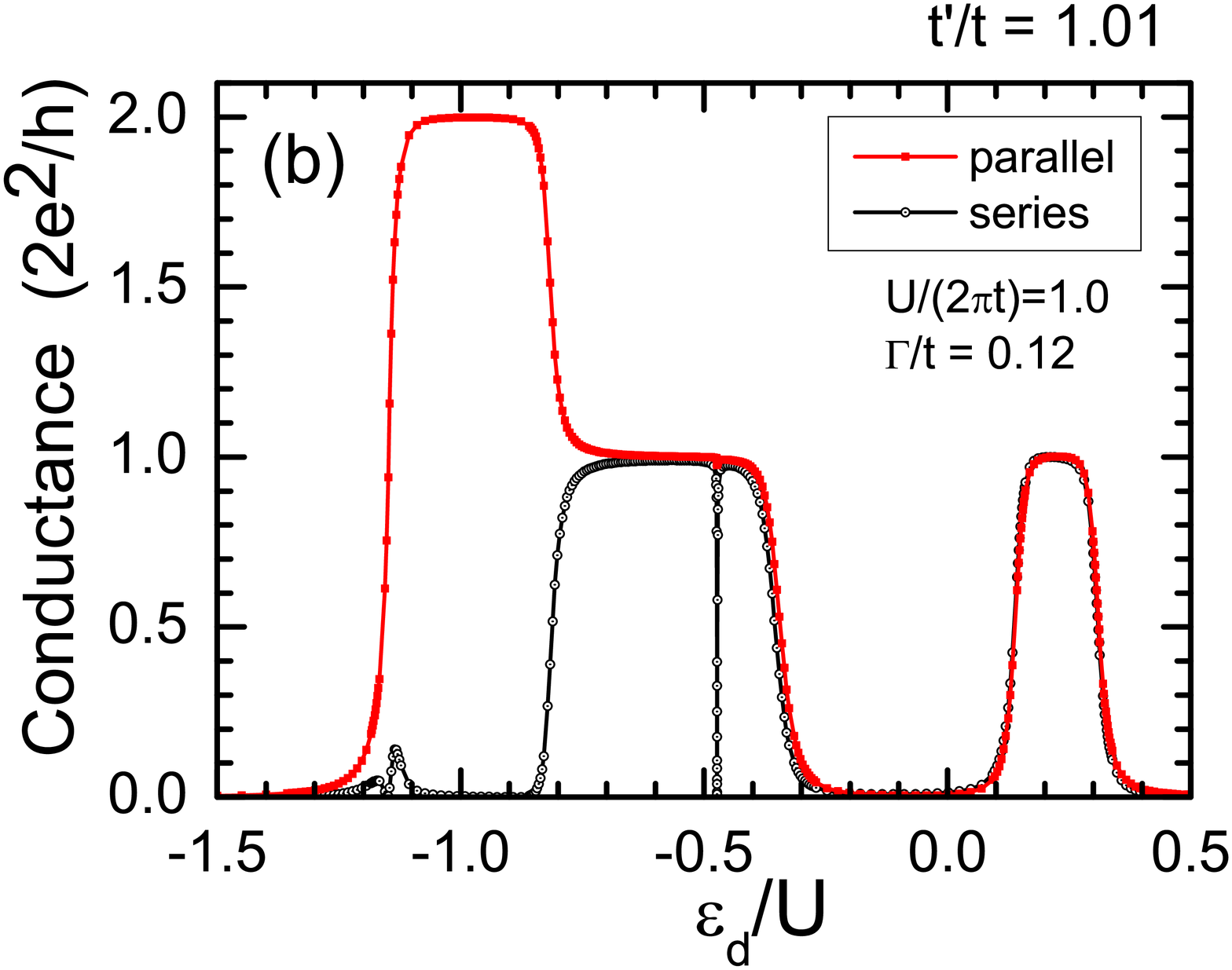}
\end{minipage}

\begin{minipage}[t]{0.49\linewidth}
 \includegraphics[width=1.0\linewidth]{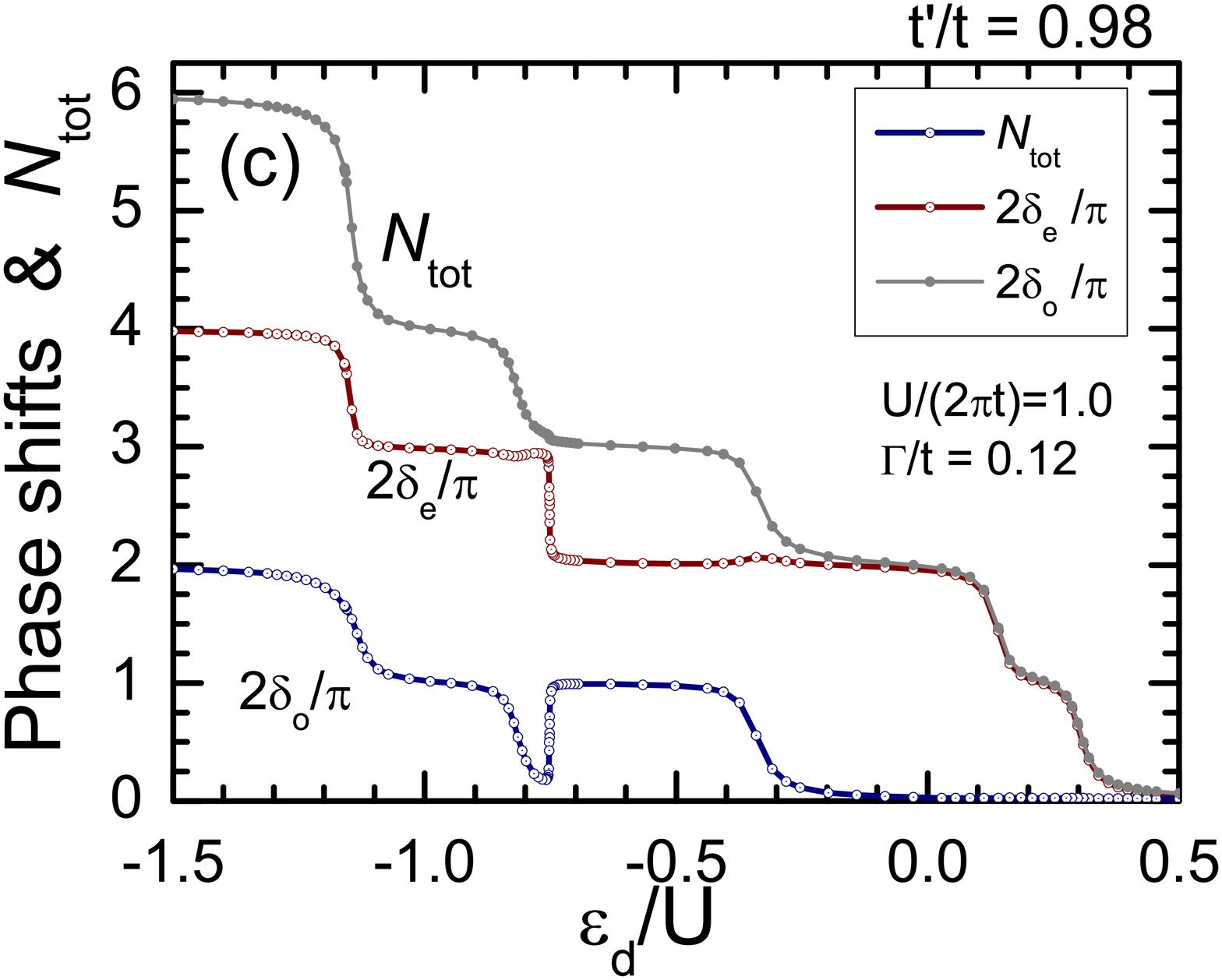}
\end{minipage}
\begin{minipage}[t]{0.49\linewidth}
 \includegraphics[width=1.0\linewidth]{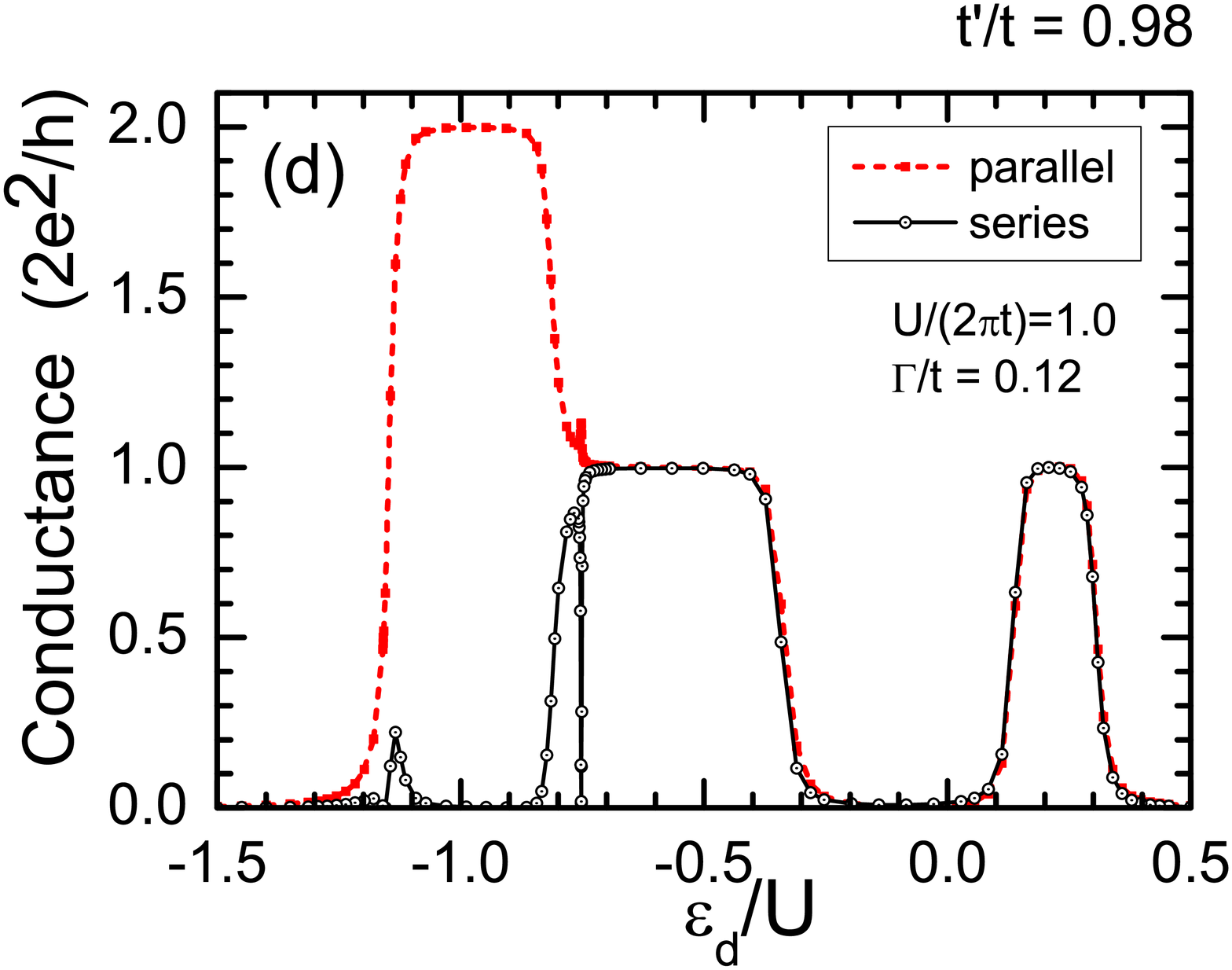}
\end{minipage}
 \caption{(Color online) 
 Phase shifts and conductances as functions of $\,\epsilon_d$, for 
 (upper panels) $t' = 1.01 t\,$ and (lower panels) $t' = 0.98 t$, 
with $U/(2\pi t) = 1.0$, $\Gamma/t=0.12$  
and $\epsilon_\mathrm{apex}= \epsilon_d$.    
}
 \label{fig:cond_tp101_098}
\end{figure}

We have examined also the effects of  small deformations caused by 
the hopping matrix elements $\delta t = t'-t$,
choosing the level positions to be uniform $\,\delta \epsilon =0$. 
In Fig.\ \ref{fig:cond_tp101_098},  
the conductances and 
phase shifts are plotted as functions of $\epsilon_d$ 
for $t'=1.01t$ (upper panels), and for $t'=0.98t$ (lower panels). 
Besides huge similarities to 
the above results for the case of $\delta \epsilon$, 
one notable difference is that no pronounced structures 
evolve at $N_\mathrm{tot} = 5.0$ for small $\delta t$.
In the cluster limit $\Gamma =0$, 
the five-electron charge state can become the ground state 
for finite $\delta t$, 
but in a very narrow region near $\epsilon_d \simeq -1.16U$, 
the width of which $\Delta \epsilon_d$ is calculated to be   
 $\Delta \epsilon_d \simeq 0.004U$ for $t'/t = 0.98$, 
and $\Delta \epsilon_d \simeq 0.002U$ for $t'/t = 1.01$. 
This energy scale is much smaller than 
the hybridization strength $\Gamma = 0.019 U$, 
and thus quantitatively the five-electron structure 
was not seen in the present situation.

In the other regions of the filling,  however, 
the results obtained for $\delta t > 0$ and $\delta t < 0$  resemble 
those for $\delta \epsilon < 0$ and $\delta \epsilon > 0$, respectively.
Particularly, 
the structures of the series conductance dip 
and the phase shifts at $N_\mathrm{tot} \simeq 3.0$ are 
similar qualitatively to those seen in the upper panels of  
 Figs.\  \ref{fig:cond_eapex_m005_m01} 
and \ref{fig:cond_eapex_p005_p01}.
These features are determined essentially by 
the way that the orbital degeneracy between 
the even and odd orbitals is lifted, namely    
$E_{\mathrm{e},+}^{(1)}< E_{\mathrm{o}}^{(1)}$ 
or $E_{\mathrm{e},+}^{(1)} > E_{\mathrm{o}}^{(1)}$.
The SU(4) Kondo effect is sensitive to relatively 
small deformations, 
because it relies on the orbital degeneracy at three-electron filling. 
Nevertheless, the sharp structure of the dip 
does not collapse  immediately under an infinitesimal perturbation, 
and thus a perfect $C_{3v}$ symmetry is not 
necessary to observe the SU(4) Kondo behavior in the TTQD.

The $S=1$ Kondo behavior 
at the four-electron filling is not sensitive to $\delta t$, 
as that in the case of small $\delta \epsilon$.  
We have confirmed also that the Nagaoka state remains as 
a ground state of the isolated TTQD cluster 
near $\epsilon_d/U \simeq -1.0$ 
also in a wide of range of the off-diagonal inhomogeneity 
 $0.30 <t'/t<2.87$ for $U/(2\pi t)=1.0$. 
The $S=1$ Kondo behavior is 
robust also against the deformations by $\,\delta t$,
and it is mainly due to the finite-size energy 
separation of $\Delta E^{(4)}$.


\section{Summary}
\label{sec:summary}

We have discussed first of all in Sec.\ \ref{sec:zero_point_SU4} 
that the fixed-point Hamiltonian which describes 
the low-lying energy excitation from a local 
Fermi-liquid ground state has an SU(4) symmetry, 
consisting of the channel and spin degrees of freedom,    
at zero points of the two-terminal conductance, 
or equivalently at the points  where 
the two phase shifts satisfy 
the condition $\delta_\mathrm{e}-\delta_\mathrm{o} = n \pi$ 
for $n = 0,\pm 1, \pm 2, \ldots$.  
This symmetry property holds at low energies 
quite generally for the systems  which can be  
described by the Hamiltonian $\mathcal{H}$ given in Eq.\ \eqref{eq:H}, 
for any numbers of quantized levels $N_D$, 
in the case that the system has time-reversal 
and inversion symmetries.
Specifically, 
the Hamiltonian $\mathcal{H}$ does not necessarily have  
an SU(4) symmetry on the global energy scale.   

In the main part of the paper 
we have discussed the Kondo effects taking place  
in the triangular triple quantum dot connected to 
two non-interacting leads, as illustrated in Fig.~\ref{fig:system}.
We have deduced from the difference between the 
two phase shifts shown in Fig.~\ref{fig:diff_phase} 
that the two-terminal conductance of this system can have 
three zero points as the level position $\epsilon_d$ varies.
The low-energy properties at these zero points 
can be characterized by the SU(4) symmetric local Fermi-liquid theory. 
Particularly, the zero point at three-electron filling 
appears as a sharp dip in the middle of the conductance plateau 
seen in Fig.~\ref{fig:cond_u1}. At this zero point 
the phase shifts  satisfy the conditions 
 $\delta_\mathrm{e}-\delta_\mathrm{o}=\pi$ and 
 $\delta_\mathrm{e}+\delta_\mathrm{o}\simeq 3\pi/2$, 
which means that 
 $\delta_\mathrm{e}\simeq 5\pi/4$ and $\delta_\mathrm{o}\simeq \pi/4$.
Furthermore, at this filling, the system has 
an SU(4) symmetry also at high energies,  
which is caused by the four-fold degeneracy 
in the local states of an isolated triangular cluster. 
These symmetry properties at two opposite limits,
at low and high energies, cause a single-stage screening  
observed in temperature dependence of the entropy 
shown in Fig.~\ref{fig:S_Tchi3}. 

For the values of $\epsilon_d$ away from the dip, 
the system  no longer has the low-energy SU(4) symmetry,
and the low-lying energy excited states have only the SU(2) symmetry 
of the spin. We have confirmed also that the charge distribution 
in the even and odd orbitals, 
the geometrical feature of which is illustrated in Fig.\ \ref{fig:even_odd}, 
changes significantly at the dip  
as shown in Fig.\ \ref{fig:even_odd_occupation}.
The inhomogeneous charge distribution 
in the even-odd basis also causes the two-stage 
Kondo behavior of the $S=1$ local moment 
which emerges at  
four-electron fillings due to Nagaoka ferromagnetic 
mechanism. The partial moment which is distributed 
away from the two leads makes the Kondo temperature 
for the second stage very small. 
Nevertheless, the Kondo temperature can be raised by increasing 
the coupling $\Gamma$ between the TTQD and leads.
Therefore, in order to observe the second screening stage, 
$\Gamma$ should be large. 
At the same time, however, $\Gamma$ has to be smaller 
than $\Delta E^{(4)} \simeq \max (U/3,t)$ so that 
the hybridization effects should not smear 
the typical structures of the $S=1$ Kondo behavior.

We have studied also the effects of the deformations,  
which break the full triangular symmetry $C_{3v}$ 
and lift the orbital degeneracy in the one-particle excited states.
Specifically, the deformations which are 
caused by inhomogeneity in the onsite potentials 
$\,\delta \epsilon \equiv \epsilon_\mathrm{apex}-\epsilon_d\,$ 
and the inter-dot couplings  $\,\delta t \equiv t'-t$ 
have been examined. 
We found that  the series conductance dip 
does still exist stably at three-electron filling,
even in the presence of infinitesimal deformations.
The position of the dip, however, shifts sensitively 
from the middle of the Kondo plateau for $N_\mathrm{tot} \simeq 3.0$.
In the other region of the fillings, 
small deformations do not affect the Kondo behavior significantly, 
and just cause some gradual and quantitative changes. 
These results support our belief that the various Kondo effects 
in the TTQD could be observed in future.

\acknowledgements
We would like to thank S.~Amaha, 
J.~Bauer, V.~Meden, Yoichi Tanaka and Y.~Utsumi   
for valuable discussions.
This work is supported by JSPS Grant-in-Aid 
for Scientific Research (C).
One of us (ACH) acknowledges 
the support of a grant from the EPSRC  (G032181/1).
Numerical computation was partly carried out 
in Yukawa Institute Computer Facility.

\appendix

\section{Nagaoka state in a triangle}
\label{sec:nagaoka_eigenstate}

The eigenvector of the Nagaoka state 
$|\Phi^{(4)}_{S=1}\rangle$
for the isolated triangle with $t'=t$ and $\epsilon_\mathrm{apex}=\epsilon_d$ 
can be constructed simply, 
by filling first the one-particle state of $k=0$ 
defined in \eqref{eq:tight_bind_triangle} 
by two electrons, 
and then filling each of the two excited states with  
$k=+{2\pi}/{3}$ and $k=-{2\pi}/{3}$ by 
a single up-spin electron.
This state can be expressed,
using the even and orbitals described  
in Eq.\ \eqref{eq:even_odd_start} and Fig.\ \ref{fig:even_odd}, 
in the form 
\begin{align} 
& 
|\Phi^{(4)}_{S=1}\rangle 
=  
\  
\alpha_0\, 
|\mathrm{I} \rangle  
+ \alpha_1\,
|\mathrm{II} \rangle  
\;, \\
& 
\quad |\mathrm{I} \rangle  =   
  b^{\dagger}_{1\uparrow}
 a^{\dagger}_{0\uparrow}
 a^{\dagger}_{1\uparrow}
 a^{\dagger}_{0\downarrow} 
|0\rangle , 
 \quad   
|\mathrm{II} \rangle  
 =  b^{\dagger}_{1\uparrow}
 a^{\dagger}_{0\uparrow}
 a^{\dagger}_{1\uparrow}
 a^{\dagger}_{1\downarrow} 
|0\rangle 
. 
\label{eq:Nagaoka_state_cluster}
\end{align} 
Here, $\alpha_0 = \sqrt{1/3}$, $\,\alpha_1 = \sqrt{2/3}$, 
 $|0\rangle$ is a vacuum, and  
the eigenvalue is give by $E^{(4)}_{S=1}=-2t +U +4\epsilon_d$.  
This state has a finite total-spin $z$ component of $S_z =+1$,
and breaks the spin rotational symmetry. 
In this state, 
the partial moments for even orbitals are given by 
\begin{align} 
m_{a,i} \equiv &\ 
\langle \Phi^{(4)}_{S=1}| 
\, \frac{1}{2}
\left(
a^{\dagger}_{i\uparrow}
a^{}_{i\uparrow}
-a^{\dagger}_{i\downarrow}
a^{}_{i\downarrow}
\right)
|\Phi^{(4)}_{S=1}\rangle 
\nonumber \\
= &
\ \, \frac{1}{2} \left(
1-|\alpha_i|^2 
\right) .
\end{align} 
Thus, 
$m_{a,0}=1/3$,  $\, m_{a,1}=1/6$,  
and the partial moment in the odd orbital 
$b^{}_1$ is given by 
$m_{b,1} =1/2$. 
Furthermore, 
the occupation number of each orbital 
takes the value   
$\langle n_{a,0} \rangle =4/3$,
$\langle n_{a,1} \rangle =5/3$, and
$\langle n_{b,1} \rangle =1$.

In the subspace with $N_\mathrm{tot}=4$ and $S=1$,
the matrix elements of the interaction Hamiltonian has a diagonal form
\begin{align}
\langle \nu |  
\mathcal{H}_\mathrm{dot}^U  
|\nu' \rangle  \,=\, U\, \delta_{\nu,\nu'}\;, 
\end{align}
for $\nu, \,\nu' =  \mathrm{I},\, \mathrm{II}$. 
This can be verified directly using Eq.\ \eqref{eq:even_odd}.  
For this reason, 
the wavefunction is determined by the noninteracting Hamiltonian, 
which can be rewritten generally in the following 
form (see also  Fig.\ \ref{fig:even_odd})
\begin{align}
\mathcal{H}_\mathrm{dot}^0 \, = & \ \    
\epsilon_\mathrm{apex}\, n_{a,0} +
\left(\epsilon_{d} - t' \right)  n_{a,1} 
+\left(\epsilon_{d} + t' \right)  n_{b,1} 
\nonumber \\
& \ - \sqrt{2} \, t \sum_{\sigma}  
 \left(
  a^{\dagger}_{0\sigma}a^{\phantom{\dagger}}_{1\sigma}  
+ a^{\dagger}_{1\sigma}a^{\phantom{\dagger}}_{0\sigma}  
\right)  .
\label{eq:H_0_dot_even_odd}
\end{align}
Therefore, the coefficients $\alpha_0$ and $\alpha_1$ defined 
in Eq.\ \eqref{eq:Nagaoka_state_cluster}
depend crucially on  the potential difference 
between the $a_0^{}$ and  $a_1^{}$ orbitals, 
which correspond to the first and second terms 
in the right-hand side of Eq.\ \eqref{eq:H_0_dot_even_odd}. 
This potential profile in the even-odd basis 
play an important role on the two-stage Kondo effects, 
which take place in the case that the two leads 
are connected to the $a_1^{}$ and $b_1^{}$ orbitals 
in the way that is shown  in Fig.\ \ref{fig:even_odd}.

 \section{Many-body phase shifts}

\label{sec:app_green}

 The phase shifts for interacting electrons can be 
defined, using the Green's function 
\begin{equation} 
G_{ij}(i\omega_n) 
\, =\,  
-    \int_0^{\beta} \! d\tau \,
   \left \langle  T_{\tau} \,  
   d^{\phantom{\dagger}}_{i \sigma} (\tau) 
   \, d^{\dagger}_{j \sigma} (0) 
     \right \rangle  \, e^{i \omega_n \tau} \;,
\label{eq:G_Matsubara}
\end{equation} 
where $\beta= 1/T$,  
$d_{j \sigma}(\tau) = 
e^{\tau  {\cal H}} d_{j \sigma} e^{- \tau  {\cal H}}$, 
and $\langle {\cal O} \rangle =
\mbox{Tr} \left[ \, e^{-\beta  {\cal H} }\, {\cal O}
\,\right]/\mbox{Tr} \, e^{-\beta  {\cal H} }$. 
The retarded and advanced Green's functions correspond to 
 the analytic continuations 
 $G_{ij}^{\pm}(\omega) =  G_{ij}(\omega \pm i\, 0^+)$. 
The interaction $\mathcal{H}_\mathrm{dot}^U$ 
between electrons in the dots 
at $1\leq i \leq N_D$ ($N_D=3$ for the triple dot) 
causes the self-energy correction $\Sigma_{ij}(z)$,
and the Dyson equation is given by 
\begin{equation} 
  G_{ij}(z)    =   G^0_{ij}(z) 
    + 
\sum_{i'=1}^{N_D}
\sum_{j'=1}^{N_D}
\,G^0_{ii'}(z)\,  \Sigma_{i'j'}(z)  
   \, G_{j'j}(z) \;.    
  \label{eq:Dyson}   
\end{equation}   
Here, $G^0_{ij}(z)$ is the non-interacting Green's function 
corresponding to 
$\mathcal{H}_0 \equiv  \mathcal{H}_\mathrm{dot}^0 
+  \mathcal{H}_\mathrm{mix}  +  \mathcal{H}_\mathrm{lead}$.

At zero temperature $T=0$, the series $g_\mathrm{s}$ and 
parallel $g_\mathrm{p}$ conductances 
can be expressed in terms of the Green's function at 
 the Fermi level $\omega=0$, in the case the system 
has a time-reversal symmetry\cite{aoQuasi,ONH}
\begin{align}
g_\mathrm{s}^{\phantom{0}}\, = & \  \frac{2 e^2}{h}  
    4\Gamma_R \Gamma_L \left| G_{N_D 1}^{+}(0)\right|^2  
         \;,  
\label{eq:cond_s} 
\\
g_\mathrm{p}^{\phantom{0}} =&  \   
{2 e^2 \over h} 
 \, 
\Bigl[\,
\Gamma_L^2 \, \left|G_{11}^+(0) \right|^2
+ 
\Gamma_R^2 \, \left|G_{N_D N_D}^+(0) \right|^2
\nonumber
\\
& \qquad \ \ 
+ \,  2 
\Gamma_L 
\Gamma_R \, \left|G_{N_D 1}^+(0) \right|^2
\,\Bigr] 
\\
= & \ 
{2 e^2 \over h} 
 \, 
\Bigl[\,
-\Gamma_L\, \mathrm{Im}\, G_{11}^+(0)
-\Gamma_R\, \mathrm{Im}\, G_{N_DN_D}^+(0)
\,\Bigr] \;.
\label{eq:def_g_p_Green}
\end{align}
Note that the contributions of the vertex correction do not appear here 
due to the property that the imaginary part of the self-energy vanishes 
 $\mathrm{Im}\,\Sigma_{ij}^{\pm}(0)=0$ 
 at $T=0$ and $\omega=0$.\cite{aoFermi}
Therefore the value of $G_{ij}^+(0)$  appearing in the above equations  
can be reproduced correctly by  
a free quasi-particle Hamiltonian 
$\mathcal{H}_\mathrm{eff} \equiv 
\widetilde{\mathcal{H}}_\mathrm{dot} 
+  \mathcal{H}_\mathrm{mix}  +  
\mathcal{H}_\mathrm{lead}$,
\begin{align}
\widetilde{\mathcal{H}}_\mathrm{dot}
\equiv & \  
 - \sum_{i,j=1}^{N_D}
\sum_{\sigma} \widetilde {t}_{ij} \, 
 d^{\dagger}_{i\sigma}d^{\phantom{\dagger}}_{j\sigma}  
\;.
\label{eq:H_d_effective}
\end{align}
Here,  $-\widetilde{t}_{ij} 
= -t_{ij} + \epsilon_{d,i}\, \delta_{ij} 
+ \mathrm{Re}\, \Sigma_{ij}^+(0)\,$ 
can be regarded as the hopping matrix element  
for free quasi-particles. 
\cite{aoQuasi,aoFermi} 

Furthermore, in the case the system has also 
the inversion symmetry $\Gamma_L = \Gamma_R$ ($\equiv \Gamma$), 
the Green's functions take the form 
\begin{align}
G_{11}^+(0)  = G_{N_DN_D}^+(0) 
    \,=& \  \frac{1}{2 \Gamma} 
    \left[\, 
    \frac{1}{\kappa_\mathrm{e} + {i}}
    \,+\, \frac{1}{\kappa_\mathrm{o} + i}
    \,\right] \;, 
\label{eq:G_11_sym}
\\
G_{N_D 1}^+(0)  
   \,=& \  \frac{1}{2 \Gamma} 
   \left[\, 
   \frac{1}{\kappa_\mathrm{e} + i}
   \,-\, \frac{1}{\kappa_\mathrm{o} + i}
   \,\right] \;. 
\label{eq:G_N1_sym}
\end{align}
Here,  $\kappa_\mathrm{e}   =   - \cot \delta_\mathrm{e}$ 
and $\kappa_\mathrm{o} = - \cot \delta_\mathrm{o}$ 
include all the many-body corrections through 
the real part of the self-energy $\mathrm{Re}\,\Sigma_{ij}^+(0)$.\cite{ONH}  
Equations \eqref{eq:gs}--\eqref{eq:gp} follow 
from Eqs.\ \eqref{eq:cond_s}--\eqref{eq:G_N1_sym}. 
One can state immediately from Eq.\ \eqref{eq:G_N1_sym}  
that the series conductance $g_\mathrm{s}$ becomes zero 
in the case  $\kappa_\mathrm{e}=\kappa_\mathrm{o}$,
or equally for 
$\delta_\mathrm{e}-\delta_\mathrm{o} = n \pi$  for $n=0,\pm 1, \pm 2,
\ldots$.

\section{NRG approach}
\label{sec:NRG_approach}

In the NRG approach the non-interacting leads are transformed 
into the chains carrying out the logarithmic discretization 
with the parameter $\Lambda$, and a sequence of 
the Hamiltonian $H_N$ in the following form is introduced,\cite{KWW,KWW2}
\begin{align}
&H_N \ =  \Lambda^{(N-1)/2} 
\left( 
\mathcal{H}_\mathrm{dot}^0 +
\mathcal{H}_\mathrm{dot}^U 
 +  H_\mathrm{mix}^{\phantom{0}}  +   H_\mathrm{lead}^{(N)}
 \right) ,
\label{eq:H_N} 
\\
&H_\mathrm{mix}^{\phantom{0}}  \, = \, \bar{v} \, 
       \sum_{\sigma}
\left(\,
f^{\dagger}_{0,L\sigma} d^{\phantom{\dagger}}_{ 1,\sigma}
\,+\, 
d^{\dagger}_{ 1,\sigma}  f^{\phantom{\dagger}}_{0,L\sigma} 
     \right)
     \nonumber \\
 & \qquad \quad  
+  \bar{v}\, 
       \sum_{\sigma} 
       \left(\,
f^{\dagger}_{0,R \sigma} d^{\phantom{\dagger}}_{N_C, \sigma} 
   \,+\, 
d^{\dagger}_{N_C, \sigma} f^{\phantom{\dagger}}_{0,R \sigma} 
         \,  \right)  \;,
\label{eq:H_mix_NRG}
\\
& H_\mathrm{lead}^{(N)} \,=\,
D\,{1+1/\Lambda \over 2} \,
\sum_{\nu=L,R}
\sum_{\sigma}
\sum_{n=0}^{N-1} 
\, \xi_n\, \Lambda^{-n/2}
\nonumber \\
& \qquad \qquad \times  
\left(\,
  f^{\dagger}_{n+1,\nu\sigma}\,f^{\phantom{\dagger}}_{n,\nu\sigma}
  +  
 f^{\dagger}_{n,\nu\sigma}\, f^{\phantom{\dagger}}_{n+1,\nu\sigma}
 \,\right) \;.
\label{eq:H_lead_NRG}
\end{align}
Here,  $D$ is the half-width of the conduction band, and 
the other parameters are given by\cite{KWW,KWW2}
\begin{align}
  \bar{v}
&\,=\, \sqrt{ \frac{2D\,\Gamma A_{\Lambda}}{\pi} }
\;,
\qquad
A_{\Lambda}  \,=\,  \frac{1}{2}\, 
 {1+1/\Lambda \over 1-1/\Lambda }
\,\log \Lambda
\;, 
\label{eq:A_lambda}
\\
\xi_n &\,=\,    
{ 1-1/\Lambda^{n+1}  
\over  \sqrt{1-1/\Lambda^{2n+1}}  \sqrt{1-1/\Lambda^{2n+3}} 
} 
\;.
\label{eq:xi_n}
\end{align}
The factor $A_{\Lambda}$ represents 
a correction to the discrete model for comparison 
 with results in the continuum limit $\Lambda \to 1$.
\cite{KWW,BullaCostiPruchke}
We have carried out the iterative 
diagonalization of $H_N$ using the even-odd basis 
defined in Sec.\ \ref{subsec:even_odd}.


\end{document}